\documentclass[11pt]{article}

\usepackage[preprint]{acl}

\definecolor{submission}{HTML}{000099}
\usepackage{times}
\usepackage{latexsym}

\usepackage[T1]{fontenc}
\usepackage[utf8]{inputenc}

\usepackage{microtype}

\usepackage{inconsolata}

\usepackage{graphicx}

\usepackage{multicol}
\usepackage{algorithm}
\usepackage[noend]{algpseudocode}
\usepackage{arydshln}
\usepackage{booktabs}
\usepackage{makecell}
\usepackage{amsmath, amssymb}
\usepackage{multirow}
\usepackage{adjustbox}
\usepackage{tabularx}
\usepackage{enumitem}
\usepackage{mdframed}
\usepackage[table]{xcolor}
\usepackage[most]{tcolorbox}
\usepackage{minted} 
\usepackage{url}

\title{Chain of Retrieval: Multi-Aspect Iterative Search Expansion \\ and Post-Order Search Aggregation for Full Paper Retrieval}%Depth-Aware Search Space Exploitation for Multi-Aspect-Guided Chain of Document to Document Retrieval

\author{
Sangwoo Park\textsuperscript{1} \quad
Jinheon Baek\textsuperscript{1} \quad
Soyeong Jeong\textsuperscript{1} \quad
Sung Ju Hwang\textsuperscript{1,2} \\
\textsuperscript{1}KAIST\quad
\textsuperscript{2}DeepAuto.ai\\
\texttt{\{jackson0530, jinheon.baek, starsuzi, sungju.hwang\}@kaist.ac.kr}
}

\begin{document}
\maketitle

\begin{abstract}
Scientific paper retrieval, particularly framed as document-to-document retrieval, aims to identify relevant papers in response to a long-form query paper, rather than a short query string. Previous approaches to this task have largely focused on abstracts, embedding them into dense vectors as surrogates for full documents and calculating similarity between them. Yet, abstracts offer only sparse and high-level summaries, and such methods primarily optimize one-to-one similarity, overlooking the dynamic relations that emerge across relevant papers during the retrieval process. To address this, we propose Chain of Retrieval (\textsc{CoR}), a novel iterative framework for full-paper retrieval. Specifically, \textsc{CoR} decomposes each query paper into multiple aspect-specific views, matches them against segmented candidate papers, and iteratively expands the search by promoting top-ranked results as new queries, thereby forming a tree-structured retrieval process. The resulting retrieval tree is then aggregated in a post-order manner: descendants are first combined at the query level, then recursively merged with their parent nodes, to capture hierarchical relations across iterations. To validate this, we present \textsc{SciFullBench}, a large-scale benchmark providing both complete and segmented contexts of full papers for queries and candidates, and results show that \textsc{CoR} significantly outperforms existing retrieval baselines. Our code and dataset is available at \url{https://github.com/psw0021/Chain-of-Retrieval-Official}.

\end{abstract}

\section{Introduction}
Information Retrieval (IR) is the task of searching for query-relevant documents from a large external corpus, evolving from sparse keyword matching~\citep{sparck1972tfidf, robertson1995okapi} to dense representation-based similarity~\citep{karpukhin-etal-2020-dense, izacard2021contriever}. Notably, in the era of Large Language Models (LLMs)~\citep{achiam2023gpt4, team2023gemini, Dubey2024llama3, DeepSeekAI2025DeepSeekR1IR}, IR has become increasingly important, which allows LLMs to utilize up-to-date external information~\citep{lewis2020rag}.

In contrast to conventional retrieval tasks, whose queries are short (such as questions or keywords), scientific paper retrieval poses unique challenges. Specifically, queries are structured, long-form documents that encapsulate diverse aspects, ranging from research motivation and proposed methodology to experimental design and empirical findings. Also, relevance in this setting is inherently multi-faceted, as a paper may be considered relevant for various reasons, such as pursuing similar research objectives or employing comparable methods. 

% \begin{figure*}
%     \centering
%     \vspace{-0.1in}
%     \includegraphics[width=0.975\textwidth, ]{Figures/Figures/Concept_Figure.pdf}
%     %\captionsetup{justification=justified, singlelinecheck=false}
%     \vspace{-0.25in}
%     \caption{\small \small Conceptual illustration of our retrieval method compared to prior retrieval approaches.}
%     \label{fig:conceptual_figure_pdf}
%     \vspace{-0.1in}
% \end{figure*}

\begin{figure*}
    \centering
    \includegraphics[width=0.95\textwidth, ]{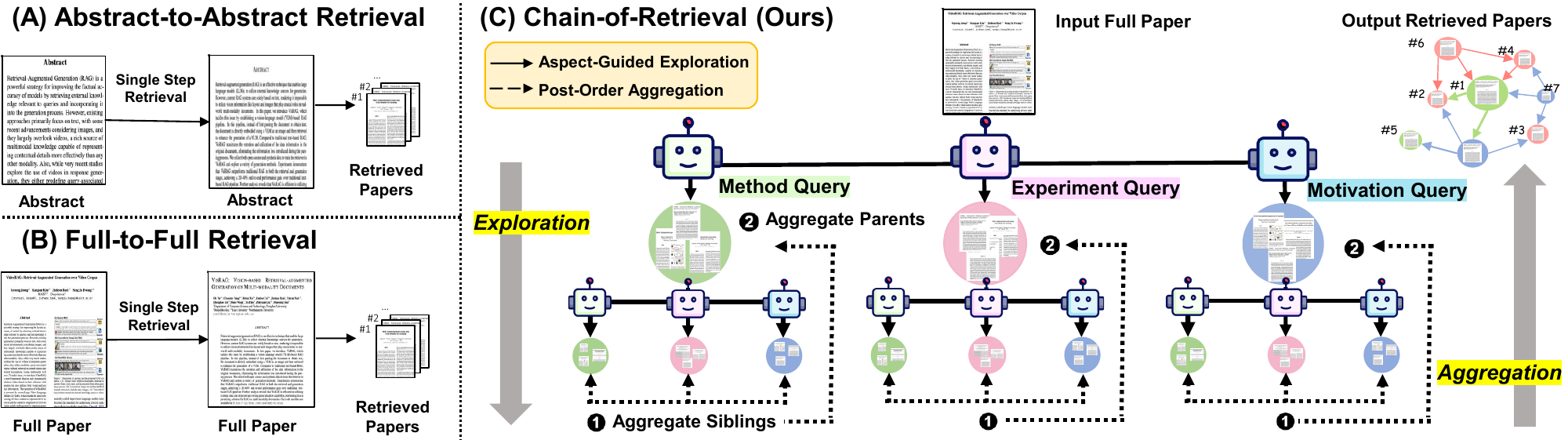}
    %\captionsetup{justification=justified, singlelinecheck=false}
    %\caption{\small \small Conceptual illustration of our retrieval method (C) compared to prior retrieval approaches (A and B).}
    \caption{Conceptual illustration of our retrieval method (C) compared to prior retrieval approaches (A and B).}
    \label{fig:conceptual_figure_pdf}
    \vspace{-0.1in}
\end{figure*}

However, previous work has focused on abstract-level representations as a proxy for full papers~\citep{cohan-etal-2020-specter, yasunaga-etal-2022-linkbert, ostendorff2022neighborhood, singh-etal-2023-scirepeval, zhang-etal-2023-scimult}, fine-tuning models on citation-linked pairs of papers, each represented by its abstract and title. While they are effective to some extent, abstracts are inherently sparse and offer only high-level summaries, making them insufficient to capture the nuanced, multi-faceted relationships between scientific works. As a result, abstract-based approaches are limited to shallow similarity and struggle with tasks requiring deeper contextual understanding of full papers beyond surface-level abstracts, such as identifying complementary works, synthesizing literature reviews, or generating novel research directions~\cite{asai2024openscholar, baek2024researchagent, chamoun-etal-2024-swift, jin2024agentreview, d2024marg}.

Nevertheless, it is non-trivial to represent and retrieve full scientific papers. On the one hand, full papers often exceed 100K tokens in length, far surpassing the context limits of most embedding models. Even if encoded into a single vector, representing its diverse aspects (e.g., motivation, methods, and experiments) within a single embedding may lead to oversimplification and blur fine-grained distinctions, particularly when relevance is tied to a specific aspect. Moreover, existing approaches typically optimize semantic similarity between isolated pairs of papers, overlooking the broader, collective relations that emerge across relevant works during retrieval. This design stands in contrast to human information-seeking behavior in the information foraging theory~\citep{pirolli1995information, pirolli1999information}, which is dynamic and iterative: humans continuously refine their understanding by comparing newly encountered works with prior knowledge.

To address this, we propose Chain of Retrieval (\textsc{CoR}), a novel framework designed to capture dynamic inter-paper relevance while leveraging the full paper context, operated through a two-phase process: \textit{Exploration} and \textit{Aggregation}, illustrated in Figure~\ref{fig:conceptual_figure_pdf}. In the \textit{Exploration} phase, the query paper is decomposed into multiple aspect-specific views, and each aspect is handled by a specialized query optimizer, which is further trained via Direct Preference Optimization (DPO)~\citep{rafailov2023direct} on a self-generated preference to mitigate the shortage of realistic supervision for multi-retrieval systems. After retrieving with these aspect-centric queries, the retrieved (non-duplicate) papers most semantically aligned with the original query are promoted for the next iteration, recursively expanding the search with a tree structure.

In the \textit{Aggregation} phase, results across different depths are merged to capture higher-order relations among retrieved works. Specifically, inspired by the notion of triadic closure in the network theory~\citep{granovetter1973strength}, we design an algorithm that first merges top-$k$ results among sibling nodes (papers retrieved from the same parent), then combines these aggregated results with those of their parent node, and continues this process bottom-up until convergence at the root. This hierarchical aggregation reinforces strong multi-hop connections while naturally attenuating weaker, indirect signals, yielding a robust ranking of relevant papers.

We evaluate \textsc{CoR} by extending existing paper retrieval benchmarks, as they are originally designed for abstract-only retrieval, adapting them to include full papers and more recent publications (to mitigate the potential model contamination) in the ML and NLP domains, which we refer to as \textsc{SciFullBench}. Across experiments on \textsc{SciFullBench}, \textsc{CoR} consistently outperforms retrieval baselines (that either rely on abstracts or naively encode full papers), while remaining compatible with diverse, domain-agnostic embedding models, highlighting both the robustness and generality of our \textsc{CoR}.

\vspace{-0.025in}
\section{Related Work}

\vspace{-0.025in}
\paragraph{Scientific Paper Retrieval} 
Scientific paper retrieval aims to identify relevant works given a query, which may take the form of short queries~\citep{ajith-etal-2024-litsearch}, research proposals~\citep{garikaparthi-etal-2025-mir}, or full papers, and the latter being the focus of this work. Pioneering work in scientific paper-to-paper retrieval used bibliometric statistics, such as influence scores~\citep{zhou2012quantifying, mohapatra2019go}, co-citations~\citep{small1973cocitation, haruna2018citation-based-recommender}, and bibliographic coupling~\citep{kessler1963bibliographic}. Recently, thanks to the capability of neural models, many studies have focused on calculating semantic similarities between abstracts of respective documents~\citep{bhagavatula-etal-2018-content, ostendorff2020contextual}, with the embedding models optimized to this domain. For example, \citet{cohan-etal-2020-specter} and \citet{ostendorff2022neighborhood} fine-tune BERT-based models~\citep{devlin-etal-2019-bert} using abstract pairs extracted from the citation graph, and \citet{mysore-etal-2022-multi} further consider the Wasserstein distance between sentence segments within abstract pairs. Moreover, recent studies target multiple tasks (such as paper classification and citation prediction in addition to paper retrieval) in a unified framework by considering their respective representations~\citep{singh-etal-2023-scirepeval, zhang-etal-2023-scimult}. Complementary to these, hybrid approaches integrate bibliometric signals with vector-based retrieval, reweighting ranks via post-hoc adjustments~\citep{hamedani2016simcc, guo2022hybrid}. However, such statistical indicators (e.g., citations) primarily reflect long-term scholarly accumulation and tend to capture retrospective importance rather than the contextual information expressed in the documents themselves. In contrast, we model paper relations dynamically during inference using aspect-driven queries generated from the full content of scientific papers.

\paragraph{Query Optimization with LLMs} 
Effective formulation of queries is central to retrieval performance, and has long been studied from classical relevance feedback approaches~\citep{rocchio1971relevance, salton1990improving} to query expansion techniques~\citep{kuzi2016query, nogueira2019document}. More recent methods either leverage LLMs themselves to reformulate queries~\citep{yu2023generate, wang-etal-2023-query2doc, gao-etal-2023-precise}, or further augment them with external query-relevant information retrieved in an auxiliary step~\citep{yu2305improving, lameR, park2024conversational, CSQE}. Another line of work seeks to reduce query ambiguity by decomposing a complex query into smaller queries~\citep{zheng2024take, DBLP:conf/ir-rag/KorikovSBKSS24}. We note that our approach aligns with this decomposition paradigm but extends it to a more challenging setting: instead of handling short, user-issued queries, we decompose long-form scientific documents into multiple aspect-specific subqueries.

% \begin{figure*}
%     \centering
%     \includegraphics[width=0.975\textwidth]{Figures/Figures/Main_Figure.pdf}
%     \vspace{-1.0in}
%     %\captionsetup{justification=justified, singlelinecheck=false}
%     \caption{\small \small Illustration of our Chain-of-Retrieval (CoR) framework, consisting of (A) Exploration and (B) Aggregation.}
%     \label{fig:main_figure_pdf}
%     \vspace{-0.05in}
% \end{figure*}

\begin{figure*}
    \centering
    \includegraphics[width=0.95\textwidth]{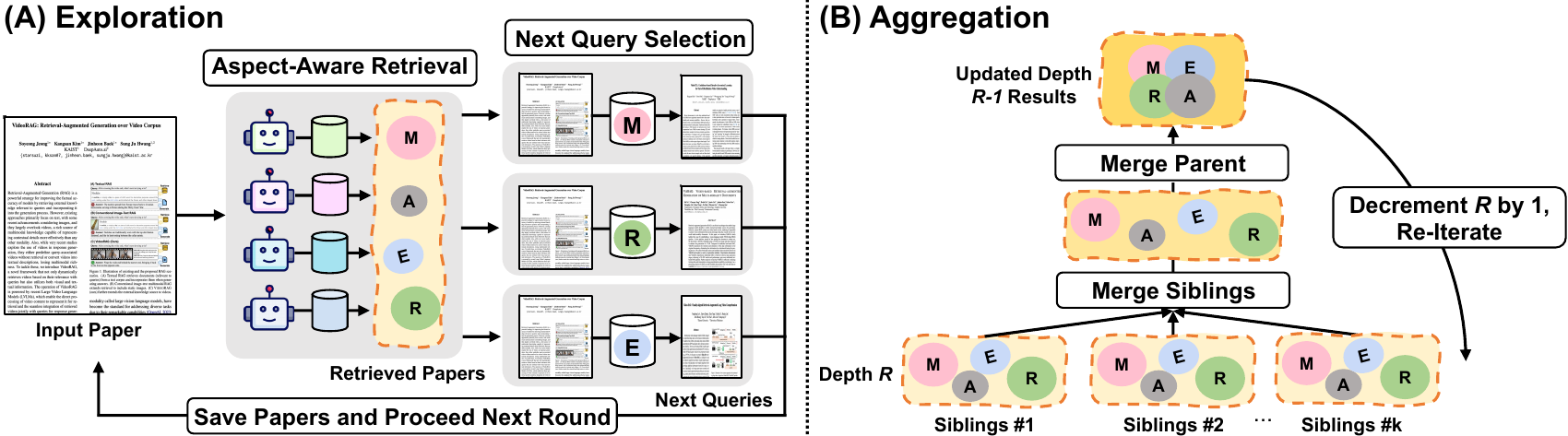}
    %\vspace{-1.0in}
    %\captionsetup{justification=justified, singlelinecheck=false}
    %\caption{\small \small Overview of Chain-of-Retrieval (CoR), consisting of (A) Exploration and (B) Aggregation (formalized in Algorithm~\ref{alg:Main_Algorithm}).}
    \caption{Overview of Chain-of-Retrieval (CoR), consisting of (A) Exploration and (B) Aggregation (Algorithm~\ref{alg:Main_Algorithm}).}
    \label{fig:main_figure_pdf}
    \vspace{-0.1in}
\end{figure*}

\section{Methodology}
We define the paper retrieval problem and present Chain of Retrieval (\textsc{CoR}), illustrated in Figure~\ref{fig:main_figure_pdf}.

\subsection{Preliminary}

\paragraph{Paper-to-Paper Retrieval}
Given a query paper $D$ (with its abstract $D_{\text{abstract}}$), the goal of paper retrieval is to return a ranked list of relevant papers from the corpus $\mathcal{C}$, where relevance could be defined by signals, such as citations, co-citations, and human labels. In contrast to existing studies that use $D_{\text{abstract}}$, we utilize its complete version $D$ for query formulation and corpus construction. 

\subsection{Aspect-Aware Multi-Vector Retrieval}
\label{method:aspect}

\paragraph{Aspect-Aware Query Optimization} 
Scientific papers encapsulate multiple facets, such as research motivation, method design, and experimental validation following~\citet{baek2024researchagent} and~\citet{ScholarEval}, which might not be represented by a single query. To capture this diversity, we formulate the query optimization process that transforms the full paper $D$ into a set of aspect-specific queries. Formally, we define a family of query optimization functions: $\mathcal{F} = \{f_\text{R},f_\text{M}, f_\text{E}\}$, where each function $f$ maps $D$ to a query $q = f(D)$ that targets a specific aspect. Notably, each function is instantiated with an LLM guided by an aspect-specific template (see Appendix~\ref{appendix:Prompts}). In addition to these fine-grained queries, we consider the abstract $D_{\text{abstract}}$, since it reflects a broad view of the overall content (that can complement specialized views), yielding the final query set: $\mathcal{Q} = \{ f_i(D) \; | \; f_i \in \mathcal{F}) \} \cup \{ D_{\text{abstract}} \}$.

\paragraph{Retrieval with Multi-View Corpora} 
Once the optimized query set $\mathcal{Q}$ is formulated, we perform retrieval individually for each query $q \in \mathcal{Q}$. Specifically, for the abstract-based query $D_{\text{abstract}}$, we use a corpus $\mathcal{C}_{\text{abstract}}$ that contains candidate abstracts (as they are comparable in length and structure). For the other aspect-specific queries derived from the full paper, we use a segmented version of the full corpus: $\mathcal{C}_{\text{chunked}}$, where each paper is split into fixed-length chunks (likely to capture its specific aspect) and indexed using multi-vector representations~\citep{khattab2020colbert, santhanam2021colbertv2}. Each query then retrieves its top-$k$ relevant segments, which are mapped back to their source papers via the mapping function $h(x)$, as follows: $\mathcal{R}_{q} = \{ h(x) \; | \; x \in \texttt{Retrieve}(q, \mathcal{C}, k) \}$. Finally, the resulting ranked lists of candidate papers from all queries are aggregated to form the multi-view retrieval pool, denoted as follows: $\mathcal{R} = \bigcup_{q \in \mathcal{Q}} \mathcal{R}_{q}$.

\subsection{Iterative Chain-of-Retrieval}
\label{method:iterative_chain}
\paragraph{Iterative Aspect-Aware Search Expansion} 
The exploration phase aims to iteratively expand the search space by promoting promising retrieved results as new queries. Specifically, starting from the root query paper, we branch into aspect-specific queries (namely, motivation, methods, and experiments), using the process described in Section~\ref{method:aspect}, while excluding the abstract view. Then, each query retrieves a set of top-$k$ candidate papers from the corpus, which serve as the first layer of expansion. At each subsequent depth $r$, every aspect-specific branch selects a single most-relevant representative work from its top-$k$ results and promotes it as the next query paper, ensuring continuity of semantic relevance while avoiding divergence from the original query, and this query paper is further decomposed into aspect-specific queries for the next iteration. At the end, this procedure grows a tree-structured retrieval space: after $r$ iterations, the search expands into $4(3)^{r-1}$ paths, each corresponding to an aspect-driven exploration chain.

To avoid redundancy and ensure efficiency, we store previously selected papers in aspect-specific caches: $\textsc{Cache}[a]$ for $a \in \{\mathrm{R},\mathrm{M}, \mathrm{E}\}$, where each cache records documents with respect to the initial branching aspects of the root query paper, so that any paper stored in the corresponding cache is excluded from future expansions within that branch (to avoid repeated visits to the same works). Further, every retrieved top-$k$ set across depths is preserved in a memory $M$, which later supports aggregation across multiple levels of the search tree. 

%Using the Aspect-Aware document retrieval module \textsc{OneHopRet} from \textbf{Algorithm} \ref{alg:Main_Algorithm}, we iteratively expand the search space for $R$ rounds, selecting single paper from the results of previous rounds for respective aspects $\forall a \in \{\mathrm{M},\mathrm{E}, \mathrm{R}\}$, excluding those of $D_{\text{abstract}}$, where $4(3)^{r-1}$ independent top k candidates are retrieved at depth $r$, forming a tree-like retrieval structure. To choose a single document for subsequent retrieval, from each top $k$ set, our selection algorithm iterates through each each previously retrieved results starting from the $\gamma$-th highest-ranked result and select a single paper with highest rank(closest semantic distance with input query), ensuring that the semantics of searched results do not diverge. Meanwhile, papers already stored in the aspect-aware cache are excluded, to restrain redundant exploration on already-seen papers. The cache, $\textsc{Cache}[a] \gets [\,]\quad \forall a \in \{\mathrm{M},\mathrm{E}, \mathrm{R}\}$ records documents with respect to the initial branching aspects of the root queried document. For example, if the aspect branching path follows Method → Experiment → Research Question, the Method cache is used to save all selected papers during the exploration phase. Moreover, every top k result retrieved from depth $r$ is saved in structured memory $M$, which later is going to be used to merge the retrieved candidates across multiple depths.%

\paragraph{Recursive Post-Order Aggregation} 
\label{method:aggregation}
After $R$ rounds of expansion, we aggregate results in a bottom-up, post-order fashion. Starting from the leaf nodes (i.e., the final retrieved sets at depth $R$), we first merge sibling results that share the same parent using Reciprocal Rank Fusion (RRF)~\cite{Cormack2009ReciprocalRF}, defined as follows: $\text{RRF}(P) = \sum_{q \in \mathcal{Q}} \frac{1}{k + \text{rank}_q(P)}$, where $\text{rank}_q(P)$ denotes the position of paper $P$ in the ranked list retrieved by query $q$, and $k$ is a smoothing constant. This step (\textsc{MergeSiblings} in Algorithm~\ref{alg:Main_Algorithm}) emphasizes documents consistently ranked highly across different aspect queries. The merged sibling sets are then recursively combined upward with their parent results (\textsc{MergeEdges}), gradually consolidating evidence across the retrieval tree until the root.

It is worth noting that, through this recursive process, rank signals from deeper layers are naturally decayed: RRF suppresses the influence of lower-ranked items, and thus papers retrieved only at later depths contribute less to the final ranking. Formally, the effective weight of ranks from depth $R$ decreases at a rate proportional to $\sigma (2\sigma)^{R-1}$, where $\sigma$ is the number of query sets per retrieval. As a result, papers strongly and repeatedly supported across multiple aspects and iterations are reinforced, while weaker or spurious signals are attenuated, yielding a robust ranking for the query.

\begin{figure}[t]
\vspace{-0.1in}
\begin{algorithm}[H]
\caption{Chain-of-Retrieval (\textsc{CoR}).\\Please see Appendix~\ref{appendix:algorithm_details} for the detailed algorithms.}
\label{alg:Main_Algorithm}
\small
\begin{algorithmic}[1]
\Require Root paper $D$; Max depth $R$; Top-$k$ per query $K$; Corpora $\mathcal{C}=\{\mathcal{C}_{\text{abstract}},\mathcal{C}_{\text{chunked}}\}$
\Ensure Final top-$k$ candidates for $D$
\vspace{0.25em}
\State $Q \gets [(D,\textsc{Root})]$
\State $\mathcal{M} \gets [\,]$ 
\State $\textsc{Cache}[a] \gets \emptyset \;\; \forall a \in \{\mathrm{R},\mathrm{M},\mathrm{E}\}$
\vspace{0.25em}
\For{$r = 0$ \textbf{to} $R-1$} \Comment{\textbf{Exploration}}
  \State $\mathcal{M}[r] \gets [\,]$, \; $Q_{\text{next}} \gets [\,]$
  \ForAll{$(d,\textsc{parent}) \in Q$}
    \State $P \gets \Call{OneHopRetrieval}{d,\textsc{parent},\mathcal{C},K}$
    \State $\mathcal{M}[r].\Call{Extend}{P}$
    \State $Q_{\text{next}} \gets \Call{SelectNextQuery}{P,\textsc{Cache},\gamma}$
  \EndFor
  \State $Q \gets Q_{\text{next}}$
\EndFor
\vspace{0.25em}
\For{$r = R-1$ \textbf{downto} $1$} \Comment{\textbf{Aggregation}}
  \State $S \gets \Call{MergeSiblings}{\mathcal{M}[r]}$
  \State $\mathcal{M}[r-1] \gets \Call{MergeEdges}{\mathcal{M}[r-1], S}$
\EndFor
\State \Return $\Call{MergeSiblings}{\mathcal{M}[0]}$
\end{algorithmic}
\end{algorithm}
\vspace{-0.25in}
\end{figure}

%More specific implementation details of our algorithm are in Appendix \ref{appendix:Experiment_Details} and Appendix \ref{appendix:algorithm_details}.

\subsection{Multi-Aspect Preference Optimization}

\paragraph{Offline Policy Exploration} 
To improve the quality of aspect-specific queries, we further train each query-generation agent $f$ with the Direct Preference Optimization (DPO) algorithm~\cite{rafailov2023direct}. A key challenge, however, lies in constructing reliable agent-specific preference datasets, as collecting human-labeled preferences for multi-agent systems is both costly and impractical. Moreover, recent studies emphasize that initializing DPO with the Supervised Fine-Tuned (SFT) model is crucial to mitigate distributional shift issues~\citep{feng2024towards, xu2024cpo, meng2024simpo}.

To address these challenges, we follow the synthetic preference construction procedure of \citet{meng2024simpo}, treating an off-the-shelf instruction-tuned LLM ($\mathcal{P}_\theta$) as the SFT policy ($\mathcal{P}_{\text{SFT}}$). Specifically, for each input paper $D$, we generate a set of $k$ rolled-out candidate queries from every functional agent: $Q_d^{\text{Exp}} = \bigcup_{f \in \{f_\text{R}, f_\text{M}, f_\text{E}\}} \{ q_{(D,f)}^{(1)}, \dots, q_{(D,f)}^{(k)} \}$. After that, among these candidates, we identify the best and worst queries according to a reward function: 
$q^{\text{Best}}_{(D,f)} = \arg\max_{j} \,\texttt{Reward}(q^{(j)}_{(D,f)})$ and $q^{\text{Worst}}_{(D,f)} = \arg\min_{j} \,\texttt{Reward}(q^{(j)}_{(D,f)})$. In addition, a preference pair is retained only if the reward gap between those two exceeds a margin $\tau$, ensuring that $\mathcal{D}_{\text{pref}}$ contains discriminative signals for training.

\paragraph{Reward Formulation} 
Designing an effective reward function is crucial for preference optimization, as it determines how query-generation agents are guided to improve. Our principle is to provide reward signals that are reliable and reproducible, cost-efficient to compute, and directly aligned with retrieval quality; thus, instead of relying on qualitative judgments from LLMs (LLMs-as-judge), which are costly and potentially inconsistent, we leverage citation links as environmental feedback, providing direct, objective reward signals (where we use Recall@$k$ in practice). Also, for each agent $f \in \{f_\text{R}, f_\text{M}, f_\text{E}\}$, we define an independent reward function $\texttt{Reward}_f$, such that the reward for one agent does not depend on the outputs of others, under the assumption that reinforcing the quality of aspect-specific queries individually is sufficient, since the final retrieval results are aggregated across agents via a linear combination of their outputs.

\section{Experiments}

\subsection{Experimental Setup}
We describe here the setup used to validate our method. Further implementation details and training procedures are in Appendix~\ref{appendix:Training Procedures} and Appendix~\ref{appendix:Experiment_Details}.

\paragraph{Datasets}
To evaluate full paper-to-paper retrieval, we build upon and extend existing abstract-only benchmarks, which we refer to as \textsc{SciFullBench}. Specifically, it preserves the original retrieval formulation and relevance definition, while augmenting each query-candidate pair with full-text content and refreshing the corpus with recent publications. Notably, we also evaluate \textsc{CoR} under abstract-only configurations on \textsc{SciFullBench} (Table~\ref{tab:abstract-to-abstract retrieval generalization SciFullBench.}) and on an existing benchmarks that use other relevance signals (Tables~\ref{tab:SciDocs_Cocite} and~\ref{tab:CFSCube_results}), as well as in the full-paper-to-full-paper variant of the same benchmark (Table~\ref{tab:SciDocs_Full}), to ensure that our gains are not specific to the full-paper scenario and generalize across retrieval regimes under diverse relevance signals, on which \textsc{CoR} consistently outperforms baselines.

To build \textsc{SciFullBench}, we follow prior work on scientific paper retrieval~\citep{singh-etal-2023-scirepeval, medic-snajder-2022-MDCR, cohan-etal-2020-specter} and define relevance using neighboring relationships in the academic citation graph, including both incoming citations and outgoing references. Query papers are collected from major ML (NeurIPS, ICLR) and NLP (ACL, EMNLP) venues using OpenReview API\footnote{\url{https://openreview.net/}}, ACL-Anthology\footnote{\url{https://aclanthology.org/}}, and SEA~\cite{yu-etal-2024-automated}. We then retain papers with at least ten relevant neighbors and randomly sample 400 query papers per venue. For the retrieval corpus, we collect CS-related papers from arXiv (2020-2025)\footnote{\url{https://arxiv.org/}}, where full text is publicly available, resulting in approximately 40K papers. Lastly, we post-process papers by removing reference sections, filtering citation markers, and eliminating their in-text mentions, to avoid hindsight leakage. Further dataset construction details are provided in Appendix~\ref{appendix:SciFullBench}.

\begin{table*}[t!]
%\caption{Main results of various domain-specific and domain-agnostic retrievers on \textsc{SciFullBench}.}
%\vspace{-0.1in}
\renewcommand{\arraystretch}{1.5}
\resizebox{\linewidth}{!}{
\begin{tabular}{llcccccccc}

\midrule \toprule \rule{0pt}{3.5ex}%
& & \multicolumn{4}{c}{\fontsize{22pt}{22pt}\selectfont\ \bf ICLR-NeurIPS} & \multicolumn{4}{c}{\fontsize{22pt}{22pt}\selectfont\ \bf ACL-EMNLP} \\
\cmidrule(l{2pt}r{2pt}){3-6} \cmidrule(l{2pt}r{2pt}){7-10}
\rule{0pt}{3.5ex}%
& & \multicolumn{2}{c}{\fontsize{22pt}{22pt}\selectfont \bf References}& \multicolumn{2}{c}{\fontsize{22pt}{22pt}\selectfont\ \bf Citations} 
  & \multicolumn{2}{c}{\fontsize{22pt}{22pt}\selectfont \bf References} & \multicolumn{2}{c}{\fontsize{22pt}{22pt}\selectfont \bf Citations} \\
\cmidrule(l{2pt}r{2pt}){3-4} \cmidrule(l{2pt}r{2pt}){5-6}
\cmidrule(l{2pt}r{2pt}){7-8} \cmidrule(l{2pt}r{2pt}){9-10}
& \fontsize{22pt}{20pt}\selectfont\ \textbf{Methods} 
& \fontsize{20pt}{20pt}\selectfont\ nDCG@300 & 
\fontsize{20pt}{20pt}\selectfont Recall@300 & 
\fontsize{20pt}{20pt}\selectfont\ nDCG@300 &
\fontsize{20pt}{20pt}\selectfont\ Recall@300 & 
\fontsize{20pt}{20pt}\selectfont\ nDCG@300 & 
\fontsize{20pt}{20pt}\selectfont\ Recall@300 & 
\fontsize{20pt}{20pt}\selectfont\ nDCG@300 & 
\fontsize{20pt}{20pt}\selectfont\ Recall@300 \\
\midrule
\noalign{\vskip 0.5ex} 
& \multicolumn{9}{l}{\textbf{\fontsize{20pt}{20pt}\selectfont Lexical-Based Retrievers}} \\
\rule{0pt}{2ex}%
& \fontsize{20pt}{20pt}\selectfont\ BM-25 (A2A) & \fontsize{21pt}{21pt}\selectfont\ 28.29  & \fontsize{21pt}{21pt}\selectfont\ 38.21 & \fontsize{21pt}{21pt}\selectfont\ 27.52 & \fontsize{21pt}{21pt}\selectfont\ 37.41 & \fontsize{21pt}{21pt}\selectfont\ 21.02 & \fontsize{21pt}{21pt}\selectfont\ 30.79 & \fontsize{21pt}{21pt}\selectfont\ 20.86 & \fontsize{21pt}{21pt}\selectfont\ 29.37 \\
& \fontsize{20pt}{20pt}\selectfont\ BM-25 (F2F) & \fontsize{21pt}{21pt}\selectfont\ 26.34 & \fontsize{21pt}{21pt}\selectfont\ 37.02 & \fontsize{21pt}{21pt}\selectfont\ 36.48 & \fontsize{21pt}{21pt}\selectfont\ 47.47 & \fontsize{21pt}{21pt}\selectfont\ 23.38 & \fontsize{21pt}{21pt}\selectfont\ 35.10 & \fontsize{21pt}{21pt}\selectfont\ 28.08 & \fontsize{21pt}{21pt}\selectfont\ 38.58 \\
\midrule\midrule
& \multicolumn{9}{l}{\textbf{\fontsize{22pt}{22pt}\selectfont Domain-Specific Retrievers (\textbf{A2A})}} \\
\rule{0pt}{2ex}%
& \fontsize{20pt}{20pt}\selectfont\ SciNCL & \fontsize{21pt}{21pt}\selectfont\ 36.24 & \fontsize{21pt}{21pt}\selectfont\ 51.80 & \fontsize{21pt}{21pt}\selectfont\ 33.07 & \fontsize{21pt}{21pt}\selectfont\ 47.90 & \fontsize{21pt}{21pt}\selectfont\ 26.12 & \fontsize{21pt}{21pt}\selectfont\ 40.71 & \fontsize{21pt}{21pt}\selectfont\ 25.58 & \fontsize{21pt}{21pt}\selectfont\ 38.91 \\

& \fontsize{20pt}{20pt}\selectfont\ SPECTER2-Base & \fontsize{21pt}{21pt}\selectfont\ 35.24 & \fontsize{21pt}{21pt}\selectfont\ 50.41 & \fontsize{21pt}{21pt}\selectfont\ 34.48 & \fontsize{21pt}{21pt}\selectfont\ 49.09 & \fontsize{21pt}{21pt}\selectfont\ 25.07 & \fontsize{21pt}{21pt}\selectfont\ 39.02 & \fontsize{21pt}{21pt}\selectfont\ 26.85 & \fontsize{20pt}{20pt}\selectfont\ 40.21 \\

& \fontsize{20pt}{20pt}\selectfont\ SPECTER2-Adapter-MTL CTRL & \fontsize{21pt}{21pt}\selectfont\ 36.22 & \fontsize{20pt}{20pt}\selectfont\ 51.07 & \fontsize{21pt}{21pt}\selectfont\ 33.12 & \fontsize{20pt}{20pt}\selectfont\ 47.38 & \fontsize{21pt}{21pt}\selectfont\ 25.32 & \fontsize{20pt}{20pt}\selectfont\ 38.86 & \fontsize{21pt}{21pt}\selectfont\ 25.48 & \fontsize{20pt}{20pt}\selectfont\ 38.54 \\

& \fontsize{20pt}{20pt}\selectfont\ SciMult-MHAExpert & \fontsize{21pt}{21pt}\selectfont\ 30.95 & \fontsize{21pt}{21pt}\selectfont\ 45.04 & \fontsize{21pt}{21pt}\selectfont\ 28.32 & \fontsize{21pt}{21pt}\selectfont\ 41.07 & \fontsize{21pt}{21pt}\selectfont\ 23.11 & \fontsize{21pt}{21pt}\selectfont\ 36.12 & \fontsize{21pt}{21pt}\selectfont\ 22.57 & \fontsize{21pt}{21pt}\selectfont\ 33.79 \\

\midrule\midrule
\noalign{\vskip 0.5ex} 
& \multicolumn{9}{l}{\textbf{\fontsize{20pt}{20pt}\selectfont Jina-Embeddings-v2-BASE-EN}} \\
\multirow{11}{*}{\raisebox{-9ex}[0pt][0pt]{\rotatebox[origin=c]{90}{\textbf{\fontsize{25pt}{25pt}\selectfont Domain-Agnostic Retrievers}}}} & 
\fontsize{20pt}{20pt}\selectfont\ Abstract-to-Abstract (A2A)& \fontsize{20pt}{21pt}\selectfont\ 36.78  & \fontsize{21pt}{21pt}\selectfont\ 49.92 & \fontsize{21pt}{21pt}\selectfont\ 33.85 & \fontsize{21pt}{21pt}\selectfont\ 47.89 & \fontsize{21pt}{21pt}\selectfont\ 24.96 & \fontsize{21pt}{21pt}\selectfont\ 37.90 & \fontsize{21pt}{21pt}\selectfont\ 26.45 & \fontsize{21pt}{21pt}\selectfont\ 38.92 \\

& \fontsize{20pt}{20pt}\selectfont\ Full-to-Full (F2F) & \fontsize{21pt}{21pt}\selectfont\ 37.05 & \fontsize{21pt}{21pt}\selectfont\ 50.60 & \fontsize{21pt}{21pt}\selectfont\ 35.80 & \fontsize{21pt}{21pt}\selectfont\ 49.63 & \fontsize{21pt}{21pt}\selectfont\ 27.55 & \fontsize{21pt}{21pt}\selectfont\ 41.69 & \fontsize{21pt}{21pt}\selectfont\ 28.34 & \fontsize{21pt}{21pt}\selectfont\ 40.65 \\

\noalign{\vskip 0.25ex}\cdashline{2-10}\noalign{\vskip 1.5ex}
& \cellcolor{blue!5} \fontsize{20pt}{20pt}\selectfont\ \textbf{\textsc{CoR} w/ Llama-3.2-3B-Instruct} & \cellcolor{blue!5} \fontsize{21pt}{21pt}\selectfont\ \textbf{38.88} & \cellcolor{blue!5} \fontsize{21pt}{21pt}\selectfont\ \textbf{55.50} & \cellcolor{blue!5} \fontsize{21pt}{21pt}\selectfont\ \textbf{38.58}  & \cellcolor{blue!5} \fontsize{21pt}{21pt}\selectfont\ \textbf{56.21} & \cellcolor{blue!5} \fontsize{21pt}{21pt}\selectfont\ \textbf{28.34} & \cellcolor{blue!5} \fontsize{21pt}{21pt}\selectfont\ \textbf{44.44} & \cellcolor{blue!5} \fontsize{21pt}{21pt}\selectfont\ \textbf{31.80} & \cellcolor{blue!5} \fontsize{21pt}{21pt}\selectfont\ \textbf{47.03} \\

& \cellcolor{blue!5} \fontsize{20pt}{20pt}\selectfont\ \textbf{\textsc{CoR} w/ QWEN-2.5-3B-Instruct}   & \cellcolor{blue!5} \fontsize{21pt}{21pt}\selectfont\ \textbf{38.91} & \cellcolor{blue!5} \fontsize{21pt}{21pt}\selectfont\ \textbf{55.78} & \cellcolor{blue!5} \fontsize{21pt}{21pt}\selectfont\ \textbf{39.36} & \cellcolor{blue!5} \fontsize{21pt}{21pt}\selectfont\ \textbf{56.97} & \cellcolor{blue!5} \fontsize{21pt}{21pt}\selectfont\ \textbf{28.21} & \cellcolor{blue!5} \fontsize{21pt}{21pt}\selectfont\ \textbf{44.83} & \cellcolor{blue!5} \fontsize{21pt}{21pt}\selectfont\ \textbf{32.54} & \cellcolor{blue!5} \fontsize{21pt}{21pt}\selectfont\ \textbf{48.30} \\

\noalign{\vskip 0.75ex}\cline{2-10}\noalign{\vskip 1.5ex} 
&\multicolumn{9}{l}{\textbf{\fontsize{20pt}{20pt}\selectfont BGE-M3 }} \\
\rule{0pt}{2ex}%
& \fontsize{20pt}{20pt}\selectfont\ Abstract-to-Abstract (A2A) & \fontsize{21pt}{21pt}\selectfont\ 32.08 & \fontsize{21pt}{21pt}\selectfont\ 44.09 & \fontsize{21pt}{21pt}\selectfont\ 30.31 & \fontsize{21pt}{21pt}\selectfont\ 42.85 & \fontsize{21pt}{21pt}\selectfont\ 22.95 & \fontsize{21pt}{21pt}\selectfont\ 34.55 & \fontsize{21pt}{21pt}\selectfont\ 24.41 & \fontsize{21pt}{21pt}\selectfont\ 36.28 \\

& \fontsize{20pt}{20pt}\selectfont\ Full-to-Full (F2F) & \fontsize{21pt}{21pt}\selectfont\ 33.71 & \fontsize{21pt}{21pt}\selectfont\ 45.84 & \fontsize{21pt}{21pt}\selectfont\ 32.22 & \fontsize{21pt}{21pt}\selectfont\ 44.01 & \fontsize{21pt}{21pt}\selectfont\ 24.13 & \fontsize{21pt}{21pt}\selectfont\ 35.88 & \fontsize{21pt}{21pt}\selectfont\ 24.85 & \fontsize{21pt}{21pt}\selectfont\ 35.87 \\

\noalign{\vskip 0.25ex}\cdashline{2-10}\noalign{\vskip 1.5ex}
& \cellcolor{blue!5} \fontsize{20pt}{20pt}\selectfont\ \textbf{\textsc{CoR} w/ Llama-3.2-3B-Instruct} & \cellcolor{blue!5} \fontsize{21pt}{21pt}\selectfont\ \textbf{33.81} & \cellcolor{blue!5} \fontsize{21pt}{21pt}\selectfont\ \textbf{49.39} & \cellcolor{blue!5} \fontsize{21pt}{21pt}\selectfont\ \textbf{34.54} & \cellcolor{blue!5} \fontsize{21pt}{21pt}\selectfont\ \textbf{49.27} & \cellcolor{blue!5} \fontsize{21pt}{21pt}\selectfont\ \textbf{24.62} & \cellcolor{blue!5} \fontsize{21pt}{21pt}\selectfont\ \textbf{39.22} & \cellcolor{blue!5} \fontsize{21pt}{21pt}\selectfont\ \textbf{28.87} & \cellcolor{blue!5} \fontsize{21pt}{21pt}\selectfont\ \textbf{42.67} \\

& \cellcolor{blue!5} \fontsize{20pt}{20pt}\selectfont\ \textbf{\textsc{CoR} w/ QWEN-2.5-3B-Instruct} & \cellcolor{blue!5} \fontsize{21pt}{21pt}\selectfont\ \textbf{34.84}  & \cellcolor{blue!5} \fontsize{21pt}{21pt}\selectfont\ \textbf{50.43} & \cellcolor{blue!5} \fontsize{21pt}{21pt}\selectfont\ \textbf{35.72} & \cellcolor{blue!5} \fontsize{21pt}{21pt}\selectfont\ \textbf{50.77} & \cellcolor{blue!5} \fontsize{21pt}{21pt}\selectfont\ \textbf{25.22} & \cellcolor{blue!5} \fontsize{21pt}{21pt}\selectfont\ \textbf{39.84} & \cellcolor{blue!5} \fontsize{21pt}{21pt}\selectfont\
\textbf{29.33} & \cellcolor{blue!5} \fontsize{21pt}{21pt}\selectfont\ \textbf{43.72} \\

\noalign{\vskip 0.75ex}\cline{2-10}\noalign{\vskip 1.5ex} 
&\multicolumn{9}{l}{\textbf{\fontsize{20pt}{20pt}\selectfont Inf-Retriever-v1-1.5B }} \\
\rule{0pt}{2ex}%
& \fontsize{20pt}{20pt}\selectfont\ Abstract-to-Abstract (A2A) & \fontsize{21pt}{21pt}\selectfont\ 45.84 & \fontsize{21pt}{21pt}\selectfont\ 61.72 & \fontsize{21pt}{21pt}\selectfont\ 38.61 & \fontsize{21pt}{21pt}\selectfont\ 54.63 & \fontsize{21pt}{21pt}\selectfont\ 33.43 & \fontsize{21pt}{21pt}\selectfont\ 50.10 & \fontsize{21pt}{21pt}\selectfont\ 30.94 & \fontsize{21pt}{21pt}\selectfont\ 45.53 \\

& \fontsize{20pt}{20pt}\selectfont\ Full-to-Full (F2F) & \fontsize{21pt}{21pt}\selectfont\ 31.09 & \fontsize{21pt}{21pt}\selectfont\ 42.27 & \fontsize{21pt}{21pt}\selectfont\ 35.58 & \fontsize{21pt}{21pt}\selectfont\ 48.79 & \fontsize{21pt}{21pt}\selectfont\ 32.86 & \fontsize{21pt}{21pt}\selectfont\ 48.00 & \fontsize{21pt}{21pt}\selectfont\ 30.31 & \fontsize{21pt}{21pt}\selectfont\ 43.64 \\

\noalign{\vskip 0.25ex}\cdashline{2-10}\noalign{\vskip 1.5ex}
& \cellcolor{blue!5} \fontsize{20pt}{20pt}\selectfont\ \textbf{\textsc{CoR} w/ Llama-3.2-3B-Instruct} & \cellcolor{blue!5} \fontsize{21pt}{21pt}\selectfont\ \textbf{47.31}  & \cellcolor{blue!5} \fontsize{21pt}{21pt}\selectfont\ \textbf{65.31} & \cellcolor{blue!5} \fontsize{21pt}{21pt}\selectfont\ \textbf{43.66} & \cellcolor{blue!5} \fontsize{21pt}{21pt}\selectfont\ \textbf{61.54} & \cellcolor{blue!5} \fontsize{21pt}{21pt}\selectfont\ \textbf{34.48} & \cellcolor{blue!5} \fontsize{21pt}{21pt}\selectfont\ \textbf{53.63} & \cellcolor{blue!5} \fontsize{21pt}{21pt}\selectfont\ \textbf{36.18} & \cellcolor{blue!5} \fontsize{21pt}{21pt}\selectfont\ \textbf{52.95} \\

& \cellcolor{blue!5} \fontsize{20pt}{20pt}\selectfont\ \textbf{\textsc{CoR} w/ QWEN-2.5-3B-Instruct} & \cellcolor{blue!5} \fontsize{21pt}{21pt}\selectfont\ \textbf{47.21} & \cellcolor{blue!5} \fontsize{21pt}{21pt}\selectfont\ \textbf{64.78} & \cellcolor{blue!5} \fontsize{21pt}{21pt}\selectfont\ \textbf{44.21} & \cellcolor{blue!5} \fontsize{21pt}{21pt}\selectfont\ \textbf{62.06} & \cellcolor{blue!5} \fontsize{21pt}{21pt}\selectfont\ \textbf{34.41} & \cellcolor{blue!5} \fontsize{21pt}{21pt}\selectfont\ \textbf{53.30} & \cellcolor{blue!5} \fontsize{21pt}{21pt}\selectfont\ \textbf{36.87} & \cellcolor{blue!5} \fontsize{21pt}{21pt}\selectfont\ \textbf{53.59} \\

\midrule
\bottomrule
\end{tabular}
}
\vspace{-0.1in}
\caption{Main results of various domain-specific and domain-agnostic retrievers on \textsc{SciFullBench}.}
\label{tab:main_final}
\vspace{-0.18in}
\end{table*}

\paragraph{Retrieval Models}
We compare our method with existing retrievers developed for scientific paper retrieval, as follows: \textbf{SPECTER2 Base}~\citep{singh-etal-2023-scirepeval}; \textbf{SciNCL} \citep{ostendorff2022neighborhood}; \textbf{SciMult-MHAExpert}~\citep{zhang-etal-2023-scimult}; \textbf{SPECTER2 Adapters + MTL CTRL}~\citep{singh-etal-2023-scirepeval}. We also consider off-the-shelf general-purpose embedding models, such as \textbf{Jina-Embeddings-V2-Base-EN}~\citep{gunther2023jina-embeddings}, \textbf{BGE-M3}~\citep{bge-m3}, and \textbf{Inf-Retriever-v1-1.5B}~\citep{infly-ai_2025} as our primary experiments, while also utilizing \textbf{Granite-Embeddings-English-R2}~\citep{awasthy2025graniteembeddingmodels}, \textbf{QWEN3-Embedding-0.6B}~\citep{qwen3embedding}, and \textbf{Dewey-EN-Beta}~\citep{zhang2025deweylongcontextembedding} to broaden evaluation in Appendix \ref{appendix:Supplementary_Experiments} and \ref{appendix:Robustness_Experiments}.

\paragraph{Retrieval Units}
We consider various retrieval units, where we denote \textbf{A} for \textit{Abstract}, \textbf{F} for \textit{Full paper}\footnote{Since full papers often exceed the context length of embedding models, we truncate them up to the maximum length.}, and \textbf{C} for \textit{chunked context} (i.e., segmented units of the full paper, each capped at 3K tokens). Using them, we consider four retrieval setups: \textbf{A2A} (Abstract-to-Abstract), \textbf{F2F} (Full-to-Full), \textbf{A2C} (Abstract-to-Chunk), and \textbf{F2C} (Full-to-Chunk).

\paragraph{Query Optimizers}
We instantiate query optimizers using LLMs: the open-weight instruct models Llama-3.2-3B-Instruct~\cite{meta2024llama3.2} and QWEN-2.5-3B-Instruct~\cite{qwen2, qwen2.5}. For reproducibility, we set the temperature to 0 and the maximum generation tokens to 2,000. 

\paragraph{Training and Inference Configuration} 
For each query, we retrieve the top-300 candidate papers. To prevent repeatedly selecting the same candidate as the next query, we always skip the highest-ranked results (selected previously) and instead promote the next-best. During preference pair construction, we generate 16 candidate queries per aspect and evaluate them based on Recall@30, while excluding pairs whose reward margin is smaller than 3\%.

\begin{figure}[t!]
    \centering
    \includegraphics[width=0.975\columnwidth]{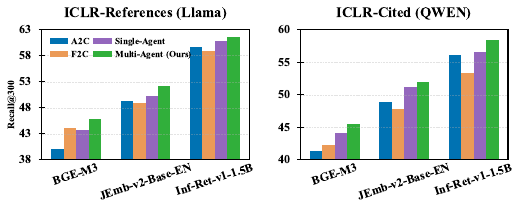}
    \captionsetup{justification=justified, singlelinecheck=false}
    \vspace{-0.1in}
    %\caption{\small Retrieval performance comparing different query formulations against our multi-agent \textsc{CoR} framework (after a single depth of retrieval and with untrained query optimizers).}
    \caption{Retrieval performance of different query formulations against multi-agent \textsc{CoR} framework after one retrieval depth, using untrained query optimizers.}
    \label{fig:effect_of_multiagent_query_optimizer}
    \vspace{-0.17in}
\end{figure}

\subsection{Experimental Results and Analysis}
\label{main:experiment}
We now provide an in-depth analysis of the experimental results. Please refer to Appendix \ref{appendix:Supplementary_Experiments} and Appendix \ref{appendix:Robustness_Experiments} for more experiments and analyses.

\paragraph{Main Results}
Table~\ref{tab:main_final} presents the main results, where \textsc{CoR} outperforms all baselines across various settings, validating the effectiveness of our proposed framework for full paper-to-paper retrieval\footnote{We report results on lower @K values and other relevance signals in Appendix~\ref{appendix:Robustness_Experiments}, where \textsc{CoR} yields consistent gains.}. Notably, when using the same domain-agnostic retriever, \textsc{CoR} surpasses abstract-to-abstract (A2A) baselines by an average of 6.37\% in Recall, demonstrating that simply relying on abstracts is suboptimal compared to our aspect-driven approach. Also, interestingly, general-purpose embedding models such as Jina-Embeddings-V2-Base-EN and BGE-M3 (whose A2A performance lags behind representative domain-specific retrievers) achieve competitive or superior performance once integrated into \textsc{CoR}, emphasizing that \textsc{CoR} delivers strong performance regardless of the retrievers by serving as a plug-and-play framework. Lastly, when compared against F2F retrieval with long-context embedding models, \textsc{CoR} achieves average gains of 7.82\% in Recall, confirming that simply embedding entire papers is insufficient, and that structured, iterative search expansion with post-order aggregation is crucial for effective full paper-to-paper retrieval.

\begin{figure}[t!]
    \centering
    \includegraphics[width=0.975\columnwidth]{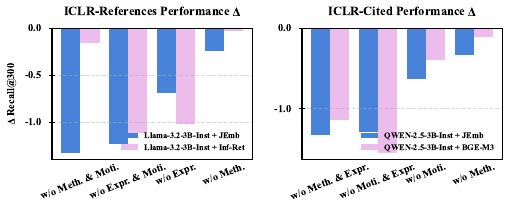}
    \captionsetup{justification=justified, singlelinecheck=false}
    \vspace{-0.120in}
    %\caption{\small Change ($\Delta$) in retrieval performance (relative to using all aspects) when excluding individual scientific aspects (Moti = Motivation, Meth = Method, Expr = Experiment).}
    \caption{Performance change ($\Delta$) when excluding each scientific aspect, relative to using all aspects (Moti = Motivation, Meth = Method, Expr = Experiment).}
    \label{fig:effect_of_multiple_aspects}
    \vspace{-0.2in}
\end{figure}

\paragraph{Effectiveness of Specialized Query Optimizers}
Our results in Figure~\ref{fig:effect_of_multiagent_query_optimizer} confirm the effectiveness of using multiple specialized LLM agents, each dedicated to a single aspect. More specifically, using multiple specialized agents outperforms a single base agent tasked with generating multiple queries (without aspect separation). Also, when contrasted with setups that use either full papers or abstracts as queries (matched with the same chunked corpus) and that retrieve the same number of total candidates as \textsc{CoR} of a single retrieval round, our multi-aspect queries generated with specialized agents achieve gains of 3.36\% and 3.33\%, respectively, highlighting the advantage of decomposing noisy, monolithic queries into aspect-aware subqueries.

\begin{table}[t!]
%\captionsetup{justification=justified, singlelinecheck=false}
%\caption{\small Performance comparison with and without the Aspect-Aware Cache in the selection algorithm, reported after retrieval depths of three using DPO-trained query optimizers. We
%report the mean and standard deviation over three independent runs, and \underline{underline} statistically significant improvements.}
%\vspace{-0.1in}
%\label{tab:effect-of-aspect-aware cache}
\resizebox{\linewidth}{!}{
\renewcommand{\arraystretch}{0.95}
\begin{tabular}{llcc}
\toprule
& & \textbf{ICLR-References} & \textbf{ICLR-Citations} \\
\cmidrule(lr){3-3} \cmidrule(lr){4-4}
& \textbf{\{Optimizer\} + \{Retriever\}} & Recall@300 & Recall@200 \\
\midrule
\multicolumn{4}{l}{\textbf{Llama-3.2-3B-Inst + JEmb-v2}} \\
&w/o Aspect-Aware Cache & 53.63 $\pm$  \textbf{\scriptsize{0.11}} & 46.80 $ \pm$ \textbf{\scriptsize{0.03}} \\
& \textbf{w/ Aspect-Aware Cache} & \textbf{54.15} $\pm$ \textbf{\scriptsize{0.39}} & \underline{\textbf{47.08} $\pm$ \textbf{\scriptsize{0.04}}}\\
\midrule
\multicolumn{4}{l}{\textbf{Llama-3.2-3B-Inst + BGE-M3}} \\
&w/o Aspect-Aware Cache & 46.88 $\pm$ \textbf{\scriptsize{0.16}} & 40.82 $\pm$ \textbf{\scriptsize{0.18}} \\
& \textbf{w/ Aspect-Aware Cache} & \underline{\textbf{47.35} $\pm$ \textbf{\scriptsize{0.13}}} & \textbf{41.16} $\pm$ \textbf{\scriptsize{0.27}} \\
\midrule
\multicolumn{4}{l}{\textbf{Llama-3.2-3B-Inst + Inf-Ret-v1}} \\
&w/o Aspect-Aware Cache & 62.65 $\pm$ \textbf{\scriptsize{0.10}} & 51.83 $\pm$ \textbf{\scriptsize{0.05}} \\
&\textbf{w/ Aspect-Aware Cache} & \underline{\textbf{62.98} $\pm$ \textbf{\scriptsize{0.19}}} & \underline{\textbf{52.31} $\pm$ \textbf{\scriptsize{0.11}}} \\
\bottomrule
\end{tabular}
}
\vspace{-0.1in}
\captionsetup{justification=justified, singlelinecheck=false}
%\caption{\small Performance comparison with and without the Aspect-Aware Cache in the selection algorithm, reported after retrieval depths of three using DPO-trained query optimizers. We report the mean and standard deviation over three independent runs, and \underline{underline} statistically significant improvements.}
\caption{Effect of the Aspect-Aware Cache after three retrieval depths, using DPO-trained query optimizers. We report the mean ± std over three runs; \underline{underline} indicates statistically significant improvements.}
\label{tab:effect-of-aspect-aware cache}
\vspace{-0.18in}
\end{table}

\paragraph{Importance of Multi-Aspect Coverage}
To examine the role of aspect diversity in query formulation, we analyze how the retrieval performance changes as we vary the set of participating query optimizer agents. As shown in Figure~\ref{fig:effect_of_multiple_aspects}, excluding any single aspect leads to measurable performance degradation, which supports the necessity of capturing the multifaceted nature of scientific papers. Not all aspects, however, contribute equally: excluding experimental or research-motivation queries causes larger drops in performance than omitting methodological queries. We attribute this to the inherent characteristics of each aspect. Specifically, experimental and research-motivation views often yield stronger semantic overlap across related papers, grounded in shared entities such as tasks, baselines, and datasets. In contrast, methodological details are more heterogeneous and less entity-driven, demanding deeper semantic abstraction and reasoning to identify commonalities.
\vspace{-0.03in}
\paragraph{Role of Aspect-Aware Cache}
Table~\ref{tab:effect-of-aspect-aware cache} shows the benefits of incorporating the aspect-aware cache into our \textit{Next Query Selection} strategy (described in Section~\ref{method:iterative_chain}), which enables preventing redundant exploration within each aspect branch and thus encouraging more diverse search trajectories. From this, \textsc{CoR} with the aspect-aware cache consistently outperforms those without it, demonstrating its effectiveness in steering the retrieval process toward broader coverage and stronger overall performance.

\paragraph{Efficacy of Query Preference Optimization}
To assess the contribution of reinforcing aspect-aware query optimizers with DPO, we conduct an ablation study isolating the DPO training. As shown in Table~\ref{tab:DPO_Ablation}, integrating DPO consistently improves performance across diverse LLM-retriever configurations in both the reference and citation setups. Notably, although the preference pairs are constructed in the Reference (outgoing-links) setting using Recall as the reward signal, the gains transfer to other evaluation setups of Citation (incoming-links).
\begin{table}[t!]
%\captionsetup{justification=justified, singlelinecheck=false}
%\caption{\small Performance comparison of specialized agents trained with DPO versus untrained query optimizers, evaluated after a single round of retrieval (i.e., one depth). We report the mean and standard deviation across three independent trials, and \underline{underline} statistically significant improvements.}
%\vspace{-0.1in}
\centering
\resizebox{\linewidth}{!}{
\renewcommand{\arraystretch}{0.90}
\begin{tabular}{clcccc}
\toprule
& & \multicolumn{2}{c}{\textbf{ICLR-References}} & \multicolumn{2}{c}{\textbf{NeurIPS-Citations}} \\
\cmidrule(lr){3-4} \cmidrule(lr){5-6}
& \textbf{\{Optimizer\} + \{Retriever\}} 
& Recall@100 & Recall@300 
& Recall@100 & Recall@300 \\
\midrule

\multicolumn{6}{l}{\textbf{Llama-3.2-3B-Inst + JEmb-v2}} \\
& w/o DPO      & 36.20 $\pm$ \textbf{\scriptsize{0.00}} & 52.14 $\pm$ \textbf{\scriptsize{0.00}} & 41.70 $\pm$ \textbf{\scriptsize{0.00}} & 57.36 $\pm$ \textbf{\scriptsize{0.00}}\\
& \textbf{w/ DPO}   & \underline{\textbf{37.10} $\pm$ \textbf{\scriptsize{0.12}}} & \underline{\textbf{52.60} $\pm$ \textbf{\scriptsize{0.05}}} & 41.33 $\pm$ \textbf{\scriptsize{0.06}} & \underline{\textbf{57.95} $\pm$ \textbf{\scriptsize{0.10}}} \\
\midrule

\multicolumn{6}{l}{\textbf{Llama-3.2-3B-Inst + BGE-M3}} \\
& w/o DPO      & 31.32 $\pm$ \textbf{\scriptsize{0.00}} &  45.61 $\pm$ \textbf{\scriptsize{0.00}} & 36.24 $\pm$ \textbf{\scriptsize{0.00}} & 50.85 $\pm$ \textbf{\scriptsize{0.00}} \\
& \textbf{w/ DPO}   & \underline{\textbf{31.59} $\pm$ \textbf{{\scriptsize{0.02}}}} & \underline{\textbf{45.91} $\pm$  \textbf{{\scriptsize{0.02}}}} & \underline{\textbf{36.55} $\pm$ \textbf{{\scriptsize{0.07}}}} & 50.49 $\pm$ \textbf{{\scriptsize{0.09}}} \\
\midrule

\multicolumn{6}{l}{\textbf{Llama-3.2-3B-Inst + Inf-Ret-v1}} \\
& w/o DPO      & 45.15 $\pm$ \textbf{\scriptsize{0.05}} & 61.75 $\pm$ \textbf{\scriptsize{0.05}} & 47.91 $\pm$ \textbf{\scriptsize{0.00}} & 63.04 $\pm$ \textbf{\scriptsize{0.00}}\\
& \textbf{w/ DPO}   & \underline{\textbf{45.92} $\pm$ \textbf{\scriptsize{0.06}}} & \underline{\textbf{62.15} $\pm$ \textbf{\scriptsize{0.06}}} & \underline{\textbf{48.57} $\pm$ \textbf{\scriptsize{0.05}}} & \underline{\textbf{63.61} $\pm$ \textbf{\scriptsize{0.07}}} \\
\midrule

\multicolumn{6}{l}{\textbf{QWEN-2.5-3B-Inst + JEmb-v2}} \\
& w/o DPO      & 36.72 $\pm$ \textbf{\scriptsize{0.00}} &  52.43 $\pm$ \textbf{\scriptsize{0.00}} & 41.82 $\pm$ \textbf{\scriptsize{0.00}} & 58.20 $\pm$ \textbf{\scriptsize{0.00}} \\
& \textbf{w/ DPO}   & \underline{\textbf{37.41} $\pm$ \textbf{\scriptsize{0.17}}} & \underline{\textbf{53.00} $\pm$ \textbf{\scriptsize{0.10}}} & \underline{\textbf{42.21} $\pm$ \textbf{\scriptsize{0.15}}} & \underline{\textbf{58.64} $\pm$ \textbf{\scriptsize{0.12}}}\\
\midrule

\multicolumn{6}{l}{\textbf{QWEN-2.5-3B-Inst + BGE-M3}} \\
& w/o DPO      & 30.94 $\pm$ \textbf{\scriptsize{0.00}} & 45.68 $\pm$ \textbf{\scriptsize{0.00}} & 37.09 $\pm$ \textbf{\scriptsize{0.00}} & 51.23 $\pm$ \textbf{\scriptsize{0.00}}\\
& \textbf{w/ DPO}   & \underline{\textbf{31.52} $\pm$ \textbf{\scriptsize{0.14}}} & \underline{\textbf{46.60} $\pm$ \textbf{\scriptsize{0.04}}} & \underline{\textbf{37.75} $\pm$ \textbf{\scriptsize{0.15}}} & \underline{\textbf{52.08} $\pm$ \textbf{\scriptsize{0.21}}}\\
\bottomrule
\end{tabular}
}
\vspace{-0.1in}
\captionsetup{justification=justified, singlelinecheck=false}
%\caption{\small Performance comparison of specialized agents trained with DPO versus untrained query optimizers, evaluated after a single round of retrieval (i.e., one depth). We report the mean and standard deviation across three independent trials, and \underline{underline} statistically significant improvements.}
\caption{Performance of \textsc{CoR} with DPO-trained specialized optimizers and untrained query optimizers after one retrieval depth. We report the mean ± std over three runs; \underline{underline} indicates statistically significant improvements.}
\label{tab:DPO_Ablation}
\vspace{-0.09in}
\end{table}

\begin{figure}[t!]
    \centering
        \includegraphics[width=0.975\columnwidth]{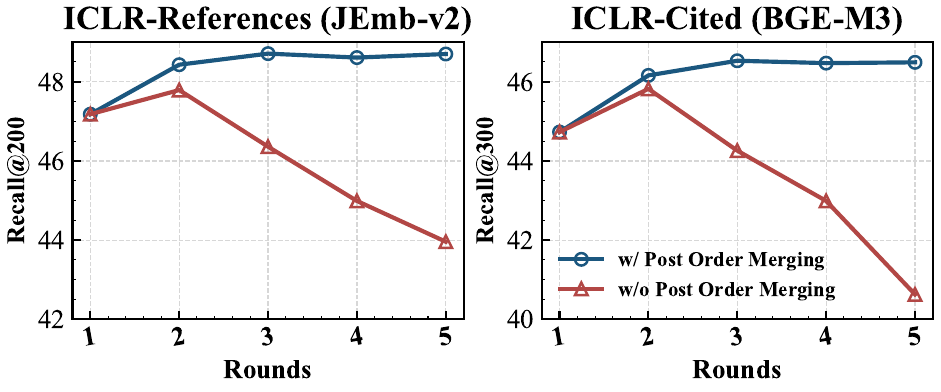}
    \captionsetup{justification=justified, singlelinecheck=false}
    \vspace{-0.15in}
    %\caption{\small Retrieval results as a function of retrieval depths, with DPO-trained Llama-3.2-3B-Instruct as optimizers.}
    \caption{Retrieval performance across depths using DPO-trained Llama-3.2-3B-Instruct query optimizers.}
    \label{fig:RetrievalRoundAnalysis}
    \vspace{-0.2in}
\end{figure}

\paragraph{Analysis on Post-Order Aggregation}
To verify whether our aggregation mechanism is responsible for performance improvements beyond those obtained from multi-aspect expansion, we conduct an analysis. As shown in Figure~\ref{fig:RetrievalRoundAnalysis}, \textsc{CoR} yields performance gains across successive retrieval rounds, with the gap over baselines widening as depth increases. This confirms the effectiveness of our post-order aggregation, which progressively attenuates weak signals from distant connections and thereby enhances robustness against noise in deeper rounds. By contrast, the baseline strategy (merging results from all branches at once) fails to adaptively penalize such signals, leading to steadily diminishing performance. However, improvements plateau beyond depth 3, which suggests diminishing returns, likely due to added noise with deeper expansions.

\begin{table}[t!]
%\captionsetup{justification=justified, singlelinecheck=false}
%\caption{\small Analysis on the semantic coverage of the Top-300 retrieved documents over ground truth documents with Convex Hull Volume. The results of \textsc{CoR} are reported after 3 rounds (i.e., depth) with Llama-3.2-3B-Inst as query optimizers.}
%\vspace{-0.1in}
%\label{tab:embedding_diversity_analysis}
\resizebox{\linewidth}{!}{
\renewcommand{\arraystretch}{0.87}
\begin{tabular}{llcc}
\toprule
& & \textbf{ACL (Citations)} & \textbf{NeurIPS (Citations)} \\
\cmidrule(lr){3-3} \cmidrule(lr){4-4}
& \textbf{\{Type\} w/ \{Retriever\}} & CHV Ratio ($\Delta$) & CHV Ratio ($\Delta$) \\
\midrule
\multicolumn{4}{l}{\textbf{Relative Coverage over A2A}} \\
& F2F w/ JEmb-v2 & 1.126 (+12.6\%) & 1.135 (+13.5\%) \\
& F2F w/ BGE-M3 & 1.088 (+8.8\%) & 1.069 (+6.9\%) \\
& F2F w/ Inf-Ret-v1 & 1.128 (+12.8\%) & 1.097 (+9.7\%) \\
\midrule
\midrule
\multicolumn{4}{l}{\textbf{Relative Coverage over A2A}} \\
& \textbf{\textsc{CoR} w/ JEmb-v2} & \textbf{1.190 (+19.0\%)} & \textbf{1.198 (+19.8\%)} \\
&\textbf{\textsc{CoR} w/ BGE-M3} & \textbf{1.107 (+10.7\%)} & \textbf{1.097 (+9.7\%)} \\
& \textbf{\textsc{CoR} w/ Inf-Ret-v1} & \textbf{1.167 (+16.7\%)} & \textbf{1.140 (+14.0\%)}\\
\midrule
\midrule
\multicolumn{4}{l}{\textbf{Relative Coverage over F2F}} \\
& \textbf{\textsc{CoR} w/ JEmb-v2} & \textbf{1.095 (+9.5\%)} & \textbf{1.078 (+7.8\%)} \\
& \textbf{\textsc{CoR} w/ BGE-M3} & \textbf{1.044 (+4.4\%)} & \textbf{1.042 (+4.2\%)} \\
& \textbf{\textsc{CoR} w/ Inf-Ret-v1} & \textbf{1.060 (+6.0\%)} & \textbf{1.072 (+7.2\%)} \\

\bottomrule
\end{tabular}
}
\vspace{-0.1in}
\captionsetup{justification=justified, singlelinecheck=false}
%\caption{\small Analysis on the semantic coverage of the Top-300 retrieved documents over ground truth documents with Convex Hull Volume. The results of \textsc{CoR} are reported after 3 rounds (i.e., depth) with Llama-3.2-3B-Inst as query optimizers.}
\caption{Semantic coverage of the Top-300 retrieved documents relative to ground truth, measured by Convex Hull Volume. \textsc{CoR} results are reported at retrieval depth of three, using Llama-3.2-3B-Instruct optimizers.}
\label{tab:embedding_diversity_analysis}
\vspace{-0.15in}
\end{table}

\paragraph{Diversity Analysis}
Recall that \textsc{CoR} is designed to expand the search space iteratively across multiple aspects of the query paper, and we further analyze this by measuring the semantic coverage of retrieved candidates over ground-truth in the embedding space using the Convex Hull Volume Ratio (CHV Ratio), which compares coverage between two configurations, with the same set (see Appendix~\ref{appendix:metrics} for details). As shown in Table~\ref{tab:embedding_diversity_analysis}, \textsc{CoR} consistently improves the diversity of semantic coverage among the retrieved documents: relative to the baseline A2A setup, it achieves a 14.98\% gain in CHV Ratio, surpassing the 10.72\% gain of the F2F setup. Moreover, when directly compared against F2F, \textsc{CoR} still provides an additional 6.52\% improvement. These results highlight that beyond improving retrieval accuracy, the proposed \textsc{CoR} enables broader semantic exploration, which allows discovering relevant papers that would likely be missed under the vanilla retrieval strategies. 

\paragraph{Generalization to Other Domains}
To examine the generality of \textsc{CoR} beyond scientific paper retrieval, we further evaluate its performance on a different but challenging document-to-document retrieval task: patent-to-patent retrieval, which has substantially long queries (please see Appendix~\ref{appendix:patent-to-patent retrieval} for details on task design and implementation). As shown in Table~\ref{tab:patent-to-patent retrieval}, \textsc{CoR} delivers robust gains over both the domain-agnostic and domain-specific baselines, surpassing F2F retrieval by 2.01\% and A2A retrieval by 6.96\%. Importantly, these improvements are achieved without any training (i.e., using the off-the-shelf GPT-4o for query optimization), highlighting the versatility and broad applicability of \textsc{CoR} for long document-to-document retrieval.

\begin{table}[t!]
%\caption{\small Experimental results on patent-to-patent retrieval using \textsc{PatentFullBench} (self-constructed; See Appendix~\ref{appendix:patent-to-patent retrieval}). Results for \textsc{CoR} are reported after retrieval depth of two.}
%\captionsetup{justification=justified, singlelinecheck=false}
%rounds (i.e., depth) of retrieval.}
%\vspace{-0.1in}
%\label{tab:patent-to-patent retrieval}
\resizebox{\linewidth}{!}{
\renewcommand{\arraystretch}{0.85}
\begin{tabular}{llcc}
\toprule
& & \textbf{References} & \textbf{Citations} \\
\cmidrule(lr){3-3} \cmidrule(lr){4-4}
& \textbf{Methods} & Recall@100 & Recall@100 \\
\midrule
\multicolumn{4}{l}{\textbf{Baseline}} \\
&PAT-SPECTER (A2A)~\citep{ghosh2024paecter} & 47.36 & 52.16 \\
&PaECTER (A2A)~\citep{ghosh2024paecter} & 52.77 & 56.25 \\
\midrule
&BGE-M3 (A2A) & 46.42 & 49.57 \\
&BGE-M3 (F2F) & 52.30 & 55.81 \\
\noalign{\vskip 1mm}
\cdashline{1-4}
\noalign{\vskip 1mm}
&Inf-Ret-v1 (A2A) & 52.78 & 57.94 \\
&Inf-Ret-v1 (F2F) & 56.23 & 58.59 \\
\noalign{\vskip 1mm}
\cdashline{1-4}
\noalign{\vskip 1mm}
&JEmb-v2-Base-EN (A2A) & 49.39 & 53.27 \\
&JEmb-v2-Base-EN (F2F) & 56.57 & 59.61 \\
\midrule
\midrule
\multicolumn{4}{l}{\textbf{\textsc{CoR} (Ours): \{Optimizer\} + \{Retriever\}}} \\
& \textbf{GPT-4o-Mini-2024-0718} + \textbf{BGE-M3} & \textbf{54.71} & \textbf{56.78} \\
& \textbf{GPT-4o-Mini-2024-0718} + \textbf{Inf-Ret-v1} & \textbf{58.62} & \textbf{63.06} \\
& \textbf{GPT-4o-Mini-2024-0718} + \textbf{JEmb-v2} & \textbf{57.23} & \textbf{60.74} \\

\bottomrule
\end{tabular}
}
\vspace{-0.1in}
%\caption{\small Experimental results on patent-to-patent retrieval using \textsc{PatentFullBench} (self-constructed; See Appendix~\ref{appendix:patent-to-patent retrieval}). Results for \textsc{CoR} are reported after retrieval depth of two.}
\caption{Patent-to-patent retrieval results on \textsc{PatentFullBench} (self-constructed; See Appendix~\ref{appendix:patent-to-patent retrieval}), with \textsc{CoR} evaluated at retrieval depth of two.}
\captionsetup{justification=justified, singlelinecheck=false}
\label{tab:patent-to-patent retrieval}
\vspace{-0.07in}

\end{table}
\begin{table}[t!]
%\captionsetup{justification=justified, singlelinecheck=false}
%\caption{\small Human study on the relevance of ground truth candidates. The results are reported after 3 rounds (i.e., depth) using DPO-trained QWEN-2.5-3B-Instruct models for query optimizer, and jina-embeddings-v2-base-en model for retriever.}
%\vspace{-0.1in}
\setlength{\tabcolsep}{15pt}
\resizebox{\linewidth}{!}{
\renewcommand{\arraystretch}{0.6}
\begin{tabular}{l c}
\toprule
{\textbf{Retrieval Methods}} & \textbf{Precision@5}  \\
\midrule

% \multicolumn{3}{l}{\textbf{Baseline}} \\
Abstract-to-Abstract (A2A) & 60.83 \\

Full-to-Full (F2F) & 53.33 \\

\noalign{\vskip 0.25ex}\cdashline{1-2}\noalign{\vskip 1.00ex}

% \multicolumn{3}{l}{\textbf{Ours}} \\
\textbf{Chain-of-Retrieval  (\textsc{CoR}; Ours)} & \textbf{64.17} \\

\bottomrule
\end{tabular}
}
\vspace{-0.1in}
\captionsetup{justification=justified, singlelinecheck=false}
%\caption{\small Human study on the relevance of ground truth candidates. The results are reported after 3 rounds (i.e., depth) using DPO-trained QWEN-2.5-3B-Instruct models for query optimizer, and jina-embeddings-v2-base-en model for retriever.}
\caption{Human evaluation of ground-truth relevance of \textsc{CoR} at retrieval depth of three, using DPO-trained Qwen-2.5-3B-Instruct optimizers and jina-embeddings-v2-base-en retrievers.}
\vspace{-0.2in}
\label{tab:Human_Evaluation}
\end{table}

\paragraph{Human Evaluation}
To further assess the quality of retrieved results beyond the citation-based ground truth in \textsc{SciFullBench}, we conduct a human evaluation. A total of 15 participants evaluate the relevance of the Top-5 retrieved candidates for three methods (A2A, F2F, \textsc{CoR}) across 15 query papers, assigning one of four labels to each candidate (motivation-relevant, method-relevant, experiment-relevant, or irrelevant). We then compute per-query precision by treating any of the three aspect labels as relevant and the remaining label as irrelevant. As shown in Table~\ref{tab:Human_Evaluation}, \textsc{CoR} achieves the highest precision, surpassing A2A and F2F by 3.34\% and 10.84\%, respectively, indicating that its improvements extend beyond citation-based metrics to human-perceived relevance. To ensure the reliability of human judgments, we measure inter-annotator agreement on it with Cohen’s Kappa coefficient~\citep{cohen1960coefficient}, obtaining a score of 0.43, which corresponds to moderate agreement and supports the reliability of our human evaluation. For further details, please refer to Appendix~\ref{appendix:Human_Evaluation_Setup}.

\vspace{-0.05in}
\section{Conclusion}
\vspace{-0.075in}
In this work, we introduced Chain of Retrieval (in short, \textsc{CoR}), a novel framework for paper-to-paper retrieval that iteratively expands the search space via aspect-aware query optimization (reinforced with DPO training) and recursively merges results via post-order aggregation. For evaluation, we presented \textsc{SciFullBench}, a large-scale dataset for full-context scientific retrieval, and results show that \textsc{CoR} consistently outperforms the abstract-level and full-context baselines, even with off-the-shelf embedding models and for a patent retrieval task, highlighting its robustness and generality.

\section*{Limitations}
While our work introduces a novel retrieval approach that iteratively expands the search space via multi-aspect query optimization over the full context of scientific papers, it still has room for future work. First, an adaptive branch pruning mechanism could be further explored to dynamically guide the search based on the semantic alignment of individual papers with the root query, which would be an interesting direction for future work. Second, while our framework retrieves diverse and relevant papers with computational efficiency comparable to baseline approaches, such as those that use raw full papers as queries and candidates, further reducing its computational overhead would be a valuable direction, which we leave as future work.

\section*{Ethical Considerations}
Although our Chain of Retrieval (\textsc{CoR}) framework improves retrieval performance over prior approaches, it can still retrieve irrelevant or misleading papers, which may propagate incorrect or harmful information to both human users and downstream AI models. To ensure the development of trustworthy automated systems, future work should incorporate robust verification or re-ranking mechanisms that can effectively filter irrelevant or potentially harmful documents from the retrieved set.

\bibliography{custom}
%\appendix

\clearpage
%\appendix

%\begin{center}
    %\textbf{\large Appendix Table of Contents}
%\end{center}
\section*{Appendix Table of Contents}

\vspace{0.5em}

\noindent\hyperref[appendix:SciFullBench]{\textbf{A}\hspace{0.8em} \textsc{SciFullBench}}

\hspace{0.5em}\hyperref[appendix:SciFullBenchConstruction]{\textbf{A.1}\hspace{0.8em} Construction Procedure}

\hspace{0.5em}\hyperref[appendix:SciFullBench_Statistics_Analysis]{\textbf{A.2}\hspace{0.8em} Benchmark Statistics}

\hspace{0.5em}\hyperref[appendix:comparison_with_prior_benchmarks]{\textbf{A.3}\hspace{0.8em} Comparison with Prior Benchmarks}

\noindent\hyperref[appendix:Training Procedures]{\textbf{B}\hspace{0.8em} Training Procedures}

\hspace{0.5em}\hyperref[appendix:query_corpus_construction]{\textbf{B.1}\hspace{0.8em} Query and Corpus Construction}

\hspace{0.5em}\hyperref[appendix:preference_set_construction]{\textbf{B.2}\hspace{0.8em} Preference Set Construction}

\hspace{0.5em}\hyperref[appendix:train_details_and_hyperparameters]{\textbf{B.3}\hspace{0.8em} Training Details and Hyper-Parameters}

\noindent\hyperref[appendix:Experiment_Details]{\textbf{C}\hspace{0.8em} Experiment Details}

\hspace{0.5em}\hyperref[appendix:Inference Settings]{\textbf{C.1}\hspace{0.8em} Inference Settings}

\hspace{0.5em}\hyperref[appendix:Models]{\textbf{C.2}\hspace{0.8em} Models}

\hspace{0.5em}\hyperref[appendix:metrics]{\textbf{C.3}\hspace{0.8em} Metrics}

\hspace{0.5em}\hyperref[appendix:Human_Evaluation_Setup]{\textbf{C.4}\hspace{0.8em} Human Evaluation Setup}

\noindent\hyperref[appendix:patent-to-patent retrieval]{\textbf{D}\hspace{0.8em} Patent-to-Patent Retrieval}

\hspace{0.5em}\hyperref[appendix:patentfullbench_construction]{\textbf{D.1}\hspace{0.8em} PatentFullBench Construction Workflow}

\hspace{0.5em}\hyperref[appendix:patent_experiment_details]{\textbf{D.2}\hspace{0.8em} Experiment Details}

\noindent\hyperref[appendix:Supplementary_Experiments]{\textbf{E}\hspace{0.8em} Supplementary Experiments}

\noindent\hyperref[appendix:Robustness_Experiments]{\textbf{F}\hspace{0.8em} Robustness across Benchmarks and Metrics}

\noindent\hyperref[appendix:algorithm_details]{\textbf{G}\hspace{0.6em} Algorithm Details}

\noindent\hyperref[appendix:Case_Study]{\textbf{H}\hspace{0.6em} Case Study}

\noindent\hyperref[appendix:HumanEvaluation]{\textbf{I}\hspace{1.0em} Human Evaluation Guideline}

\noindent\hyperref[appendix:Prompts]{\textbf{J}\hspace{1.0em} Prompts}

\clearpage

\appendix

\section{\textsc{SciFullBench}}
\label{appendix:SciFullBench}

In this section, we provide more details on the construction process of \textsc{SciFullBench}, its detailed statistics, and comparison against existing benchmarks. 

\subsection{Construction Procedure}
\label{appendix:SciFullBenchConstruction}
\captionsetup{justification=centering}
\begin{table*}[t!]

\centering
%\caption{\small Statistics of query papers in the \textsc{SciFullBench} benchmark.}
%\label{tab:query_paper_details}
%\vspace{-0.05in}
\resizebox{\textwidth}{!}{
\renewcommand{\arraystretch}{1.5}
\begin{tabular}{c cc cc cc cc c}
\toprule
\textbf{} 
& \multicolumn{2}{c}{\fontsize{15pt}{15pt}\selectfont\textbf{ICLR}} 
& \multicolumn{2}{c}{\fontsize{15pt}{15pt}\selectfont\textbf{NeurIPS}} 
& \multicolumn{2}{c}{\fontsize{15pt}{15pt}\selectfont\textbf{EMNLP}} 
& \multicolumn{2}{c}{\fontsize{15pt}{15pt}\selectfont\textbf{ACL}} \\

& \fontsize{15pt}{15pt}\selectfont References & \fontsize{15pt}{15pt}\selectfont Citations & \fontsize{15pt}{15pt}\selectfont References & \fontsize{15pt}{15pt}\selectfont Citations & \fontsize{15pt}{15pt}\selectfont References & \fontsize{15pt}{15pt}\selectfont Citations & \fontsize{15pt}{15pt}\selectfont References & \fontsize{15pt}{15pt}\selectfont Citations \\
\midrule

\fontsize{15pt}{15pt}\selectfont domain
& \multicolumn{2}{c}{\fontsize{14pt}{14pt}\selectfont ML} 
& \multicolumn{2}{c}{\fontsize{14pt}{14pt}\selectfont ML} 
& \multicolumn{2}{c}{\fontsize{14pt}{14pt}\selectfont ML/CL} 
& \multicolumn{2}{c}{\fontsize{14pt}{14pt}\selectfont ML/CL} \\

\fontsize{15pt}{15pt}\selectfont years included
& \multicolumn{2}{c}{\fontsize{14pt}{14pt}\selectfont 2024,2025} 
& \multicolumn{2}{c}{\fontsize{14pt}{14pt}\selectfont 2023,2024} 
& \multicolumn{2}{c}{\fontsize{14pt}{14pt}\selectfont 2023,2024} 
& \multicolumn{2}{c}{\fontsize{14pt}{14pt}\selectfont 2023,2024} \\

\fontsize{15pt}{15pt}\selectfont Maximum \# of Ground Truth Candidates
& \fontsize{14pt}{14pt}\selectfont 62 & \fontsize{14pt}{14pt}\selectfont 750 & \fontsize{14pt}{14pt}\selectfont 133 & \fontsize{14pt}{14pt}\selectfont 763 & \fontsize{14pt}{14pt}\selectfont 160 & \fontsize{14pt}{14pt}\selectfont 583 & \fontsize{14pt}{14pt}\selectfont 60 & \fontsize{14pt}{14pt}\selectfont 320 \\

\fontsize{15pt}{15pt}\selectfont \# of papers per year
& \fontsize{14pt}{14pt}\selectfont (141,259) & \fontsize{14pt}{14pt}\selectfont (324,76) & \fontsize{14pt}{14pt}\selectfont (170, 230) & \fontsize{14pt}{14pt}\selectfont (353,47) & \fontsize{14pt}{14pt}\selectfont (226, 174) & \fontsize{14pt}{14pt}\selectfont (334, 66) & \fontsize{14pt}{14pt}\selectfont (184,216) & \fontsize{14pt}{14pt}\selectfont (256,144)\\

\fontsize{15pt}{15pt}\selectfont \# of tokens per paper
& \fontsize{14pt}{14pt}\selectfont 9402.35 & \fontsize{14pt}{14pt}\selectfont 8183.37 & \fontsize{14pt}{14pt}\selectfont 11184.39 & \fontsize{14pt}{14pt}\selectfont 9035.43 & \fontsize{14pt}{14pt}\selectfont 5908.36 & \fontsize{14pt}{14pt}\selectfont 6226.45 & \fontsize{14pt}{14pt}\selectfont 6190.06 & \fontsize{14pt}{14pt}\selectfont 6614.6 \\

\fontsize{15pt}{15pt}\selectfont average \# of candidates per sample 
& \fontsize{14pt}{14pt}\selectfont 21.74 & \fontsize{14pt}{14pt}\selectfont 42.25 & \fontsize{14pt}{14pt}\selectfont 24.58 & \fontsize{14pt}{14pt}\selectfont 40.37 & \fontsize{14pt}{14pt}\selectfont 19.64 & \fontsize{14pt}{14pt}\selectfont 39.24 & \fontsize{14pt}{14pt}\selectfont 18.05 & \fontsize{14pt}{14pt}\selectfont 33.89 \\

\fontsize{15pt}{15pt}\selectfont minimum \# of candidates per sample 
& \fontsize{14pt}{14pt}\selectfont 10 & \fontsize{14pt}{14pt}\selectfont 10 & \fontsize{14pt}{14pt}\selectfont 10 & \fontsize{14pt}{14pt}\selectfont 10 & \fontsize{14pt}{14pt}\selectfont 10 & \fontsize{14pt}{14pt}\selectfont 10 & \fontsize{14pt}{14pt}\selectfont 10 & \fontsize{14pt}{14pt}\selectfont 10\\

\bottomrule

\end{tabular}
}
\vspace{-0.05in}
%\caption{\small Statistics of query papers in the \textsc{SciFullBench} benchmark.}
\caption{Statistics of query papers in the \textsc{SciFullBench} benchmark.}
\label{tab:query_paper_details}
\vspace{-0.05in}
\end{table*}
\begin{table}[t!]
\centering
\setlength{\tabcolsep}{20pt}
%\caption{\small Statistics of the corpus in the \textsc{SciFullBench}.}
%\label{tab:corpus-statistics}
%\vspace{-0.05in}
\resizebox{\linewidth}{!}{
\renewcommand{\arraystretch}{1.5}
\begin{tabular}{l c}
\toprule
\textbf{}
& \fontsize{15pt}{15pt}\selectfont \textsc{SciFullBench} \\
\midrule
\fontsize{15pt}{15pt}\selectfont Total \# of papers
& \fontsize{14pt}{14pt}\selectfont 40,782 \\
\fontsize{15pt}{15pt}\selectfont Avg.\ \# of tokens per abstract
& \fontsize{14pt}{14pt}\selectfont 200.39 \\
\fontsize{15pt}{15pt}\selectfont Avg.\ \# of tokens per paper
& \fontsize{14pt}{14pt}\selectfont 6987.67 \\
\fontsize{15pt}{15pt}\selectfont Total \# of segmented corpus
& \fontsize{14pt}{14pt}\selectfont 115,044 \\
\fontsize{15pt}{15pt}\selectfont Avg.\ \# of tokens per segment
& \fontsize{14pt}{14pt}\selectfont 2479.90 \\
\bottomrule
\end{tabular}
}
\vspace{-0.05in}
%\caption{\small Statistics of the corpus in the \textsc{SciFullBench}.}
\caption{Corpus Statistics of \textsc{SciFullBench}.}
\captionsetup{justification=justified, singlelinecheck=false}
\label{tab:corpus-statistics}
\vspace{-0.1in}
\end{table}

\paragraph{Step 1: Collect Data}
We collected research papers from leading machine learning venues including ICLR 2024, 2025, NeurIPS 2024, 2023, ACL 2024, 2023, and EMNLP 2024, 2023 for our query documents. We scraped existing publications from the main conference for ACL and EMNLP, including both long- and short- papers from the ACL-Anthology website\footnote{\url{https://aclanthology.org/}}, and used the OpenReview\footnote{\url{https://openreview.net/}} API to obtain submitted research papers for ICLR and NeurIPS, using a subset of documents uploaded by authors from SEA~\cite{yu-etal-2024-automated}, which includes both accepted and rejected papers from respective venues. Afterwards, we obtain metadata from the arXiv database uploaded on the Kaggle\footnote{\url{https://www.kaggle.com/}} website, with papers from 2020 to January of 2025 that belong in the machine learning domain. %Also, we only filtered papers that are in the machine learning domain, to those belong in cs.AI, cs.LG, cs.CL, cs.CV, cs.NE, cs.IR, cs.DS, cs.CC, cs.DL, cs.HC, cs.RO, cs.MM, cs.CG, cs.SY.

\paragraph{Step 2: Build a Raw Dataset}
We subsequently collect the information of papers that cite our potential query documents using the Semantic Scholar API\footnote{\url{https://www.semanticscholar.org}} and check the existence of a paper %with a matched title (case-insensitive) 
in the arxiv database, and rule out any potential query documents that had fewer than ten remaining candidates after such process. We incorporate citation proximity signals as a proxy for relevance signals between scientific papers, following established practices in scientific paper retrieval~\citep{cohan-etal-2020-specter, singh-etal-2023-scirepeval}.
We then use Allen AI Science Parse\footnote{\url{https://github.com/allenai/science-parse}} released under Apache 2.0 license to obtain reference information for respective documents, and filter documents based on the implemented criteria from above. Using such data, we formulate our raw (query document, gold candidate) pairs where its abstract information is intact and where more than 10 deduplicated candidates exist, organized by splits in respective years belonging to each venue. For each split, we aggregate all filtered (query, candidate) set from each venue without distinction of its published year. 
%For example, we merge all pairs in the citation (in-coming links) split for ICLR 2024 and ICLR 2025 into a unified set of ICLR-citations split. 
From this filtered pool, we randomly sampled 500 potential queries from each split and formulated a corpus containing all ground truth candidates of sampled queries, keeping our corpus at manageable size. Next, we collected the pdf files of the papers, %Since arXiv APIs do not support large-scale requests,
using the Google Cloud API\footnote{\url{https://cloud.google.com/apis?hl=ko}} to access the full paper dump directly and collect the latest updated versions for each paper. %by matching its unique ids within the arXiv database. 

\paragraph{Step 3: Filter and Sanitize Dataset}
Using the Allen AI Science Parse tool, we parse the query and candidate pdf files collected in Step 2 into a json format file containing title, abstract, main content list, and %containing dictionaries with header and contents as keys, and 
reference section along with its inline-citation mentions for both queries and candidates. Since we use citation information as the main signal for measuring document similarities, excluding the in-line citations and the reference section was crucial to validate the fairness of our benchmark. Although our parsing tool generally separates reference from main content, reference information is frequently intermixed within the main content. 

Thus, we applied an auxiliary filtering algorithm to remove such issues. Since our tool parses a pdf file into a structured json file, we were able to obtain text information segmented into multiple passages. Hence, prior to reformatting it back to complete text, we removed sections or headers that contain at least one reference title information. To mitigate concerns regarding the severe loss of information from reformatted query papers, we excluded documents in which such problematic passages made up the majority. 
Furthermore, we ensured to remove any residing reference title information within the formatted text.

\paragraph{Step 4: Finalize Benchmark}
Based on the preprocessed full document, we formulate a pseudo-definitive benchmark consisting of (query, ground-truth candidate), and target corpus set by random sampling 400 query documents per split in each venue that still 10 gold candidates residing subsequent to the filtering procedure. Ultimately, we formulate a total of 3200 query documents, along with large-scale target corpora consisting of all the gold candidates of such queries. Finally, we remove citations and interlinked mentions from the entire set of formatted papers and finalize our benchmark using the information provided from our parsing tool, %along with 10 different citation patterns
capturing the patterns using the Python regular expression. Consequently, we construct final benchmark in which all documents for both query and candidate corpus consist of title, abstracts, full-paper text, and list of paper segments.

\subsection{Benchmark Statistics} 
\label{appendix:SciFullBench_Statistics_Analysis}
Table \ref{tab:query_paper_details} presents the query paper statistics for each benchmark split (References and Citations) across venues. The statistics show that each query paper can have up to 700 candidate papers, demonstrating that \textsc{SciFullBench} covers a diverse range of potential candidates and enables a comprehensive evaluation of paper retrieval performance. Furthermore, Table \ref{tab:corpus-statistics} shows that \textsc{SciFullBench} contains approximately 40K unique candidate papers. Even when considering the maximum number of ground-truth candidates per query, more than 39K papers serve as distractors, with an even larger number when considering the average number of ground-truth candidates. These statistics demonstrate that \textsc{SciFullBench} provides a challenging and robust evaluation of a retrieval system's ability to distinguish relevant papers from irrelevant ones, with a substantially larger pool of distractors than previous benchmarks such as \citet{cohan-etal-2020-specter} and \citet{singh-etal-2023-scirepeval}, which include approximately 25 distractor papers per query. Moreover, the 40K-paper corpus approximates realistic scientific paper retrieval settings at scale, while accounting for the substantial engineering effort required to process full papers, as detailed in Appendix \ref{appendix:SciFullBenchConstruction}. Taken together, these statistics establish \textsc{SciFullBench} as a realistic and useful benchmark for evaluating scientific paper retrieval systems.

\subsection{Comparison with Prior Benchmarks}
\label{appendix:comparison_with_prior_benchmarks}
Previous benchmarks for scientific literature search, where the full context is available for both query and candidates, rarely exist. \citet{kanakia2019scalablemicrosoft} introduce expert-annotated paper recommendation benchmark using abstract and citations within Microsoft Academic Graph (MAG). \citet{cohan-etal-2020-specter} and \citet{singh-etal-2023-scirepeval} reveal an evaluation set to search relevant scientific literature within a pool of 30 candidates per query document with 5 gold labels in the computer science domain, while \citet{medic-snajder-2022-MDCR} disclose a benchmark with 60 candidates per query from 19 scientific fields sourced from MAG. \citet{brown2019Relish} presents RELISH, which also provides expert-annotated ground-truth candidates in the biomedical domain. 

In addition, \citet{mysore2021csfcube} formulates a CFSCube benchmark to evaluate fine-grained sentence-wise alignment between abstract passages. \citet{wu-etal-2024-scimmir} also presents SciMMIR, a multimodal document retrieval evaluation set to evaluate the performance of figure-wise document retrieval frameworks. However, such benchmarks do not include candidates or queries in which their entire lexical content is fully disclosed. Although \citet{zhang2023FUTEX} intend to evaluate their classifier pipeline using the complete scientific context, it contains five potential candidates per query, which is infeasible to evaluate large-scale literature retrieval frameworks. 

Meanwhile, \citet{multicite-lauscher-2021} provides full context with fine-grained citation-intent annotation on the paper's citation context (sentences), to evaluate the model's ability to accurately identify citation intents. However, its scope is inherently limited to a trivial classification task, and does not reveal metadata information (especially full context) on the cited papers nor properly filtered and preprocessed. Conversely, our proposed benchmark consists of fully-disclosed document context, along with massive number of target candidates, adhering to the realistic setting for evaluating scientific literature retrieval.

\section{Training Procedures}
\label{appendix:Training Procedures}
In this section, we provide details for the training data construction process, followed by the preference set construction and detailed settings, such as hyper-parameters and training environment.

\subsection{Query and Corpus Construction}
\label{appendix:query_corpus_construction}

\paragraph{Composition} 
Our input query document for offline rollout includes publications and submissions from leading venues in ML, with different published or submitted years from \textsc{SciFullBench}. We incorporate publications and submissions from \textbf{ICLR} 2017 to 2023, \textbf{NeurIPS} 2016 to 2022 from the dataset revealed in REVIEWER2~\citep{gao2024reviewer2}, \textbf{ACL} 2017 to 2022, \textbf{EMNLP} 2017 to 2022, and \textbf{NAACL} 2017 to 2022. As for the target candidate corpus construction, we follow a similar approach to SciFullBench in which we crawled ML papers in the arXiv database. Ultimately, we formulate 15K input query documents with at least 10 ground-truth candidates, along with 97k target corpus consisting of all the ground-truth candidates of respective input query documents.

\paragraph{Formulation Workflow} 
The formulation procedure for building the training set is similar to the process explained in Appendix~\ref{appendix:SciFullBench}, with several key differences. 

\textbf{First}, to prepare query documents that do not overlap with our \textsc{SciFullBench} benchmark (to prevent data leakage), we collected publications and submissions from different venues (or years) from ones in our evaluation set. Yet, duplicate papers still may exist in both the benchmark and train set since research papers can be resubmitted to other venues if rejected. To preclude such corner cases, we excluded query documents from our train set where its abstracts match %(e.g., using a custom matching function, including lowering case, removing empty sequences) 
the ones in query document of \textsc{SciFullBench}. %We particularly used abstracts to check duplicate papers because several queries in our \textsc{SciFullBench} may not contain title information. 

\textbf{Second}, compared to the benchmark construction process that filters query documents with at least 10 ground truth documents for each citation and reference split, we view ground truth candidates as an aggregation of both relevance signals. Moreover, since the number of our input query documents drastically exceeds those in \textsc{SciFullBench}, our target candidate corpus size inevitably becomes larger. Thus, we randomly sampled 140K samples from the initial filtered target corpus, and reformulated our queries accordingly with at least 10 candidates. 

\subsection{Preference Set Construction}
\label{appendix:preference_set_construction}
Using the query documents and the formulated corpus from Appendix~\ref{appendix:query_corpus_construction}, we exclusively used references (out-going links) as ground truth annotations. To ensure fairness in evaluating the effectiveness of our algorithm, we excluded potential query documents from our training set that exist in the corpus used for the inference setting, as the documents in the corpus are used as query for later rounds of retrieval. Moreover, to ensure that data leakage does not occur, we additionally checked the SciFullBench queries for those whose title information is available, and additionally filtered queries in our train set with similar or same titles. 

During the rollout, we set the maximum input tokens to 60k and the maximum completion tokens to 2000. Moreover, we set the temperature to 0.7 and the nucleus sampling ratio to 0.8 to promote exploration, with the branching factor of 16 per aspect-aware LLM agent, using a single 48GB NVIDIA A6000 GPU with VLLM gpu memory utilization rate of 0.8 for each respective LLM agent + Retriever setup. Note that we arbitrarily stopped exploration under our notion that enough preference data was accumulated through self-exploration to enhance training efficiency. Subsequently, we measured the recall@30 reward of generated queries and paired the query response with the highest reward and the lowest reward, sampling only those with a recall@30 difference of at least 3\%.

\subsection{Training Details and Hyper-Parameters}
\label{appendix:train_details_and_hyperparameters}
We trained two instruct models, Llama-3.2-3B-Instruct and QWEN-2.5-3B-Instruct, using the Unsloth~\citep{unsloth} library to reduce memory consumption. We trained the respective query optimizer agents with LoRA~\citep{hu2022lora} adapters and 4-bit quantization on the default models. In addition, we set the maximum total number of tokens to 4000, allocating 2000 tokens each for input and output. This ensures that the responses from our preference set match the output token limits in the inference setup, allowing us to exclusively assess the effect of training rather than other factors. Furthermore, we trained each LLM agent with a linear scheduler, a learning rate of 1e-5, a weight decay rate of 0.01, and a maximum gradient norm of 0.6 for 3 epochs, with a global batch size of 32 on a single NVIDIA A5000 GPU with 24GB size VRAM, while occasionally using NVIDIA A6000 GPU with 48GB size VRAM to speedup training process. Under such setup, it required approximately 14 hours to train a single agent. As for the LoRA configuration, we set both the rank and $\alpha$ scale to 64, and applied the adapters to both the attention layers and the feedforward network, without leveraging dropout or bias factors. 

\section{Experiment Details}
\label{appendix:Experiment_Details}
In this section, we provide detailed setup for experiments, including prompts, neural retrievers, query optimizers, experiment environment, metrics, and implementation details. We also provide a detailed description for human evaluation results in Table~\ref{tab:Human_Evaluation}.

\subsection{Inference Settings}
\label{appendix:Inference Settings}

\paragraph{Prompt Construction}
We constructed prompts for scientific paper-to-paper retrieval based on the three major aspects that comprise a scientific paper: \textbf{Methodology}, \textbf{Experiments}, and \textbf{Research Questions}, inheriting the approaches in conventional research for the AI-for-Scientific Discovery domain~\citep{baek2024researchagent, wang-etal-2024-scimon, ScholarEval}, widely regarded as the most common and foundational aspects that constitute scientific papers. This design choice was intended to capture the representative aspects of a complete scientific paper, rather than to prescribe a universally optimal set of contents, thereby providing a plug-and-play framework that can accommodate varying levels of aspect-wise decomposition. The initial raw prompts were first created using ChatGPT, and human-in-the loop revision was conducted, to devise the optimal prompt suitable for our framework. The aspect-specific prompts used in \textsc{CoR} are provided in Figures \ref{app:fig:method_focused_query_prompt}, \ref{app:fig:experiment_focused_query_prompt}, and \ref{app:fig:research_question_focused_query_prompt}. The base prompt, which incorporates all three aspects and generates three separate queries, is provided in Figure \ref{app:fig:base_query_prompt}. Finally, the prompt for generating a single comprehensive query that covers all three aspects is provided in Figure \ref{app:fig:single_comprehensive_query_prompt} in Appendix \ref{appendix:Prompts}.

\paragraph{Retrieval Setting} 
We primarily use the Euclidean distance to measure similarities between our adaptively generated queries and candidates, where we acquired candidates specifically with the minimum L2 distance. Since minimizing the L2 distance was objective for most of our baseline domain-specific embedding models such as SciNCL, SPECTER2-Base, SPECTER2-Adapter-MTL CTRL, we matched such settings when we used jina-embeddings-v2-base-en, BGE-M3, and Inf-Retriever-v1-1.5b. However, for SciMult, we measured MIPs (Maximum Inner Product), since it was trained to maximize MIPs in the original paper. In addition, we utilize the FAISS library~\citep{douze2024faiss}, which enables efficient retrieval from a large-scale corpora. We implemented an efficient L2 search using FlatL2 and FlatIP for MIPs. 

In addition, we concatenated the title and abstract information for our baseline A2A retrieval settings, following the conventional retrieval setup from~\citep{cohan-etal-2020-specter, singh-etal-2023-scirepeval} for both query and target candidates. 
As for our \textsc{CoR} framework, when using abstracts as query and candidates, we use the concatenated version of title and abstracts for both query and candidates, with the exception of root depth to balance the effect of title information of root query paper, where the query optimizers receive full paper which occasionally receives title information. 

\paragraph{Query Optimizers} 
We mainly used the VLLM~\citep{kwon2023vllm} framework for faster inference of open source models, setting the repetition penalty to 1.2. For Llama-3.2-3B-Instruct, we used the default context window size of 131072, while using YARN~\citep{peng2023yarn} to extrapolate the default context window size of QWEN-2.5-3B-Instruct from 32768 to 131072. Moreover, for inference, we used combinations of three NVIDIA A5000/RTX3090/4090 GPUs with 24GB VRAM size, along with vllm gpu memory utilization rate of 1.0, using fp16 precision for faster inference. 

As for the inference setup for untrained query optimizer models, we used a combination of single NVIDIA A6000/RTX 4090 GPU with 48GB VRAM using the same context window and fp16 precision, along with a fixed gpu memory utilization rate of 0.7. When it comes to experiments using GPT-4o-Mini-2024-0718 and GPT-4o-2024-11-20 as query optimizers, we used temperature 0 and default hyper parameters of the OpenAI client.chat.completions API\footnote{\url{https://platform.openai.com/docs/guides/}}, while completely using the default hyperparameter settings for gpt-4.1-2025-04-14 (temperature is also set to default). Moreover, 60 is used as the hyperparameter k for Reciprocal Rank Fusion. For base agent experiments in Table~\ref{fig:effect_of_multiagent_query_optimizer}, we equally set the hyper parameters to those used in its respective counterpart comparison groups, and are prompted to generate three different queries in a structured manner, using the Python Pydantic BaseModel\footnote{\url{https://docs.pydantic.dev/latest/api}} module.

\subsection{Models}
\label{appendix:Models}
\paragraph{Domain Specific Retrievers}
We compare our method with retrieval using State-of-The-Art domain-specific embedding models, to demonstrate robust improvement over approaches that seek to devise optimal representations of paper abstracts through extensive training. We mainly use SPECTER2 models, where its variant with adapters and its multitask control codes have been reported to achieve SoTA performance on the MDCR benchmark~\cite{medic-snajder-2022-MDCR}. We used proximity control code adapter, and base models on huggingface\footnote{\url{https://huggingface.co/allenai/specter2}}. Meanwhile, we also use SciNCL and SciMult models as baselines, where SciNCL\footnote{\url{https://huggingface.co/malteos/scincl}} achieves the strongest performance on the SciDocs benchmark to date, while SciMult-MHAExpert model implemented with Mixture-of-Experts~\citep{fedus2022MoE} architecture 
is reported to outperform previous models on the recommendation benchmark~\citep{kanakia2019scalablemicrosoft}. For SciMult-MHAExpert, we used the model released in the official SciMult GitHub repository \footnote{\url{https://github.com/yuzhimanhua/SciMult}}, specifically an expert model trained for link prediction. %that predicts linked documents from a source document’s abstract.

\paragraph{Domain Agnostic Retrievers}
%We particularly chose off-the shelf retrievers with long context window to ensure that we provide fair experimental setting for our baselines, most notably full document-to-full document retrieval setting.%
Since general full-documents exceed the context window of most embedding models, we specifically chose long-context window embedding models to construct a fair evaluation suite. In particular, we mainly use three different neural retrievers, Jina-Embeddings-v2-Base-EN, and BGE-M3. both highly cited general-purpose embedding models with a context window of 8192 tokens, and Inf-Retriever-v1-1.5B (context window of 32768), a lightweight version of Inf-Retriever-v1, demonstrating SoTA performance among open source embedding models for long-context retrieval (LongEmbed) in MTEB~\citep{muennighoff2022mteb}\footnote{\url{https://huggingface.co/spaces/mteb/leaderboard}} and AIR-Bench ~\citep{chen2024airbench} (general QA tasks) as of September 12, 2025.
%This further shows that our method can perform robustly with the most recent general purpose embedding models with advanced long-context handling capabilities.

In addition, we provide additional experiments using other recent general-purpose embedding models varying in context window size, such as QWEN3-Embedding-0.6B~\citep{qwen3embedding}, Dewey-EN-Beta~\citep{zhang2025deweylongcontextembedding}, and Granite-Embeddings-English-R2~\citep{awasthy2025graniteembeddingmodels}. QWEN3-Embedding-0.6B is a light weight version of QWEN3-Embedding model series, recently released model that displays SoTA performance on MTEB-multilingual evaluation split, while Granite-Embeddings-English-R2 is the most recently released model as of August 2025.  In addition, Dewey-EN-Beta is another recent model that has a context window size of 131072, which is the longest among up-to-date existing models, attaining SoTA performance on the MTEB LongEmbed benchmark.

\subsection{Metrics} 
\label{appendix:metrics}

In our experiments, we adopt four primary metrics. \textbf{Recall@K}, \textbf{Normalized Discount Cumulative Gain (nDCG)}, \textbf{Mean Average Precision (mAP)}, and \textbf{Mean Reciprocal Rank (MRR)}. Unlike the standard nDCG metric, which assigns higher weights to gold candidates with higher significance, we treat all gold candidates for a given query equally, assigning each a uniform weight of 1 for \textsc{SciFullBench} experiments.

We use large cutoff value K following the convention from previous works such as SPECTER ~\citep{cohan-etal-2020-specter}, which sets K based on the maximum number of ground-truth candidates, as well as SciMult~\citep{zhang-etal-2023-scimult}, which reports metrics with K up to 100. In our benchmark, the number of ground-truth relevant papers per query is often very large, frequently reaching several hundreds and exceeding 700 in some cases (Table \ref{tab:query_paper_details}), which reduces the bias of evaluation and robustly measures the capacity of retrieval systems. As a result, choosing a small K (e.g., K = 20) would inherently fail to cover a substantial portion of the ground-truth set, leading to artificially low recall and NDCG scores that reflect truncation effects rather than true retrieval quality. Thus, we report results with relatively high K values (e.g., 300), which covers the 99.5th percentile of ground-truth candidates per query. 

In addition, we provide \textbf{Convex Hull Volume}~\citep{graham1972convexhull} as a metric to quantitatively measure the dispersion of the embedding representations of retrieved candidates, computing the volume of the smallest possible bounding convex closure of the data points after projecting the embedding vectors to the three-dimensional vector space. 

Notably, we employ convex hull volume in two different ways. As for the analysis provided in Table~\ref{tab:embedding_diversity_analysis},we measured the average relative convex hull volume ratio of retrieved ground-truth candidates between binary samples from two distinct retrieval variants, representing the relative coverage expansion on a sample level. Meanwhile, in Figure~\ref{fig:retrieval_round_diversity_analysis}, we measured the average absolute convex hull volume of total Top@K retrieved samples per retrieval variant, to analyze the trend of semantic coverage with respect to retrieval depth progression. 

\subsection{Human Evaluation Setup}
\label{appendix:Human_Evaluation_Setup}

For sampling query papers for human study, we use NeurIPS, ACL, EMNLP split in \textsc{SciFullBench} where its query paper exists for both reference and citations split, and randomly sampled 15 different query papers. For A2A, F2F, and \textsc{CoR} variant, we used result variants where jina-embeddings-v2-base-en is utilized as backbone retriever, and DPO-trained QWEN-2.5-3B-Instruct model for query optimizers in \textsc{CoR}, with retrieval depth of 3, and utilized retrieved results from Citations (inward citation) split. When selecting the top-5 results for each variant, we excluded the 0th index result, since the same paper is mostly retrieved due to the structure of our target corpus. Moreover, we exclude ground truth candidates for reference and citations split, in order to measure the efficacy of \textsc{CoR} on candidates that are regarded as incorrect by our benchmark, and formulated top-$k$ results for qualitative evaluation. For each paper we made two duplicate tasks for evaluation and originally hired 15 human evaluators studying the field of machine learning in South Korea, who willingly consented to perform evaluation on two different query papers, for approximately 20 U.S. Dollars. During the evaluation process, each participant is provided three retrieval variants with its order randomly shuffled, each variant consisting of 15 pairs of query, retrieved paper with title and abstracts. The participant is then instructed to choose one of the following (\textbf{Method}, \textbf{Experiment}, \textbf{Motivation/Research Question}, and \textbf{Irrelevant}) for every retrieved result, checking the relevance with the query paper. The details on human evaluation guideline can be found in Figure \ref{app:fig:Human_Evalaution_Guideline} of Appendix \ref{appendix:HumanEvaluation}.

\section{Patent-to-Patent Retrieval}
\label{appendix:patent-to-patent retrieval}

In this section, we provide detailed procedures for constructing the benchmark and experimental settings for the patent-to-patent retrieval task. The statistics for \textsc{PatentFullBench} can be observed at Table~\ref{tab:patent_query_paper_details} and Table~\ref{tab:corpus-statistics_patents}.
\begin{figure*}[t!]
    \centering
    \includegraphics[width=\textwidth]{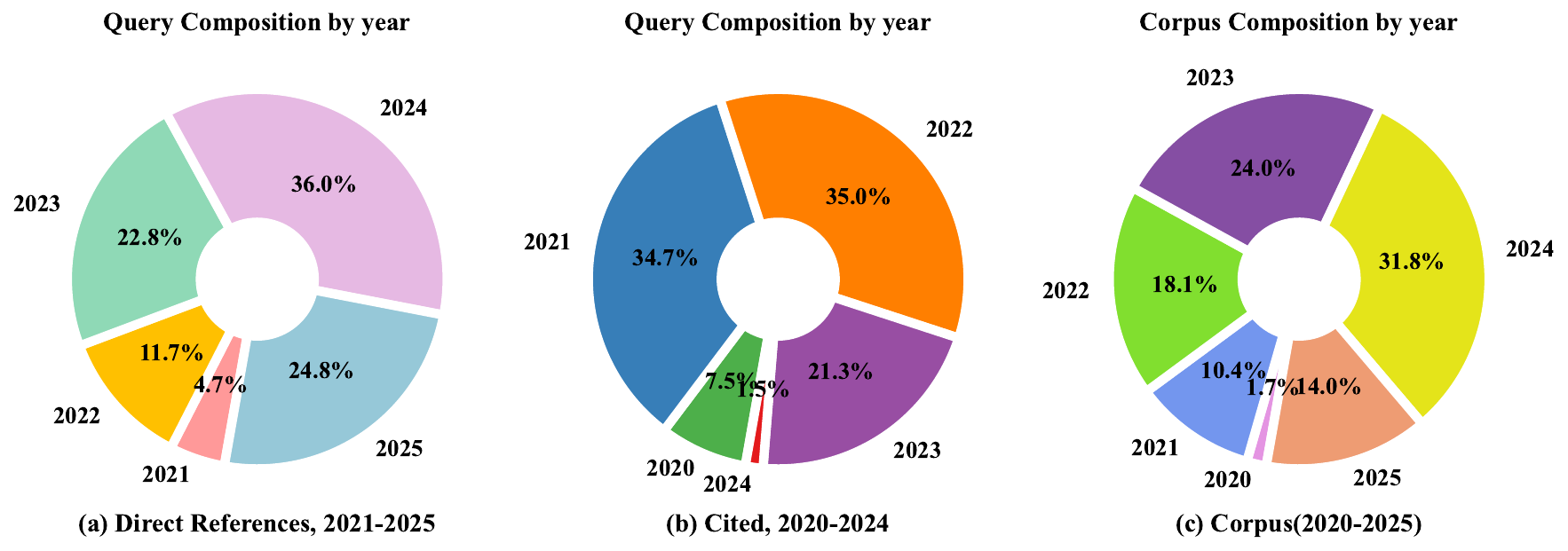}
    \vspace{-0.18in}
    %\captionsetup{justification=justified, singlelinecheck=false}
    %\caption{\small PatentFullBench composition by published year.}
    \caption{PatentFullBench composition by published year.}
    \label{fig:PatentFullBench_Composition}
    \vspace{-0.1in}
\end{figure*}

\subsection{PatentFullBench Construction Workflow}
\label{appendix:patentfullbench_construction}
\paragraph{Step 1: Collecting Raw Data:}
We initially obtained a full XML dump from the USPTO Open Data portal website\footnote{\url{https://data.uspto.gov/bulkdata/datasets}}, and initially crawled US patents from 2020 to 2025 June 17th.

\paragraph{Step 2: \textbf{Formatting Raw Data:}} 
We parsed XML dump from the USPTO Open Data portal website, obtaining metadata such as publication number, publication date, title, abstract, inventors, priority claims, claims, descriptions, references (citations from given patent). As for citations and references, we initially filtered U.S patents exclusively, especially ones where its date is over 2020, and obtained metadata such as classified country, date, document number, collecting a group of raw patents where each patent at least had 10 citations from the parsed XML dump.

\paragraph{Step 3: \textbf{Initial Gold Label Annotation and Target Corpus Construction:}} After forming a pool of raw extracted patent contents from XML dump data, we obtained patent invention title, abstracts,id, and date, using PatentsView API\footnote{\url{https://patentsview.org/apis/purpose}}. Since PatentsView API does not support full-content access, we made sure to exclude patents absent from the extracted content from \textbf{step 2}. After obtaining actual reference to full patents per input query, we attempt to further label each query document with inward citations (incoming links) by labeling with gold documents that cite the query document. We further process citation relations through traversing all the patent files, and record the query patent B that cites A patent as ground-truth candidate for inward citation of A.

Subsequent to obtaining both inward citations and outward citations (references) for respective query patent documents, we removed duplicate labels that have either the same title or abstract content.
For respective splits, we excluded query documents from our benchmark set that had less than 10 ground truth candidates, and randomly sampled 800 query documents for respective splits. Successively, we formulated our target corpus via accumulating the ground truth candidates that exist within our randomly sampled query document splits without duplicates, excluding documents that either had the exact same title and abstracts. 

Then, we removed ground-truth candidates from the initially sampled query documents that were absent from the target corpus. 
This process ensures that our potential ground-truth candidates for queries always exist in the target corpus. Finally, we randomly sampled 400 queries that had at least 10 ground-truth candidates after the above filtering step, and reformulated our target corpus, making sure that duplicate candidates do not exist. 
%\captionsetup{justification=centering}
\begin{table}[t!]

\centering
%\caption{\small Statistics of query documents in PatentFullBench.}
%\label{tab:patent_query_paper_details}
%\vspace{-0.05in}
\resizebox{\linewidth}{!}{
\renewcommand{\arraystretch}{1.5}
\begin{tabular}{c cc}
\toprule
\textbf{} 

& \fontsize{15pt}{15pt}\selectfont References & \fontsize{15pt}{15pt}\selectfont Citations \\
\midrule

\fontsize{15pt}{15pt}\selectfont Nation
& \multicolumn{2}{c}{\fontsize{14pt}{14pt}\selectfont United States} \\
\fontsize{15pt}{15pt}\selectfont \# of documents per split
& \fontsize{14pt}{14pt}\selectfont 400 & \fontsize{14pt}{14pt}\selectfont 400 \\
\fontsize{15pt}{15pt}\selectfont \# of tokens per abstract
& \fontsize{14pt}{14pt}\selectfont 119.50 & \fontsize{14pt}{14pt}\selectfont 120.31 \\
\fontsize{15pt}{15pt}\selectfont \# of tokens per document
& \fontsize{14pt}{14pt}\selectfont 190335.46 & \fontsize{14pt}{14pt}\selectfont 186504.29 \\
\fontsize{15pt}{15pt}\selectfont average \# of candidates per sample 
& \fontsize{14pt}{14pt}\selectfont 37.9 & \fontsize{14pt}{14pt}\selectfont 53.93\\

\bottomrule
\end{tabular}

}
\vspace{-0.05in}
%\caption{\small Statistics of query documents in PatentFullBench.}
\caption{Query Statistics in PatentFullBench.}
\label{tab:patent_query_paper_details}
\vspace{-0.1in}
\end{table}
\paragraph{\textbf{Step 4: Formatting Full Context of Query and Candidate Documents and Final Filtering}} 
We concatenated title, abstract, claims, and description content to formulate a full context for query and candidate patents. In order to eliminate potential cues that reveal hindsight information on cited patents, we used python regular expression to remove such signals, 
while using the citation information obtained from \textbf{step 2} to eliminate in-line citation mentions, and document numbers, along with the title and abstract information of ground truth candidates that exist in the full content of respective query patents. 

As for candidate documents, we removed the query title and abstract from candidate documents whenever the candidate served as a ground-truth document for that query, preventing candidate documents from revealing direct hints about the relevant query patents. After sanitizing the full documents, we segmented the full context of candidate documents for every 3000 tokens (based on NLTK), and eliminated the patent candidate from our corpus and ground truth labels that had at least a single duplicate partial chunk, and finalized our patent retrieval benchmark. 
\begin{table}[t!]
\centering
%\caption{\small Corpus statistics in PatentFullBench.}
%\label{tab:corpus-statistics_patents}
%\vspace{-0.05in}
\resizebox{\linewidth}{!}{
\renewcommand{\arraystretch}{1.35}
\begin{tabular}{c c}
\toprule
\textbf{} 
& \fontsize{15pt}{15pt}\selectfont \textsc{PatentFullBench} \\
\midrule
\fontsize{15pt}{15pt}\selectfont Total \# of documents
& \fontsize{14pt}{14pt}\selectfont 5150 \\
\fontsize{15pt}{15pt}\selectfont Avg.\ \# of tokens per abstract
& \fontsize{14pt}{14pt}\selectfont 120.97 \\
\fontsize{15pt}{15pt}\selectfont Avg.\ \# of tokens per document
& \fontsize{14pt}{14pt}\selectfont 29141.85 \\
\fontsize{15pt}{15pt}\selectfont Total \# of segmented corpus
& \fontsize{14pt}{14pt}\selectfont 52584 \\
\fontsize{15pt}{15pt}\selectfont Avg.\ \# of tokens per segment
& \fontsize{14pt}{14pt}\selectfont 2865.06 \\
\bottomrule
\end{tabular}
}
\vspace{-0.05in}
%\caption{\small Corpus statistics in PatentFullBench.}
\caption{Corpus statistics in PatentFullBench.}
\label{tab:corpus-statistics_patents}
\vspace{-0.1in}
\end{table}

\subsection{Experiment Details}
\label{appendix:patent_experiment_details}
\paragraph{Prompt Construction} 
We constructed prompts for patent-to-patent retrieval based on the three major aspects that comprise a patent: \textbf{Claim}, \textbf{Backgrounds}, and \textbf{Method}. This design choice was intended to capture the representative aspects of a complete patent document, rather than to prescribe a universally optimal set of contents, thereby providing a plug-and-play framework that can accommodate varying levels of aspect-wise decomposition, similar to Appendix \ref{appendix:Inference Settings}. \textbf{Method} aspect refers to technical details and implementation structure of the invention, while \textbf{Backgrounds} provide background information on the technical problems of previous works. Moreover, \textbf{Claim} aspects comprise of the scope of its legal protection. The prompt construction process took roughly one hour,
following the same procedure demonstrated in the Appendix \ref{appendix:Inference Settings}. For more details on respective aspect-specific prompts for patent-to-patent retrieval using \textsc{CoR}, please see Figures \ref{app:fig:patent_method_focused_query_prompt}, \ref{app:fig:patent_claim_focused_query_prompt}, \ref{app:fig:patent_background_focused_query_prompt} in Appendix~\ref{appendix:Prompts}.
\paragraph{Inference Setting}
We evaluate our patent to patent retrieval setting with the same domain-agnostic embedding models that we utilized for our evaluation, additionally reporting abstract-to-abstract retrieval results with domain-specific retrievers, PAT-SPECTER and PaECTER~\citep{ghosh2024paecter}. %which is the most recent model for patent retrieval trained via minimizing the triplet loss using the binary patent abstract samples, using dataset sampled from European Patent Office (EPO) \footnote{https://www.epo.org/en}. 
Moreover, we utilized proprietary LLM, GPT-4o-Mini-2024-0718 with temperature 0 for query optimizers and truncated the input document context to 120000 tokens using the tiktoken\footnote{\url{https://github.com/openai/tiktoken}} library since patent context length is significantly longer than scientific papers, occasionally exceeding the context window of GPT-4o-Mini-2024-0718 model.

\section{Supplementary Experiments}
\label{appendix:Supplementary_Experiments}
Here, we provide additional experiments to further support the results and analysis in the main paper.

\paragraph{Consistent Generalization Across Embedding Models} 
In our main results, we showed three variations of embedding models to demonstrate the effectiveness of our \textsc{CoR} framework. In this section, we further provide additional experiments on other existing long-context embedding models, namely \textbf{Granite-Embeddings-R2}, \textbf{QWEN3-0.6B-Embedding}, and \textbf{Dewey-en-Beta} in Table~\ref{tab:additional_score-reporting}. The experimental results strongly support the robustness and general applicability of our research findings, as even without a trained model, by using proprietary \textbf{GPT-4o-2024-11-20} and \textbf{GPT-4.1-2025-04-14} with just a single round (depth) of retrieval, we were able to observe consistent performance improvements compared to naive F2F and A2A retrieval setups.

The improvements achieved using Granite-Embeddings-R2 indicate that our framework can perform robustly under a context window of 8192 as repeatedly demonstrated in our main results, and is compatible with even the most recently released model. Moreover, the results from the QWEN3-0.6B-Embedding model imply that our model can be applied to embedding models with longer context windows (32768) and also Inf-Retriever-v1-1.5B in Table~\ref{tab:main_final}, as well as to popular state-of-the-art long-context embedding models for general domain-agnostic multilingual retrieval tasks in the MTEB benchmark.

Furthermore, we provide evaluation results on Dewey-En-Beta, which is an embedding model with currently the longest context window among existing models, while demonstrating State-of-the-Art performance on LongEmbed benchmark. This suggests that our paper-retrieval framework is applicable to models with the longest context windows up-to date, indicating the effectiveness and robustness of our suggested method to handle long context retrieval.
\begin{table}[t!]
\centering
%\captionsetup{justification=justified, singlelinecheck=false}
%\caption{\small Performance of \textsc{CoR} with different LLM backbones and neural retrievers under single-round retrieval.}
%\vspace{-0.075in}
%\label{tab:additional_score-reporting}
\resizebox{\linewidth}{!}{
\renewcommand{\arraystretch}{0.74}
\begin{tabular}{llcc}
\toprule
& & \textbf{ICLR-Citations} & \textbf{NeurIPS-Citations} \\
\cmidrule(lr){3-3} \cmidrule(lr){4-4}
& \textbf{Model + Retriever} & Recall@300 & Recall@100 \\
\midrule
\multicolumn{4}{l}{\textbf{Baseline}} \\
&Granite-Emb-Eng-R2 (A2A) & 50.19 &  41.39 \\
&Granite-Emb-Eng-R2 (F2F) & 54.00 &  44.37 \\
\noalign{\vskip 1mm}
\cdashline{2-4}
\noalign{\vskip 1mm}
&QWEN3-0.6B-Emb (A2A) & 51.47 & 41.76 \\
&QWEN3-0.6B-Emb (F2F) & 54.03 & 44.81 \\
\noalign{\vskip 1mm}
\cdashline{2-4}
%\noalign{\vskip 1mm}
%\cdashline{2-4}
\noalign{\vskip 1mm}
&Dewey-en-Beta (A2A) & 47.76 & 40.54 \\
&Dewey-en-Beta (F2F) & 53.15 & 42.06 \\
\midrule
\midrule
\multicolumn{4}{l}{\textbf{Ours (\textsc{CoR})}} \\
&\textbf{GPT-4o + Granite-Emb-Eng-R2} & \textbf{55.64} & \textbf{47.22} \\
& \textbf{GPT-4o + QWEN3-0.6B-Emb} & \textbf{56.94} & \textbf{47.15} \\
& \textbf{GPT-4.1 + Dewey-en-Beta} & \textbf{56.73} & \textbf{47.99} \\

\bottomrule
\end{tabular}
}
\vspace{-0.05in}
\captionsetup{justification=justified, singlelinecheck=false}
%\caption{\small Performance of \textsc{CoR} with different LLM backbones and neural retrievers under single-round retrieval.}
\caption{Performance of \textsc{CoR} with different LLM backbones and neural retrievers under a single retrieval depth.}
\label{tab:additional_score-reporting}
\end{table}

\begin{table}[t!]
%\captionsetup{justification=justified, singlelinecheck=false}
%\caption{\small Performance analysis of ours when the original abstract to abstract retrieval is not utilized in our pipeline.}
%\vspace{-0.075in}
\resizebox{0.975\linewidth}{!}{
\renewcommand{\arraystretch}{0.62}
\begin{tabular}{llcc}
\toprule
& & \textbf{ACL-Citations} & \textbf{ICLR-Citations} \\
\cmidrule(lr){3-3} \cmidrule(lr){4-4}
& \textbf{Retrieval Method} & Recall@200 & Recall@200 \\
\midrule

\multicolumn{4}{l}{\textbf{Baseline}} \\
&JEmb-v2 (A2A)  & 35.33 & 39.19 \\
&JEmb-v2 (F2F)  & 35.88 & 42.44 \\
\noalign{\vskip 1mm}
\cdashline{1-4}
\noalign{\vskip 1mm}
&BGE-M3 (A2A)  & 33.76 & 35.16 \\
&BGE-M3 (F2F)  & 32.64 & 36.06 \\
\noalign{\vskip 1mm}
\cdashline{1-4}
\noalign{\vskip 1mm}
&Inf-Ret-v1 (A2A)  & 41.77 & 45.42 \\
&Inf-Ret-v1 (F2F)  & 38.86 & 43.00 \\
\midrule
\midrule
\multicolumn{4}{l}{\textbf{Ours w/o A2A}} \\
&\textbf{QWEN-2.5-3B-Inst + JEmb-v2} & \textbf{41.13} & \textbf{44.41} \\
&\textbf{Llama-3.2-3B-Inst + BGE-M3} & \textbf{35.83} & \textbf{37.49} \\
&\textbf{Llama-3.2-3B-Inst + Inf-Ret-v1} & \textbf{47.05} & \textbf{50.65}  \\

\bottomrule
\end{tabular}
}
\vspace{-0.05in}
\captionsetup{justification=justified, singlelinecheck=false}
%\caption{\small Performance analysis of ours when the original abstract to abstract retrieval is not utilized in our pipeline.}
\caption{Performance analysis of ours when the original abstract to abstract retrieval is not utilized in our pipeline.}
\label{tab:agent_coordination_wo_abstract}
\vspace{-0.01in}
\end{table}

\paragraph{Robustness without Abstract-to-Abstract Retrieval}
Meanwhile, in Table~\ref{tab:agent_coordination_wo_abstract}, the results are reported
when abstract-to-abstract retrieval (A2A) is not utilized within \textsc{CoR}. Despite A2A setup being omitted from its original multi-aspect queries, consistent improvement over baselines can be observed, although not to the extent when abstract-abstract retrieval is integrated into our pipeline, indicating the positive, balancing effect that A2A retrieval presents, along with robustness of our method. This in turn highlights the effectiveness of our pipeline in generating multiple aspect-aware queries in the absence of complementary abstract to abstract retrieval, strongly supporting our research contributions.
\paragraph{Comparison with Naive Chaining}
Table~\ref{tab:comparison_w_naive_chain_of_retrieval} shows the benefits of employing multi-round retrieval with \textsc{CoR} compared to that of naive chain-of-retrieval methods using Full Paper-to-Full Paper and Abstract-to-Abstract retrieval setup. The results indicate that our aspect-guided chain-of-retrieval along with post-order recursive aggregation is more effective, and trivial iterative retrieval is not sufficient. Note that the baseline experiments were conducted with under the same hyper parameter setting, with a cache to avoid redundant sampling likewise our framework, retrieving 300 candidates and aggregated via Reciprocal Rank Fusion. This shows that our approach in exploration and aggregation guided by aspect-centric queries is more effective compared to iterative top-k retrieval using raw content as queries, supporting the validity of our proposed retrieval system. 

\begin{table}[t!]
%\captionsetup{justification=justified, singlelinecheck=false}
%\caption{\small Ablation study on the effect of our search framework compared to naive chain-of retrieval methods based on A2A and F2F retrieval setting.}
%\vspace{-0.1in}
%\label{tab:comparison_w_naive_chain_of_retrieval}
\resizebox{\linewidth}{!}{
\renewcommand{\arraystretch}{0.8}
\begin{tabular}{llcc}
\toprule
& & \textbf{ICLR-References} & \textbf{ICLR-Citations} \\
\cmidrule(lr){3-3} \cmidrule(lr){4-4}
& \textbf{Model + Retriever} & Recall@100 & Recall@200 \\
\midrule
\multicolumn{4}{l}{\textbf{Baseline}} \\
&JEmb-v2 (A2A)  & 33.04 & 39.10 \\
&JEmb-v2 (F2F)  & 33.19 & 41.54 \\
\noalign{\vskip 1mm}
\cdashline{1-4}
\noalign{\vskip 1mm}
&BGE-M3 (A2A)  & 28.71 & 34.72 \\
&BGE-M3 (F2F)  & 31.00 & 36.12 \\
\noalign{\vskip 1mm}
\cdashline{1-4}
\noalign{\vskip 1mm}
&Inf-Ret-v1 (A2A)  & 42.50 & 44.67 \\
&Inf-Ret-v1 (F2F)  & 36.67 & 42.38 \\
\midrule
\midrule
\multicolumn{4}{l}{\textbf{Ours (w/o DPO)}} \\
&\textbf{Llama-3.2-3B-Inst + JEmb-v2} & \textbf{38.34} &  \textbf{46.73} \\
&\textbf{Llama-3.2-3B-Inst + BGE-M3} & \textbf{32.62} &  \textbf{41.84} \\
&\textbf{Llama-3.2-3B-Inst + Inf-Ret-v1} & \textbf{46.74} & \textbf{52.42}  \\

\bottomrule
\end{tabular}
}
\vspace{-0.05in}
\captionsetup{justification=justified, singlelinecheck=false}
%\caption{\small Ablation study on the effect of our search framework compared to naive chain-of retrieval methods based on A2A and F2F retrieval setting.}
\caption{Ablation study on the effect of our search framework compared to naive chain-of retrieval methods based on A2A and F2F retrieval setting.}
\label{tab:comparison_w_naive_chain_of_retrieval}
\vspace{-0.06in}
\end{table}
\begin{figure}[t!]
    \centering
    \includegraphics[width=0.975\columnwidth]{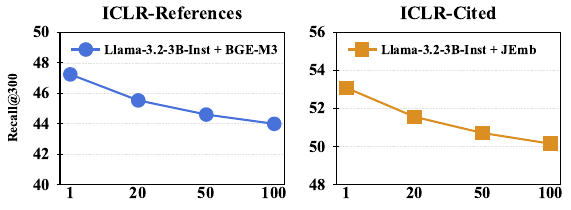}
    \captionsetup{justification=justified, singlelinecheck=false}
    \vspace{-0.05in}
    %\caption{\small Hyperparameter Ablation Study on the effect of selection starting index $\gamma$ on the overall performance. We use DPO-trained query optimizers on 3 rounds of retrieval.}
    \caption{Hyperparameter Ablation Study on the effect of selection starting index $\gamma$ on the overall performance. We use DPO-trained query optimizers on 3 rounds of retrieval.}
    \label{fig:Starting_Index_HyperParameter_Ablation}
\end{figure}
\paragraph{Starting Index Hyper Parameter Ablation}
In this section, we provide experiments on the hyper parameter selection starting index $\gamma$ that we used for \textbf{Next Query Selection Algorithm}, to choose the starting point of choosing the representative paper with highest semantic similarity. In our case, value 1 is used to report our results in the main paper (always start from the second highest rank), to select the paper with highest similarity while avoiding the same papers as input query that may exist in our corpus (which inevitably appears at rank 0, since same papers display 100\% semantic similarity). To further validate the benefits of selecting paper with highest-semantic similarity for subsequent retrieval, variations of \textsc{CoR} with different selection starting index $\gamma$ is reported in Figure~\ref{fig:Starting_Index_HyperParameter_Ablation}. The results demonstrate steady decrease in performance as selection starting index is increased, implying that \textsc{CoR} using the nearest-possible paper for next query selection is more desirable due to the semantic drift that occurs when relatively unrelated papers are selected as query for subsequent round of retrieval.

\begin{figure}[t!]
    \centering
    \includegraphics[width=0.975\columnwidth]{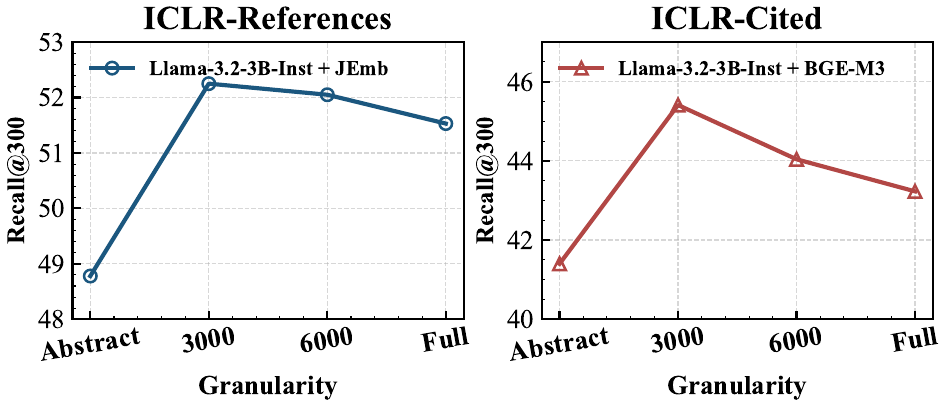}
    \vspace{-0.05in}
    %\caption{\small Results across candidate paper granularities.}
    \caption{Results across candidate paper granularities.}
    \captionsetup{justification=justified, singlelinecheck=false}
    \label{fig:corpus granularity ablation}
\end{figure}

\paragraph{Corpus Granularity Ablation Study}
Furthermore, in order to see how the granularity of candidate representations affects the performance on our multi-vector retrieval system, we conduct further analysis. The results in Figure~\ref{fig:corpus granularity ablation} demonstrate the effectiveness of our multi-vector approach: segmenting each candidate document into 3K-token chunks yields the best performance, which outperforms not only abstract-only and full-document single-vector representations but also coarser chunking strategies (such as 6K-token segments), suggesting that finer-grained representations are effective in capturing diverse and localized signals within full-length papers, which enables our query to capture a more multi-faceted dimension of target papers, resulting in an enhanced performance. 
\begin{table}[t!]
%\captionsetup{justification=justified, singlelinecheck=false}
%\caption{\small Analysis on the impact of incorporating rank fusion to attain unified results of our hybrid retrieval framework on the EMNLP-Citations split.}
%\vspace{-0.075in}
\resizebox{\linewidth}{!}{
\renewcommand{\arraystretch}{1.0}
\begin{tabular}{llcc}
\toprule
& \textbf{Retrieval Method} & \textbf{Recall@100} & \textbf{MRR@50}  \\
\midrule
\multicolumn{4}{l}{\textbf{Ours w/o RRF \& w/o A2A}} \\
\multicolumn{4}{l}{\quad \textbf{Naive Aggregation}} \\
& \quad GPT-4o-Mini-2024-0718 + BGE-M3 & 21.61 & 32.61 \\
\multicolumn{4}{l}{\quad \textbf{Late Interaction (MaxSim)}} \\
& \quad GPT-4o-Mini-2024-0718 + BGE-M3 & 26.00 & 31.53 \\
\midrule
\multicolumn{4}{l}{\textbf{Ours w/o A2A}} \\
& \cellcolor{blue!5} \textbf{GPT-4o-Mini-2024-0718 + BGE-M3} & \cellcolor{blue!5} \textbf{26.88} & \cellcolor{blue!5} \textbf{34.54} \\
\bottomrule
\end{tabular}
}
\vspace{-0.05in}
\captionsetup{justification=justified, singlelinecheck=false}
%\caption{\small Analysis on the impact of incorporating rank fusion to attain unified results of our hybrid retrieval framework on the EMNLP-Citations split.}
\caption{Analysis on the impact of incorporating rank fusion to attain unified results of our hybrid retrieval framework on the EMNLP-Citations split.}
\label{tab:rank_fusion}
\vspace{-0.05in}
\end{table}

\paragraph{Effect of Reciprocal Rank Fusion} 
We conduct an additional ablation study on the effect of RRF (Reciprocal Rank Fusion) in our hybrid retrieval system (single round of \textsc{CoR}), compared to embedding-level merging approach in prior multi-vector retrieval approaches. We primarily compare with two traditional approaches in embedding-level merging strategy, the naive aggregation strategy that forcefully computes similarities (in our case L2 distance) between all the sub-vectors of respective query and candidates and naively sums it up to acquire similarity score of original query and documents. In our experiments, to mitigate the unfair penalization due to document length, we normalize the aggregated results with the total number of subvectors used for similarity calculation. 

\begin{figure}[t!]
    \centering
    \includegraphics[width=0.975\columnwidth]{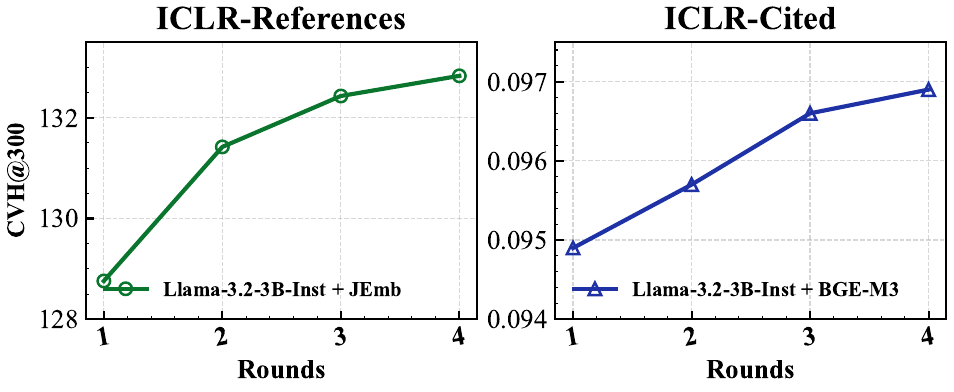}
    \captionsetup{justification=justified, singlelinecheck=false}
    %\caption{\small Diversity Analysis with retrieval depth progression. Note that we perform evaluation using DPO-trained Query Optimizers.}
    \caption{Diversity Analysis with retrieval depth progression. Note that we perform evaluation using DPO-trained Query Optimizers.}
    \vspace{-0.2in}
    \label{fig:retrieval_round_diversity_analysis}
\end{figure}
Meanwhile, we also present comparison with late interaction strategy~\citep{khattab2020colbert, santhanam2021colbertv2}, where only the maximum similarity for subquery and its corresponding sub-documents is aggregated to form original query, document similarity (in our case, minimum l2 distance). Table~\ref{tab:rank_fusion} illustrates a performance drop when queries optimized in various aspects are used as subvector representations of original documents and are used to compute a single similarity value when matched with the segmented corpus of target candidates. This supports the validity of our design choice: our ranking-merging system is more robust to noise while still retrieving candidates that are strongly aligned in at least one aspect.

\paragraph{Additional Diversity Analysis}
In this section, we provide further analysis on the improved diversity of semantics that our retrieval system presents. Figure~\ref{fig:retrieval_round_diversity_analysis} presents a linearly increasing trend where the Top@300 absolute Convex Hull Volume of 400 samples per benchmark split proportionally increases with respect to the retrieval depth, indicating that our \textsc{CoR} framework is able to diversify the semantics of search results as the rounds progress. Moreover, along with the quantitative metrics, we visualize the distribution of the embedding vectors of Top@300 retrieved documents for eight sampled queries on a two-dimensional plane using \textbf{t-distributed Stochastic Neighbor Embedding (t-SNE)}~\citep{van2008visualizingtsne} and its distribution boundaries via a two-dimensional convex hull area of the projected data points in Figure~\ref{fig:embedding_diversity_tsne_visualization}. We simultaneously visualize the distribution of t-SNE data points for both ground truth candidates, and baseline retrieved results (including A2A retrieved results and F2F results using the same embedding model), and retrieved candidates from \textsc{CoR}. The t-SNE Convex Hull visualization results also advocate our claim, where our method typically displays a more dispersed embedding distribution compared to dispersion of retrieved document embeddings from baselines.
\begin{table}[t]
\centering
\small
\renewcommand{\arraystretch}{0.8}
%\captionsetup{justification=justified, singlelinecheck=false}
%\caption{\small Wall-Clock Latency--Performance Trade-off of \textsc{CoR} across Different Retrieval Depths with asynchronous query expansion per retrieval depth, using BGE-M3 as retriever.}
%\vspace{-0.1in}
%\label{tab:latency_tradeoff_retrieval_depth}
\resizebox{\linewidth}{!}{
\setlength{\tabcolsep}{5pt}
\begin{tabular}{llcc}
\toprule
& \textbf{Methods} & \textbf{Wall Clock Latency (s)} & \textbf{Recall@200} \\
\midrule

\multicolumn{4}{l}{\textbf{Retrieval Depth 1}} \\
& A2A & 0.05 & 35.16 \\
& F2F & 1.02 & 36.06 \\
& \textbf{CoR w/ GPT-4o-Mini} & 9.74 & \textbf{41.17} \\
\midrule

\multicolumn{4}{l}{\textbf{Retrieval Depth 2}} \\
& A2A & 0.09 & 35.25 \\
& F2F & 2.01 & 36.82 \\
& \textbf{CoR w/ GPT-4o-Mini} & 21.33 & \textbf{42.01} \\
\midrule

\multicolumn{4}{l}{\textbf{Retrieval Depth 3}} \\
& A2A & 0.13 & 34.72 \\
& F2F & 3.02 & 36.12 \\
& \textbf{CoR w/ GPT-4o-Mini} & 41.69 & \textbf{42.75} \\

\bottomrule
\end{tabular}
}
\vspace{-0.05in}
\captionsetup{justification=justified, singlelinecheck=false}
%\caption{\small Wall-Clock Latency--Performance Trade-off of \textsc{CoR} across Different Retrieval Depths with asynchronous query expansion per retrieval depth, using BGE-M3 as retriever.}
\caption{Wall-Clock Latency--Performance Trade-off of \textsc{CoR} across Different Retrieval Depths with asynchronous query expansion per retrieval depth, using BGE-M3 as retriever.}
\label{tab:latency_tradeoff_retrieval_depth}
\end{table}
\begin{table}[t]
\centering
\small
\renewcommand{\arraystretch}{0.8}
%\captionsetup{justification=justified, singlelinecheck=false}
%\caption{\small Wall-Clock Latency-Performance Trade-off of \textsc{CoR} across Different Retrieval Budgets with asynchronous query expansion per retrieval budget, using BGE-M3 as retriever.}
%\vspace{-0.1in}
%\label{tab:latency_tradeoff_retrieval_budget}
\resizebox{\linewidth}{!}{
\setlength{\tabcolsep}{5pt}
\begin{tabular}{llcc}
\toprule
& \textbf{Methods} & \textbf{Wall Clock Latency (s)} & \textbf{Recall@200} \\
\midrule

\multicolumn{4}{l}{\textbf{Retrieval budget 4}} \\
& A2A & 0.17 & 34.14 \\
& F2F & 4.02 & 35.63 \\
& \textbf{CoR w/ GPT-4o-Mini} & 9.74 & \textbf{41.17} \\
\midrule

\multicolumn{4}{l}{\textbf{Retrieval budget 16}} \\
& A2A & 0.66 & 24.92 \\
& F2F & 16.16 & 26.46 \\
& \textbf{CoR w/ GPT-4o-Mini} & 21.33 & \textbf{42.01} \\
\midrule

\multicolumn{4}{l}{\textbf{Retrieval budget 52}} \\
& A2A & 2.29 & 12.95 \\
& F2F & 53.30 & 14.71 \\
& \textbf{CoR w/ GPT-4o-Mini} & 41.69 & \textbf{42.75} \\

\bottomrule
\end{tabular}
}
\vspace{-0.05in}
\captionsetup{justification=justified, singlelinecheck=false}
%\caption{\small Wall-Clock Latency-Performance Trade-off of \textsc{CoR} across Different Retrieval Budgets with asynchronous query expansion per retrieval budget, using BGE-M3 as retriever.}
\caption{Wall-Clock Latency-Performance Trade-off of \textsc{CoR} across Different Retrieval Budgets with asynchronous query expansion per retrieval budget, using BGE-M3 as retriever.}
\label{tab:latency_tradeoff_retrieval_budget}
\vspace{-0.1in}
\end{table}
\begin{table*}[t!]
%\caption{Comparison of retrieval performance under the Introduction-to-Introduction retrieval setup.}
%\vspace{-0.075in}
\setlength{\tabcolsep}{20pt}
\resizebox{\linewidth}{!}{
\renewcommand{\arraystretch}{0.8}
\begin{tabular}{llcccc}
\toprule

& & \multicolumn{2}{c}{\textbf{ACL-References}} & \multicolumn{2}{c}{\textbf{ACL-Citations}} \\
\cmidrule(lr){3-4} \cmidrule(lr){5-6}
& \textbf{Retrieval Method} & \textbf{nDCG@300} & \textbf{Recall@300} & \textbf{nDCG@300} & \textbf{Recall@300} \\

\midrule
\multicolumn{6}{l}{\textbf{JEmb-v2-Base-EN}} \\
& Intro-to-Intro & 23.06 & 34.97 & 24.32 & 34.41 \\
\cellcolor{blue!5} & \cellcolor{blue!5}\textbf{CoR w/ Llama-3.2-3B-Instruct} & \cellcolor{blue!5}\textbf{28.52} & \cellcolor{blue!5}\textbf{44.92} & \cellcolor{blue!5}\textbf{32.76} & \cellcolor{blue!5}\textbf{48.05} \\
\cellcolor{blue!5} & \cellcolor{blue!5}\textbf{CoR w/ QWEN-2.5-3B-Instruct} & \cellcolor{blue!5}\textbf{28.62} & \cellcolor{blue!5}\textbf{45.67} & \cellcolor{blue!5}\textbf{33.40} & \cellcolor{blue!5}\textbf{49.00} \\
\noalign{\vskip 0.2ex} 
\midrule

\multicolumn{6}{l}{\textbf{BGE-M3}} \\
& Intro-to-Intro & 19.64 & 30.20 & 21.48 & 30.36 \\
\cellcolor{blue!5} & \cellcolor{blue!5}\textbf{CoR w/ Llama-3.2-3B-Instruct} & \cellcolor{blue!5}\textbf{25.12} & \cellcolor{blue!5}\textbf{40.19} & \cellcolor{blue!5}\textbf{30.18} & \cellcolor{blue!5}\textbf{44.46} \\
\cellcolor{blue!5} & \cellcolor{blue!5}\textbf{CoR w/ QWEN-2.5-3B-Instruct} & \cellcolor{blue!5}\textbf{25.77} & \cellcolor{blue!5}\textbf{40.71} & \cellcolor{blue!5}\textbf{30.44} & \cellcolor{blue!5}\textbf{45.29} \\
\noalign{\vskip 0.2ex} 
\midrule

\multicolumn{6}{l}{\textbf{Inf-Retriever-v1-1.5B}} \\
& Intro-to-Intro & 34.03 & 51.94 & 31.72 & 45.89 \\
\cellcolor{blue!5} & \cellcolor{blue!5}\textbf{CoR w/ Llama-3.2-3B-Instruct} & \cellcolor{blue!5}\textbf{35.09} & \cellcolor{blue!5}\textbf{55.14} & \cellcolor{blue!5}\textbf{36.98} & \cellcolor{blue!5}\textbf{54.13} \\
\cellcolor{blue!5} & \cellcolor{blue!5}\textbf{CoR w/ QWEN-2.5-3B-Instruct} & \cellcolor{blue!5}\textbf{35.16} & \cellcolor{blue!5}\textbf{54.68} & \cellcolor{blue!5}\textbf{37.74} & \cellcolor{blue!5}\textbf{54.78} \\
\noalign{\vskip 0.2ex} 
\midrule

\bottomrule
\end{tabular}
}
\vspace{-0.075in}
%\caption{\small Comparison of retrieval performance under the Introduction-to-Introduction retrieval setup.}
\caption{Comparison of retrieval performance under the Introduction-to-Introduction retrieval setup.}
\vspace{-0.1in}
\label{tab:Intro_to_Intro_Retrieval}
\end{table*}
\paragraph{Latency Analysis} 
While the primary objective of our work is to improve retrieval effectiveness rather than efficiency, we also report the wall-clock latency of \textsc{CoR} in Tables~\ref{tab:latency_tradeoff_retrieval_depth}, \ref{tab:latency_tradeoff_retrieval_budget}, measured using asynchronous execution of query expansion along with performance compared to baseline A2A and F2F retrieval settings. Since the aspect-specific query expansions within each retrieval depth are independent, they can be executed asynchronously, so the wall-clock latency is governed primarily by the retrieval depth (a theoretical linear time complexity of $\mathcal{O}(r)$) rather than the total number of expansions, which is identical to baseline retrieval methods. The results demonstrate that under the identical retrieval depth and budgets, \textsc{CoR} is able to outperform baselines with comparable latency, where robust performance improvements can be observed proportional to retrieval depth and budget, displaying effective cost-performance tradeoff. In contrast, severe performance degradation can be observed in baselines with higher additional latency showing that \textsc{CoR} achieves a favorable effectiveness-efficiency trade-off.. Although latency reduction lies outside the scope of our work, exploring techniques such as intermediate-branching or search-space pruning would be an exciting direction for future work to further reduce latency.
\begin{table}[t!]
%\captionsetup{justification=justified, singlelinecheck=false}
%\caption{\small Results of applying \textsc{CoR} to the A2A retrieval setting on \textsc{SciFullBench} with two retrieval rounds.}
\resizebox{0.975\linewidth}{!}{
\renewcommand{\arraystretch}{1.0}
\begin{tabular}{llcc}
\toprule
& & \textbf{ICLR-References} & \textbf{ICLR-Citations} \\
\cmidrule(lr){3-3} \cmidrule(lr){4-4}
& \textbf{Retrieval Method} & Recall@300 & Recall@300 \\
\midrule

\multicolumn{4}{l}{\textbf{Baseline}} \\
&JEmb-v2 (A2A)  & 48.38 & 44.49 \\
&BGE-M3 (A2A)  & 41.95 & 39.61 \\
&Inf-Ret-v1 (A2A)  & 59.66 & 51.56 \\
\midrule
\midrule
\multicolumn{4}{l}{\textbf{Ours\textsc{ (CoR)}}} \\
&\textbf{GPT-4o-Mini + JEmb-v2} & \textbf{49.47} & \textbf{46.41} \\
&\textbf{GPT-4o-Mini + BGE-M3} & \textbf{43.12} & \textbf{41.92} \\
&\textbf{GPT-4.1 + Inf-Ret-v1} & \textbf{60.95} & \textbf{53.27} \\

\bottomrule
\end{tabular}

\vspace{-0.05in}}
\captionsetup{justification=justified, singlelinecheck=false}
%\caption{\small Results of applying \textsc{CoR} to the A2A retrieval setting on \textsc{SciFullBench} with two retrieval rounds.}
\caption{Results of applying \textsc{CoR} to the A2A retrieval setting on \textsc{SciFullBench} with two retrieval rounds.}
\label{tab:abstract-to-abstract retrieval generalization SciFullBench.}
\vspace{-0.1in}
\end{table}
\paragraph{Comparison against Section-Level-Retrieval Baselines}
Moreover, we perform comparisons using retrieval setup using the subsection of a paper as query and candidates, particularly using Introduction section of papers as query and candidates in place of full paper or abstract content. Despite the difficulty in parsing the introduction section from papers, we have made our best effort to parse the introduction section through python regular expression while making sure to use full-paper content when introduction cannot be explicitly parsed. The results in Table \ref{tab:Intro_to_Intro_Retrieval} demonstrate that \textsc{CoR} is able to obtain substantial improvements over such baselines. 

\paragraph{Generalization to the A2A Retrieval Setting}
While our framework is primarily designed for the full-paper–to-full-paper retrieval setting, we additionally evaluate whether \textsc{CoR} is applicable to the A2A setting. To this end, we report its performance on the A2A configuration of \textsc{SciFullBench} in Table~\ref{tab:abstract-to-abstract retrieval generalization SciFullBench.}, where title–abstract pairs are used as inputs and retrieval candidates, %and retrieval is performed over an abstract-only corpus
while keeping the same query-optimization prompts. The results demonstrate that \textsc{CoR} is not limited to the full-paper–to-full-paper setting, but generalizes effectively to the A2A scenario, yielding meaningful performance gains.
\begin{table}[t!]

%\captionsetup{justification=justified, singlelinecheck=false}
%\caption{\small Performance comparison against baselines augmented with metadata (i.e., author information).}

\resizebox{0.975\linewidth}{!}{
\renewcommand{\arraystretch}{0.95}
\begin{tabular}{llcc}
\toprule
& & \textbf{ACL-References} & \textbf{ACL-Citations} \\
\cmidrule(lr){3-3} \cmidrule(lr){4-4}
& \textbf{Retrieval Method} & Recall@200 & Recall@200 \\
\midrule

\multicolumn{4}{l}{\textbf{Baseline w/ Author Information}} \\
&JEmb-v2 (A2A)  & 34.25 & 34.84 \\
&JEmb-v2 (F2F)  & 36.54 & 36.04 \\
\noalign{\vskip 1mm}
\cdashline{1-4}
\noalign{\vskip 1mm}
&BGE-M3 (A2A)  & 32.16 & 33.73 \\
&BGE-M3 (F2F)  & 31.62 & 32.66 \\
\noalign{\vskip 1mm}
\cdashline{1-4}
\noalign{\vskip 1mm}
&Inf-Ret-v1 (A2A)  & 46.66 & 42.36 \\
&Inf-Ret-v1 (F2F)  & 44.31 & 38.92 \\

\midrule
\midrule
\multicolumn{4}{l}{\textbf{Ours\textsc{ (CoR)}}} \\
&\textbf{Llama-3.2-3B-Inst + JEmb-v2} & \textbf{39.81} & \textbf{42.93} \\
&\textbf{Llama-3.2-3B-Inst + BGE-M3} & \textbf{35.75} & \textbf{39.48} \\
&\textbf{Llama-3.2-3B-Inst + Inf-Ret-v1} & \textbf{49.30} & \textbf{48.23} \\

\bottomrule
\end{tabular}
}
\vspace{-0.05in}
\captionsetup{justification=justified, singlelinecheck=false}
%\caption{\small Performance comparison against baselines augmented with metadata (i.e., author information).}
\caption{Performance comparison against baselines augmented with metadata (i.e., author information).}
\label{tab:Additional_MetaData_Incorporation}
\vspace{-0.1in}
\end{table}
\paragraph{Comparison against Metadata-Augmented Baselines}
We also consider a setting where additional metadata is provided to the baselines by concatenating author information to both queries and candidate documents. As shown in Table~\ref{tab:Additional_MetaData_Incorporation}, \textsc{CoR} remains effective under this comparison and continues to outperform the baselines even when they are augmented with author metadata.
\begin{table*}[!t]
\centering
\small
\renewcommand{\arraystretch}{0.95}
%\captionsetup{justification=justified, singlelinecheck=false}
%\caption{\small Comparison of retrieval performance between the proposed method and the A2A and F2F baselines using auxiliary reranking with \texttt{bge-reranker-v2-m3} on the ICLR split of \textsc{SciFullBench}.}
%\label{tab:reranking}
\resizebox{\linewidth}{!}{
\setlength{\tabcolsep}{8pt}
\begin{tabular}{llcccc}
\toprule
& \textbf{Methods} & \textbf{nDCG@200 (Ref)} & \textbf{Recall@200 (Ref)} & \textbf{nDCG@200 (Cite)} & \textbf{Recall@200 (Cite)} \\
\midrule

\multicolumn{6}{l}{\textbf{Jina-Embeddings-v2-BASE-EN}} \\
& A2A w/ reranking & 28.95 & 42.98 & 26.25 & 38.69 \\
& F2F w/ reranking & 19.80 & 36.96 & 22.04 & 37.24 \\
& \cellcolor{blue!5}\textbf{\textsc{CoR} w/ Llama-3.2-3B-Instruct} & \cellcolor{blue!5}\textbf{35.05} & \cellcolor{blue!5}\textbf{48.71} & \cellcolor{blue!5}\textbf{33.94} & \cellcolor{blue!5}\textbf{47.12} \\
& \cellcolor{blue!5}\textbf{\textsc{CoR} w/ Qwen-2.5-3B-Instruct} & \cellcolor{blue!5}\textbf{34.64} & \cellcolor{blue!5}\textbf{48.33} & \cellcolor{blue!5}\textbf{34.16} & \cellcolor{blue!5}\textbf{47.43} \\
\midrule

\multicolumn{6}{l}{\textbf{BGE-M3}} \\
& A2A w/ reranking & 26.82 & 38.48 & 24.21 & 34.93 \\
& F2F w/ reranking & 17.60 & 33.45 & 18.62 & 31.48 \\
& \cellcolor{blue!5}\textbf{\textsc{CoR} w/ Llama-3.2-3B-Instruct} & \cellcolor{blue!5}\textbf{29.47} & \cellcolor{blue!5}\textbf{42.47} & \cellcolor{blue!5}\textbf{29.99} & \cellcolor{blue!5}\textbf{41.44} \\
& \cellcolor{blue!5}\textbf{\textsc{CoR} w/ Qwen-2.5-3B-Instruct} & \cellcolor{blue!5}\textbf{30.32} & \cellcolor{blue!5}\textbf{43.28} & \cellcolor{blue!5}\textbf{30.95} & \cellcolor{blue!5}\textbf{42.81} \\
\midrule

\multicolumn{6}{l}{\textbf{Inf-Retriever-v1-1.5B}} \\
& A2A w/ reranking & 32.55 & 50.41 & 28.39 & 43.01 \\
& F2F w/ reranking & 20.54 & 39.45 & 21.08 & 36.71 \\
& \cellcolor{blue!5}\textbf{\textsc{CoR} w/ Llama-3.2-3B-Instruct} & \cellcolor{blue!5}\textbf{42.42} & \cellcolor{blue!5}\textbf{57.14} & \cellcolor{blue!5}\textbf{38.26} & \cellcolor{blue!5}\textbf{52.33} \\
& \cellcolor{blue!5}\textbf{\textsc{CoR} w/ Qwen-2.5-3B-Instruct} & \cellcolor{blue!5}\textbf{42.54} & \cellcolor{blue!5}\textbf{57.22} & \cellcolor{blue!5}\textbf{38.83} & \cellcolor{blue!5}\textbf{52.79} \\

\bottomrule
\end{tabular}
}
\captionsetup{justification=justified, singlelinecheck=false}
%\caption{\small Comparison of retrieval performance between the proposed method and the A2A and F2F baselines using auxiliary reranking with \texttt{bge-reranker-v2-m3} on the ICLR split of \textsc{SciFullBench}.}
\vspace{-0.05in}
\caption{Comparison of retrieval performance between the proposed method and the A2A and F2F baselines using auxiliary reranking with \texttt{bge-reranker-v2-m3} on the ICLR split of \textsc{SciFullBench}.}
\label{tab:reranking}
\vspace{-0.1in}
\end{table*}
\begin{table}[t!]
%\captionsetup{justification=justified, singlelinecheck=false}
%\caption{\small Performance analysis when a single comprehensive query covering the three aspects is used.}

\resizebox{0.975\linewidth}{!}{
\renewcommand{\arraystretch}{1.05}
\begin{tabular}{llcc}
\toprule
& & \textbf{ACL-Citations} & \textbf{ICLR-Citations} \\
\cmidrule(lr){3-3} \cmidrule(lr){4-4}
& \textbf{Retrieval Method} & Recall@200 & Recall@200 \\
\midrule

\multicolumn{4}{l}{\textbf{Single Comprehensive Query w/o A2A}} \\
&QWEN-2.5-3B-Inst + JEmb-v2  & 38.66 & 41.68 \\
&Llama-3.2-3B-Inst + BGE-M3  & 32.45 & 34.03 \\
&Llama-3.2-3B-Inst + Inf-Ret-v1  & 45.36 & 48.55 \\
\midrule
\midrule
\multicolumn{4}{l}{\textbf{Ours (CoR) w/o A2A}} \\
&\textbf{QWEN-2.5-3B-Inst + JEmb-v2} & \textbf{41.13} & \textbf{44.41} \\
&\textbf{Llama-3.2-3B-Inst + BGE-M3} & \textbf{35.83} & \textbf{37.49} \\
&\textbf{Llama-3.2-3B-Inst + Inf-Ret-v1} & \textbf{47.05} & \textbf{50.65} \\

\bottomrule
\end{tabular}
}
\captionsetup{justification=justified, singlelinecheck=false}
%\caption{\small Performance analysis when a single comprehensive query covering the three aspects is used.}
\vspace{-0.05in}
\caption{Performance analysis when a single comprehensive query covering the three aspects is used.}
\vspace{-0.1in}
\label{tab:single_comprehensive_query_comparison_study}
\end{table}
\paragraph{Comparison against Re-Ranking Baselines} We further compare \textsc{CoR} against two-stage retrieval--reranking baselines for paper-to-paper retrieval. Specifically, we augment the A2A and F2F baselines with the powerful \texttt{bge-reranker-v2-m3} reranker~\citep{li2023making,bge-m3}. As shown in Table~\ref{tab:reranking}, \textsc{CoR} consistently achieves higher retrieval performance than these reranking-enhanced baselines.

\paragraph{Comparison with a Single Aspect-Aware Comprehensive Query}
To examine the effectiveness of our multiple aspect-aware query design, we compare \textsc{CoR} against a baseline that uses a single comprehensive query covering all three aspects simultaneously. As shown in Table~\ref{tab:single_comprehensive_query_comparison_study}, \textsc{CoR} with multiple aspect-specific queries outperforms the single-query baseline, indicating the benefit of separating queries by aspect in paper-to-paper retrieval.

\paragraph{Comparison against other Aggregation Algorithms} We further compare \textsc{CoR} with an alternative aggregation strategy, namely BFS aggregation, which aggregates retrieval results in the order of retrieval depth, starting from the initial retrieval and then sequentially incorporating deeper multi-hop retrieval results. In contrast, Post-Order Aggregation first aggregates deeper retrieval branches before propagating their information upward. As shown in Table~\ref{tab:bfs_postorder}, Post-Order Aggregation consistently achieves higher retrieval performance than BFS aggregation. These results suggest that aggregating deeper branches first helps alleviate the accumulation of irrelevant documents introduced at larger retrieval depths, resulting in more robust retrieval performance.
\begin{table}[t]
\centering
\small
\renewcommand{\arraystretch}{1.25}
%\captionsetup{justification=justified, singlelinecheck=false}
%\caption{\small Performance comparison of BFS Aggregation and Post-Order Aggregation using the BGE-M3 retriever on the ICLR-Citations split of \textsc{SciFullBench}.}
%\label{tab:bfs_postorder}
\resizebox{0.99\linewidth}{!}{
\setlength{\tabcolsep}{2pt}
\begin{tabular}{llc}
\toprule
\textbf{Methods} & & \textbf{Recall@300} \\
\midrule
\multicolumn{3}{l}{\textbf{BFS Aggregation}} \\
&\textsc{CoR} w/ Llama-3.2-3B-Inst w/ Depth 2 & 43.17 \\
&\textsc{CoR} w/ Llama-3.2-3B-Inst w/ Depth 3 & 38.97 \\
\midrule
\midrule
\multicolumn{3}{l}{\textbf{Post-Order Aggregation}} \\
&\textbf{\textsc{CoR} w/ Llama-3.2-3B-Inst w/ Depth 2} & \textbf{46.16} \\
&\textbf{\textsc{CoR} w/ Llama-3.2-3B-Inst w/ Depth 3} & \textbf{46.53} \\
\bottomrule
\end{tabular}
}
\captionsetup{justification=justified, singlelinecheck=false}
%\caption{\small Performance comparison of BFS Aggregation and Post-Order Aggregation in \textsc{CoR} using the BGE-M3 retriever on the ICLR-Citations split of \textsc{SciFullBench}.}
\vspace{-0.05in}
\caption{Performance comparison of BFS Aggregation and Post-Order Aggregation in \textsc{CoR} using the BGE-M3 retriever on the ICLR-Citations split of \textsc{SciFullBench}.}
\label{tab:bfs_postorder}
\vspace{-0.1in}
\end{table}
\paragraph{Case Study} 
We provide the qualitative examples of the retrieved results from \textsc{SciFullBench} and \textsc{PatentFullBench} in Table~\ref{tab:retrieved_case_analysis} and Table~\ref{tab:retrieved_case_analysis_patents}, respectively, as well as the examples of the generated queries from both scientific papers and patents in Figures \ref{app:fig:Query_Case_Analysis} and \ref{app:fig:Query_Case_Analysis_Patent}, in Appendix \ref{appendix:Case_Study}.

\section{Robustness across Benchmarks and Metrics}
\label{appendix:Robustness_Experiments}

\paragraph{Effectiveness on the Existing Benchmark (A2A)}
While our main experiments are conducted on \textsc{SciFullBench}, which provides a full-paper–to-full-paper retrieval setting with more recent publications, we also examine whether our findings hold under previously established benchmarks. To this end, we additionally evaluate \textsc{CoR} on the SciDocs-Cocite split~\cite{cohan-etal-2020-specter}, which is originally defined under an A2A retrieval setting, using 993 queries whose abstract content is available\footnote{\url{https://huggingface.co/datasets/allenai/scirepeval}}. As shown in Table~\ref{tab:SciDocs_Cocite}, \textsc{CoR} remains effective under the existing benchmark with A2A setting, yielding statistically meaningful improvements over baseline methods. In addition, in Table \ref{tab:CFSCube_results}, we report evaluation results on the CSFCube benchmark~\citep{mysore2021csfcube}, which consists of a total of 50 queries and with human-annotated ground truth signals based on three aspects (e.g., background, method, result) and demonstrate that \textsc{CoR} is able to gain robust performance improvements using diverse embedding models.

\begin{table}[t!]
%\captionsetup{justification=justified, singlelinecheck=false}
%\caption{\small Performance of \textsc{CoR} on the SciDocs benchmark under the A2A setup; statistically meaningful gains are \underline{underlined} based on paired t-tests.}
\resizebox{\linewidth}{!}{
\renewcommand{\arraystretch}{0.93}
\begin{tabular}{llccc}
\toprule
& \textbf{Retrieval Method} & \textbf{Recall@10} & \textbf{mAP@10} & \textbf{nDCG@10} \\
\midrule
\multicolumn{4}{l}{\textbf{A2A Retrieval}} \\
& JEmb-v2 (A2A) & 92.46 $\pm$ \textbf{\scriptsize{0.00}} & 85.53 $\pm$ \textbf{\scriptsize{0.00}} & 91.66 $\pm$ \textbf{\scriptsize{0.00}} \\
& BGE-M3 (A2A) & 90.36 $\pm$ \textbf{\scriptsize{0.00}} & 80.88 $\pm$ \textbf{\scriptsize{0.00}} & 88.70 $\pm$ \textbf{\scriptsize{0.00}} \\
& Inf-Ret-v1 (A2A) & 95.69 $\pm$ \textbf{\scriptsize{0.00}} & 90.02 $\pm$ \textbf{\scriptsize{0.00}} & 94.44 $\pm$ \textbf{\scriptsize{0.00}} \\
\noalign{\vskip 0.2ex} 
\midrule
\midrule
\multicolumn{4}{l}{\textbf{Ours (\textsc{CoR}) w/ A2A}} \\
& \textbf{GPT-4.1 + JEmb-v2} & \underline{\textbf{93.20} $\pm$ \textbf{\scriptsize{0.09}}} & \underline{\textbf{86.20} $\pm$ \textbf{\scriptsize{0.08}}} & \underline{\textbf{92.18} $\pm$ \textbf{\scriptsize{0.06}}} \\
& \textbf{GPT-4.1 + BGE-M3} & \textbf{90.52} $\pm$ \textbf{\scriptsize{0.07}} & \underline{\textbf{81.15} $\pm$ \textbf{\scriptsize{0.10}}} & \underline{\textbf{88.88}} $\pm$ \textbf{\scriptsize{0.04}} \\
& \textbf{GPT-4.1 + Inf-Ret-v1} & \underline{\textbf{96.16} $\pm$ \textbf{\scriptsize{0.10}}} & \underline{\textbf{90.40} $\pm$ \textbf{\scriptsize{0.03}}} & \underline{\textbf{94.80} $\pm$ \textbf{\scriptsize{0.04}}} \\
\bottomrule
\end{tabular}
}
\vspace{-0.05in}
\captionsetup{justification=justified, singlelinecheck=false}
%\caption{\small Performance of \textsc{CoR} on the SciDocs benchmark under the A2A setup; statistically meaningful gains are \underline{underlined} based on paired t-tests.}
\caption{Performance of \textsc{CoR} on the SciDocs benchmark under the A2A setup; statistically meaningful gains are \underline{underlined} based on paired t-tests.}
\label{tab:SciDocs_Cocite}
\end{table}

\begin{table}[t!]
%\captionsetup{justification=justified, singlelinecheck=false}
%\caption{\small Performance of \textsc{CoR} on the SciDocs benchmark under the F2F setup.}
\resizebox{\linewidth}{!}{
\renewcommand{\arraystretch}{0.9}
\begin{tabular}{llccc}
\toprule
& \textbf{Retrieval Method} & \textbf{Recall@5} & \textbf{Recall@10} & \textbf{Average} \\
\midrule
\multicolumn{4}{l}{\textbf{A2A Retrieval}} \\
& JEmb-v2 (A2A) & 63.61 $\pm$ \textbf{\scriptsize{0.00}} & 69.86 $\pm$ \textbf{\scriptsize{0.00}} & 66.74 \\
& BGE-M3 (A2A) & 55.56 $\pm$ \textbf{\scriptsize{0.00}} & 65.97 $\pm$ \textbf{\scriptsize{0.00}} & 60.77 \\
& Inf-Ret-v1 (A2A) & 65.28 $\pm$ \textbf{\scriptsize{0.00}} & 76.11 $\pm$ \textbf{\scriptsize{0.00}} & 70.70 \\
\noalign{\vskip 0.2ex} 
\midrule
\multicolumn{4}{l}{\textbf{F2F Retrieval}} \\
& JEmb-v2 (F2F) & 65.97 $\pm$ \textbf{\scriptsize{0.00}} & 71.94 $\pm$ \textbf{\scriptsize{0.00}} & 68.96 \\
& BGE-M3 (F2F) & 53.75 $\pm$ \textbf{\scriptsize{0.00}} & 62.92 $\pm$ \textbf{\scriptsize{0.00}} & 58.34 \\
& Inf-Ret-v1 (F2F) & 65.42 $\pm$ \textbf{\scriptsize{0.00}} & 73.89 $\pm$ \textbf{\scriptsize{0.00}} & 69.66 \\
\midrule
\midrule
\multicolumn{4}{l}{\textbf{Ours (\textsc{CoR})}} \\
& \textbf{GPT-4.1 + JEmb-v2} & \textbf{66.25} $\pm$ \textbf{\scriptsize{1.58}} & 71.53 $\pm$ \textbf{\scriptsize{1.10}} & 68.89 \\
& \textbf{GPT-4.1 + BGE-M3} & \textbf{58.33} $\pm$ \textbf{\scriptsize{0.64}} & \textbf{66.81} $\pm$ \textbf{\scriptsize{0.96}} & \textbf{62.57} \\
& \textbf{GPT-4.1 + Inf-Ret-v1} & \textbf{66.25} $\pm$ \textbf{\scriptsize{1.46}} & \textbf{77.50} $\pm$ \textbf{\scriptsize{0.96}} & \textbf{71.88} \\
\bottomrule
\end{tabular}
}
\vspace{-0.05in}
\captionsetup{justification=justified, singlelinecheck=false}
%\caption{\small Performance of \textsc{CoR} on the SciDocs benchmark under the F2F setup.}
\caption{Performance of \textsc{CoR} on the SciDocs benchmark under the F2F setup.}
\label{tab:SciDocs_Full}
\end{table}

\paragraph{Effectiveness on the Existing Benchmark (F2F)}
In addition to evaluating \textsc{CoR} on the existing A2A benchmark setting of SciDocs, we further extend this benchmark to a F2F configuration. For this purpose, we construct a SciDocs-Full variant from the Cocite, CoRead, and CoView splits of SciDocs~\cite{cohan-etal-2020-specter}, where both queries and candidate documents contain full-paper content, following the same preprocessing pipeline as \textsc{SciFullBench} introduced in Appendix~\ref{appendix:SciFullBench}. As shown in Table~\ref{tab:SciDocs_Full}, \textsc{CoR} yields consistent improvements over baselines on this extended version of the benchmark, indicating that our method remains also effective across different retrieval configurations within the established benchmarks.

\paragraph{Effectiveness on Other Metrics} In Table \ref{tab:main_result_w_lower_k_metrics}, we show the effectiveness of \textsc{CoR} under lower-$k$ cutoff values on SciFullBench, where results are averaged between respective splits of two venues (ACL, EMNLP). Meanwhile, Tables \ref{tab:Human_Evaluation}, \ref{tab:SciDocs_Cocite}, \ref{tab:SciDocs_Full}, \ref{tab:CFSCube_results}, and \ref{tab:main_final_w_additional_metrics} report results at relatively lower $k$ metrics as well. Overall, the results indicate that \textsc{CoR} achieves consistent performance gains even at lower $k$ values, demonstrating its robustness under user-relevant evaluation settings.

\begin{table}[t!]
%\captionsetup{justification=justified, singlelinecheck=false}
%\caption{\small Aggregated Performance of \textsc{CoR} on the CSFCube benchmark under the A2A setup using GPT-4.1 as query optimizers. We report the mean and standard deviation across three independent trials.}

\setlength{\tabcolsep}{9pt}
\resizebox{\linewidth}{!}{
\renewcommand{\arraystretch}{0.6}
\begin{tabular}{llccc}
\toprule

\midrule
& \textbf{Retrieval Method} & \textbf{nDCG@20} & \textbf{mAP@20} & \textbf{Recall@20} \\
\midrule
\multicolumn{4}{l}{\textbf{JEmb-v2-Base-EN}} \\
& A2A & 60.26 $\pm$ \textbf{\scriptsize{0.00}} & 30.67 $\pm$ \textbf{\scriptsize{0.00}} & 39.48 $\pm$ \textbf{\scriptsize{0.00}} \\
& \textbf{\textsc{CoR} w/ A2A} & \textbf{60.73} $\pm$ \textbf{\scriptsize{0.14}} & \textbf{31.89} $\pm$ \textbf{\scriptsize{0.22}} & \textbf{40.31} $\pm$ \textbf{\scriptsize{0.21}} \\
\noalign{\vskip 0.2ex} 
\midrule
\multicolumn{4}{l}{\textbf{BGE-M3}} \\
& A2A & 57.83 $\pm$ \textbf{\scriptsize{0.00}} & 28.78 $\pm$ \textbf{\scriptsize{0.00}} & 37.45 $\pm$ \textbf{\scriptsize{0.00}} \\
& \textbf{\textsc{CoR} w/ A2A} & \textbf{59.07} $\pm$ \textbf{\scriptsize{0.07}} & \textbf{30.37} $\pm$ \textbf{\scriptsize{0.23}} & \textbf{38.96} $\pm$ \textbf{\scriptsize{0.23}} \\
\noalign{\vskip 0.2ex} 
\midrule
\multicolumn{4}{l}{\textbf{Inf-Ret-v1-1.5B}} \\
& A2A & 63.80 $\pm$ \textbf{\scriptsize{0.00}} & 33.77 $\pm$ \textbf{\scriptsize{0.00}} & 41.86 $\pm$ \textbf{\scriptsize{0.00}} \\
& \textbf{\textsc{CoR} w/ A2A} & \textbf{65.26} $\pm$ \textbf{\scriptsize{0.41}} & \textbf{35.26} $\pm$ \textbf{\scriptsize{0.29}} & \textbf{43.44} $\pm$ \textbf{\scriptsize{0.35}} \\
\noalign{\vskip 0.2ex} 
\midrule
\multicolumn{4}{l}{\textbf{Granite-Emb-Eng-R2}} \\
& A2A & 64.37 $\pm$ \textbf{\scriptsize{0.00}} & 33.52 $\pm$ \textbf{\scriptsize{0.00}} & 41.98 $\pm$ \textbf{\scriptsize{0.00}} \\
& \textbf{\textsc{CoR} w/ A2A} & \textbf{66.53} $\pm$ \textbf{\scriptsize{0.36}} & \textbf{35.55} $\pm$ \textbf{\scriptsize{0.15}} & \textbf{43.57} $\pm$ \textbf{\scriptsize{0.27}} \\
\noalign{\vskip 0.2ex} 
\midrule
\multicolumn{4}{l}{\textbf{QWEN3-0.6B-Emb}} \\
& A2A & 64.13 $\pm$ \textbf{\scriptsize{0.00}} & 33.54 $\pm$ \textbf{\scriptsize{0.00}} & 42.35 $\pm$ \textbf{\scriptsize{0.00}} \\
& \textbf{\textsc{CoR} w/ A2A} & \textbf{65.28} $\pm$ \textbf{\scriptsize{0.49}} & \textbf{35.46} $\pm$ \textbf{\scriptsize{0.03}} & \textbf{43.59} $\pm$ \textbf{\scriptsize{0.20}} \\
\noalign{\vskip 0.2ex} 
\midrule
\multicolumn{4}{l}{\textbf{Dewey-EN-Beta}} \\
& A2A & 61.61 $\pm$ \textbf{\scriptsize{0.00}} &  31.95 $\pm$ \textbf{\scriptsize{0.00}} & 40.05 $\pm$ \textbf{\scriptsize{0.00}} \\
& \textbf{\textsc{CoR} w/ A2A} & \textbf{63.77} $\pm$ \textbf{\scriptsize{0.55}} & \textbf{33.46} $\pm$ \textbf{\scriptsize{0.23}} & \textbf{41.76} $\pm$ \textbf{\scriptsize{0.09}} \\
\noalign{\vskip 0.2ex} 
\midrule

\bottomrule
\end{tabular}
}
\vspace{-0.05in}
\captionsetup{justification=justified, singlelinecheck=false}
%\caption{\small Aggregated Performance of \textsc{CoR} on the CSFCube benchmark under the A2A setup using GPT-4.1 as query optimizers. We report the mean and standard deviation across three independent trials.}
\caption{Aggregated Performance of \textsc{CoR} on the CSFCube benchmark under the A2A setup using GPT-4.1 as query optimizers. We report the mean and standard deviation across three independent trials.}
\vspace{-0.1in}
\label{tab:CFSCube_results}
\end{table}

\clearpage
\begin{table*}[t!]
%\caption{Main results with additional metrics.}
%\vspace{-0.1in}
\renewcommand{\arraystretch}{1.6}
\resizebox{\linewidth}{!}{
\begin{tabular}{llcccccccc}

\midrule \toprule \rule{0pt}{3.5ex}%
& & \multicolumn{4}{c}{\fontsize{22pt}{22pt}\selectfont\ \bf ICLR} & \multicolumn{4}{c}{\fontsize{22pt}{22pt}\selectfont\ \bf NeurIPS} \\
\cmidrule(l{2pt}r{2pt}){3-6} \cmidrule(l{2pt}r{2pt}){7-10}
\rule{0pt}{3.5ex}%
& & \multicolumn{2}{c}{\fontsize{22pt}{22pt}\selectfont \bf References}& \multicolumn{2}{c}{\fontsize{22pt}{22pt}\selectfont\ \bf Citations} 
  & \multicolumn{2}{c}{\fontsize{22pt}{22pt}\selectfont \bf References} & \multicolumn{2}{c}{\fontsize{22pt}{22pt}\selectfont \bf Citations} \\
\cmidrule(l{2pt}r{2pt}){3-4} \cmidrule(l{2pt}r{2pt}){5-6}
\cmidrule(l{2pt}r{2pt}){7-8} \cmidrule(l{2pt}r{2pt}){9-10}
& \fontsize{22pt}{20pt}\selectfont\ \textbf{IR Method} 
& \fontsize{20pt}{20pt}\selectfont\ nDCG@300 & 
\fontsize{20pt}{20pt}\selectfont Recall@300 & 
\fontsize{20pt}{20pt}\selectfont\ nDCG@300 &
\fontsize{20pt}{20pt}\selectfont\ Recall@300 & 
\fontsize{20pt}{20pt}\selectfont\ nDCG@300 & 
\fontsize{20pt}{20pt}\selectfont\ Recall@300 & 
\fontsize{20pt}{20pt}\selectfont\ nDCG@300 & 
\fontsize{20pt}{20pt}\selectfont\ Recall@300 \\
\midrule
\noalign{\vskip 0.5ex} 
& \multicolumn{9}{l}{\textbf{\fontsize{20pt}{20pt}\selectfont Lexical-Based Retrievers}} \\
\rule{0pt}{2ex}%
& \fontsize{20pt}{20pt}\selectfont\ BM-25 (A2A) & \fontsize{21pt}{21pt}\selectfont\ 24.78 & \fontsize{21pt}{21pt}\selectfont\ 34.69 & \fontsize{21pt}{21pt}\selectfont\ 23.35 & \fontsize{21pt}{21pt}\selectfont\ 32.59 & \fontsize{21pt}{21pt}\selectfont\ 31.79  & \fontsize{21pt}{21pt}\selectfont\ 41.73 & \fontsize{21pt}{21pt}\selectfont\ 31.68 & \fontsize{21pt}{21pt}\selectfont\ 42.22 \\
& \fontsize{20pt}{20pt}\selectfont\ BM-25 (F2F) & \fontsize{21pt}{21pt}\selectfont\ 31.09 & \fontsize{21pt}{21pt}\selectfont\ 44.54 & \fontsize{21pt}{21pt}\selectfont\ 34.57 & \fontsize{21pt}{21pt}\selectfont\ 46.28 & \fontsize{21pt}{21pt}\selectfont\ 21.59 & \fontsize{21pt}{21pt}\selectfont\ 29.49 & \fontsize{21pt}{21pt}\selectfont\ 38.38 & \fontsize{21pt}{21pt}\selectfont\ 48.65 \\
\midrule\midrule
& \multicolumn{9}{l}{\textbf{\fontsize{20pt}{20pt}\selectfont Domain-Specific Retriever}} \\
\rule{0pt}{2ex}%
& \fontsize{20pt}{20pt}\selectfont\ SciNCL-\textbf{A2A} & \fontsize{21pt}{21pt}\selectfont\ 34.04 & \fontsize{21pt}{21pt}\selectfont\ 50.51 & \fontsize{21pt}{21pt}\selectfont\ 29.57 & \fontsize{21pt}{21pt}\selectfont\ 44.13 & \fontsize{21pt}{21pt}\selectfont\ 38.44 & \fontsize{21pt}{21pt}\selectfont\ 53.08 & \fontsize{21pt}{21pt}\selectfont\ 36.57 & \fontsize{21pt}{21pt}\selectfont\ 51.67 \\

& \fontsize{20pt}{20pt}\selectfont\ SPECTER2-Base-\textbf{A2A} & \fontsize{21pt}{21pt}\selectfont\ 32.71 & \fontsize{21pt}{21pt}\selectfont\ 48.99 & \fontsize{21pt}{21pt}\selectfont\ 30.97  & \fontsize{21pt}{21pt}\selectfont\ 45.23 & \fontsize{21pt}{21pt}\selectfont\ 37.77 & \fontsize{21pt}{21pt}\selectfont\ 51.82 & \fontsize{21pt}{21pt}\selectfont\ 37.99 & \fontsize{20pt}{20pt}\selectfont\ 52.95 \\

& \fontsize{20pt}{20pt}\selectfont\ SPECTER2-Adapter-MTL CTRL-\textbf{A2A} & \fontsize{21pt}{21pt}\selectfont\ 33.91 & \fontsize{20pt}{20pt}\selectfont\ 49.98 & \fontsize{21pt}{21pt}\selectfont\ 29.49 & \fontsize{20pt}{20pt}\selectfont\ 43.29& \fontsize{21pt}{21pt}\selectfont\ 38.54 & \fontsize{20pt}{20pt}\selectfont\ 52.16 & \fontsize{21pt}{21pt}\selectfont\ 36.75 & \fontsize{20pt}{20pt}\selectfont\ 51.46 \\

& \fontsize{20pt}{20pt}\selectfont\ SciMult-MHAExpert-\textbf{A2A} & \fontsize{21pt}{21pt}\selectfont\ 28.26 & \fontsize{21pt}{21pt}\selectfont\ 42.56 & \fontsize{21pt}{21pt}\selectfont\ 24.61 & \fontsize{21pt}{21pt}\selectfont\ 36.88 & \fontsize{21pt}{21pt}\selectfont\ 33.64 & \fontsize{21pt}{21pt}\selectfont\ 47.52 & \fontsize{21pt}{21pt}\selectfont\ 32.03 & \fontsize{21pt}{21pt}\selectfont\ 45.25 \\

\midrule\midrule
\noalign{\vskip 0.5ex} 
& \multicolumn{9}{l}{\textbf{\fontsize{20pt}{20pt}\selectfont Jina-Embeddings-v2-BASE-EN}} \\
\multirow{12}{*}{\raisebox{-9ex}[0pt][0pt]{\rotatebox[origin=c]{90}{\textbf{\fontsize{26pt}{26pt}\selectfont Domain-Agnostic Retriever}}}} & 
\fontsize{20pt}{20pt}\selectfont\ A2A (Abstract-to-Abstract)& \fontsize{20pt}{21pt}\selectfont\ 34.43 & \fontsize{21pt}{21pt}\selectfont\ 48.38 & \fontsize{21pt}{21pt}\selectfont\ 30.47 & \fontsize{21pt}{21pt}\selectfont\ 44.49 & \fontsize{21pt}{21pt}\selectfont\ 39.12 & \fontsize{21pt}{21pt}\selectfont\ 51.45 & \fontsize{21pt}{21pt}\selectfont\ 37.22 & \fontsize{21pt}{21pt}\selectfont\ 51.28 \\

& \fontsize{20pt}{20pt}\selectfont\ F2F (Full-to-Full) & \fontsize{21pt}{21pt}\selectfont\ 35.27 & \fontsize{21pt}{21pt}\selectfont\ 49.58 & \fontsize{21pt}{21pt}\selectfont\ 33.71 & \fontsize{21pt}{21pt}\selectfont\ 47.85 & \fontsize{21pt}{21pt}\selectfont\ 38.82 & \fontsize{21pt}{21pt}\selectfont\ 51.62 & \fontsize{21pt}{21pt}\selectfont\ 37.89 & \fontsize{21pt}{21pt}\selectfont\ 51.41 \\

\noalign{\vskip 0.25ex}\cdashline{2-10}\noalign{\vskip 1.5ex}
& \cellcolor{blue!5} \fontsize{20pt}{20pt}\selectfont\ \textbf{CoR w/ Llama-3.2-3B-Instruct (w/ DPO)}  & \cellcolor{blue!5} \fontsize{21pt}{21pt}\selectfont\ \textbf{37.19} & \cellcolor{blue!5} \fontsize{21pt}{21pt}\selectfont\ \textbf{54.60} & \cellcolor{blue!5} \fontsize{21pt}{21pt}\selectfont\ \textbf{36.23} & \cellcolor{blue!5} \fontsize{21pt}{21pt}\selectfont\ \textbf{53.08} & \cellcolor{blue!5} \fontsize{21pt}{21pt}\selectfont\ \textbf{40.56} & \cellcolor{blue!5} \fontsize{21pt}{21pt}\selectfont\ \textbf{56.40} & \cellcolor{blue!5} \fontsize{21pt}{21pt}\selectfont\ \textbf{40.93} & \cellcolor{blue!5} \fontsize{21pt}{21pt}\selectfont\ \textbf{59.33} \\

& \cellcolor{blue!5} \fontsize{20pt}{20pt}\selectfont\ \textbf{CoR w/ QWEN-2.5-3B-Instruct (w/ DPO)} & \cellcolor{blue!5} \fontsize{21pt}{21pt}\selectfont\ \textbf{36.82} & \cellcolor{blue!5} \fontsize{21pt}{21pt}\selectfont\ \textbf{54.37} & \cellcolor{blue!5} \fontsize{21pt}{21pt}\selectfont\ \textbf{36.55} & \cellcolor{blue!5} \fontsize{21pt}{21pt}\selectfont\ \textbf{53.59} & \cellcolor{blue!5} \fontsize{21pt}{21pt}\selectfont\ \textbf{41.00} & \cellcolor{blue!5} \fontsize{21pt}{21pt}\selectfont\ \textbf{57.19}  & \cellcolor{blue!5} \fontsize{21pt}{21pt}\selectfont\ \textbf{42.17} & \cellcolor{blue!5} \fontsize{21pt}{21pt}\selectfont\ \textbf{60.35} \\

\noalign{\vskip 0.75ex}\cline{2-10}\noalign{\vskip 1.5ex} 
&\multicolumn{9}{l}{\textbf{\fontsize{20pt}{20pt}\selectfont BGE-M3 }} \\
\rule{0pt}{2ex}%
& \fontsize{20pt}{20pt}\selectfont\ A2A (Abstract-to-Abstract) & \fontsize{21pt}{21pt}\selectfont\ 29.51 & \fontsize{21pt}{21pt}\selectfont\ 41.95 & \fontsize{21pt}{21pt}\selectfont\ 27.15 & \fontsize{21pt}{21pt}\selectfont\ 39.61 & \fontsize{21pt}{21pt}\selectfont\ 34.65 & \fontsize{21pt}{21pt}\selectfont\ 46.22 & \fontsize{21pt}{21pt}\selectfont\ 33.48 & \fontsize{21pt}{21pt}\selectfont\ 46.08 \\

& \fontsize{20pt}{20pt}\selectfont\ F2F (Full-to-Full) & \fontsize{21pt}{21pt}\selectfont\ 31.78 & \fontsize{21pt}{21pt}\selectfont\ 44.10 & \fontsize{21pt}{21pt}\selectfont\ 29.06 & \fontsize{21pt}{21pt}\selectfont\ 41.13 & \fontsize{21pt}{21pt}\selectfont\ 35.63 & \fontsize{21pt}{21pt}\selectfont\ 47.57 & \fontsize{21pt}{21pt}\selectfont\ 35.38 & \fontsize{21pt}{21pt}\selectfont\ 46.89 \\

\noalign{\vskip 0.25ex}\cdashline{2-10}\noalign{\vskip 1.5ex}
& \cellcolor{blue!5} \fontsize{20pt}{20pt}\selectfont\ \textbf{CoR w/ Llama-3.2-3B-Instruct (w/ DPO)} & \cellcolor{blue!5} \fontsize{21pt}{21pt}\selectfont\ 31.21 & \cellcolor{blue!5} \fontsize{21pt}{21pt}\selectfont\ \textbf{47.24} & \cellcolor{blue!5} \fontsize{21pt}{21pt}\selectfont\ \textbf{31.96} & \cellcolor{blue!5} \fontsize{21pt}{21pt}\selectfont\ \textbf{46.53} & \cellcolor{blue!5} \fontsize{21pt}{21pt}\selectfont\ \textbf{36.40} & \cellcolor{blue!5} \fontsize{21pt}{21pt}\selectfont\ \textbf{51.53} & \cellcolor{blue!5} \fontsize{21pt}{21pt}\selectfont\ \textbf{37.12} & \cellcolor{blue!5} \fontsize{21pt}{21pt}\selectfont\ \textbf{52.01} \\

& \cellcolor{blue!5} \fontsize{20pt}{20pt}\selectfont\ \textbf{CoR w/ QWEN-2.5-3B-Instruct (w/ DPO)} & \cellcolor{blue!5} \fontsize{21pt}{21pt}\selectfont\ \textbf{32.10}  & \cellcolor{blue!5} \fontsize{21pt}{21pt}\selectfont\ \textbf{48.16} & \cellcolor{blue!5} \fontsize{21pt}{21pt}\selectfont\ \textbf{32.94} & \cellcolor{blue!5} \fontsize{21pt}{21pt}\selectfont\ \textbf{47.97} & \cellcolor{blue!5} \fontsize{21pt}{21pt}\selectfont\ \textbf{37.58} & \cellcolor{blue!5} \fontsize{21pt}{21pt}\selectfont\ \textbf{52.69} & \cellcolor{blue!5} \fontsize{21pt}{21pt}\selectfont\ \textbf{38.50} & \cellcolor{blue!5} \fontsize{21pt}{21pt}\selectfont\ \textbf{53.57} \\

\noalign{\vskip 0.75ex}\cline{2-10}\noalign{\vskip 1.5ex} 
&\multicolumn{9}{l}{\textbf{\fontsize{20pt}{20pt}\selectfont Inf-Retriever-v1-1.5B }} \\
\rule{0pt}{2ex}%
& \fontsize{20pt}{20pt}\selectfont\ A2A (Abstract-to-Abstract) & \fontsize{21pt}{21pt}\selectfont\ 43.03 & \fontsize{21pt}{21pt}\selectfont\ 59.66 & \fontsize{21pt}{21pt}\selectfont\ 35.40 & \fontsize{21pt}{21pt}\selectfont\ 51.56 & \fontsize{21pt}{21pt}\selectfont\ 48.64 & \fontsize{21pt}{21pt}\selectfont\ 63.77 & \fontsize{21pt}{21pt}\selectfont\ 41.81 & \fontsize{21pt}{21pt}\selectfont\ 57.70 \\

& \fontsize{20pt}{20pt}\selectfont\ F2F (Full-to-Full) & \fontsize{21pt}{21pt}\selectfont\ 39.01 & \fontsize{21pt}{21pt}\selectfont\ 53.85 & \fontsize{21pt}{21pt}\selectfont\ 34.06 & \fontsize{21pt}{21pt}\selectfont\ 47.89 & \fontsize{21pt}{21pt}\selectfont\ 23.17 & \fontsize{21pt}{21pt}\selectfont\ 30.69 & \fontsize{21pt}{21pt}\selectfont\ 37.09 & \fontsize{21pt}{21pt}\selectfont\ 49.69 \\

\noalign{\vskip 0.25ex}\cdashline{2-10}\noalign{\vskip 1.5ex}
& \cellcolor{blue!5} \fontsize{20pt}{20pt}\selectfont\ \textbf{CoR w/ Llama-3.2-3B-Instruct (w/ DPO)} & \cellcolor{blue!5} \fontsize{21pt}{21pt}\selectfont\ \textbf{44.65} & \cellcolor{blue!5} \fontsize{21pt}{21pt}\selectfont\ \textbf{63.19} & \cellcolor{blue!5} \fontsize{21pt}{21pt}\selectfont\ \textbf{40.66} & \cellcolor{blue!5} \fontsize{21pt}{21pt}\selectfont\ \textbf{58.41} & \cellcolor{blue!5} \fontsize{21pt}{21pt}\selectfont\ \textbf{49.96} & \cellcolor{blue!5} \fontsize{21pt}{21pt}\selectfont\ \textbf{67.43} &  \cellcolor{blue!5} \fontsize{21pt}{21pt}\selectfont\ \textbf{46.66} & \cellcolor{blue!5} \fontsize{21pt}{21pt}\selectfont\ \textbf{64.66} \\

& \cellcolor{blue!5} \fontsize{20pt}{20pt}\selectfont\ \textbf{CoR w/ QWEN-2.5-3B-Instruct (w/ DPO)} & \cellcolor{blue!5} \fontsize{21pt}{21pt}\selectfont\ \textbf{44.65} & \cellcolor{blue!5} \fontsize{21pt}{21pt}\selectfont\ \textbf{62.93} & \cellcolor{blue!5} \fontsize{21pt}{21pt}\selectfont\ \textbf{41.36} & \cellcolor{blue!5} \fontsize{21pt}{21pt}\selectfont\ \textbf{59.24} & \cellcolor{blue!5} \fontsize{21pt}{21pt}\selectfont\ \textbf{49.77} & \cellcolor{blue!5} \fontsize{21pt}{21pt}\selectfont\ \textbf{66.62} & \cellcolor{blue!5} \fontsize{21pt}{21pt}\selectfont\ \textbf{47.06} & \cellcolor{blue!5} \fontsize{21pt}{21pt}\selectfont\ \textbf{64.88} \\
\midrule\midrule
& & \multicolumn{4}{c}{\fontsize{22pt}{22pt}\selectfont\ \bf ACL} & \multicolumn{4}{c}{\fontsize{22pt}{22pt}\selectfont\ \bf EMNLP} \\
\cmidrule(l{2pt}r{2pt}){3-6} \cmidrule(l{2pt}r{2pt}){7-10}
\rule{0pt}{3.5ex}%
& & \multicolumn{2}{c}{\fontsize{22pt}{22pt}\selectfont \bf References}& \multicolumn{2}{c}{\fontsize{22pt}{22pt}\selectfont\ \bf Citations} 
  & \multicolumn{2}{c}{\fontsize{22pt}{22pt}\selectfont \bf References} & \multicolumn{2}{c}{\fontsize{22pt}{22pt}\selectfont \bf Citations} \\
\cmidrule(l{2pt}r{2pt}){3-4} \cmidrule(l{2pt}r{2pt}){5-6}
\cmidrule(l{2pt}r{2pt}){7-8} \cmidrule(l{2pt}r{2pt}){9-10}
& \fontsize{22pt}{20pt}\selectfont\ \textbf{IR Method} 
& \fontsize{20pt}{20pt}\selectfont\ nDCG@300 & 
\fontsize{20pt}{20pt}\selectfont Recall@300 & 
\fontsize{20pt}{20pt}\selectfont\ nDCG@300 &
\fontsize{20pt}{20pt}\selectfont\ Recall@300 & 
\fontsize{20pt}{20pt}\selectfont\ nDCG@300 & 
\fontsize{20pt}{20pt}\selectfont\ Recall@300 & 
\fontsize{20pt}{20pt}\selectfont\ nDCG@300 & 
\fontsize{20pt}{20pt}\selectfont\ Recall@300 \\
\midrule
\noalign{\vskip 0.5ex} 
& \multicolumn{9}{l}{\textbf{\fontsize{20pt}{20pt}\selectfont Lexical-Based Retrievers}} \\
\rule{0pt}{2ex}%
& \fontsize{20pt}{20pt}\selectfont\ BM-25 (A2A) & \fontsize{21pt}{21pt}\selectfont\ 21.47 & \fontsize{21pt}{21pt}\selectfont\ 31.80 & \fontsize{21pt}{21pt}\selectfont\ 21.90 & \fontsize{21pt}{21pt}\selectfont\ 30.74 & \fontsize{21pt}{21pt}\selectfont\ 20.57 & \fontsize{21pt}{21pt}\selectfont\ 29.78 & \fontsize{21pt}{21pt}\selectfont\ 19.82 & \fontsize{21pt}{21pt}\selectfont\ 28.00 \\
& \fontsize{20pt}{20pt}\selectfont\ BM-25 (F2F) & \fontsize{21pt}{21pt}\selectfont\ 22.77 & \fontsize{21pt}{21pt}\selectfont\ 34.64 & \fontsize{21pt}{21pt}\selectfont\ 27.47 & \fontsize{21pt}{21pt}\selectfont\ 37.95 & \fontsize{21pt}{21pt}\selectfont\ 24.00 & \fontsize{21pt}{21pt}\selectfont\ 35.55 & \fontsize{21pt}{21pt}\selectfont\ 28.69 & \fontsize{21pt}{21pt}\selectfont\ 39.21 \\
\midrule\midrule
& \multicolumn{9}{l}{\textbf{\fontsize{20pt}{20pt}\selectfont Domain-Specific Retriever}} \\
\rule{0pt}{2ex}%
& \fontsize{20pt}{20pt}\selectfont\ SciNCL-\textbf{A2A} & \fontsize{21pt}{21pt}\selectfont\ 26.72 & \fontsize{21pt}{21pt}\selectfont\ 41.63 & \fontsize{21pt}{21pt}\selectfont\ 26.29 & \fontsize{21pt}{21pt}\selectfont\ 39.79 & \fontsize{21pt}{21pt}\selectfont\ 25.52 & \fontsize{21pt}{21pt}\selectfont\ 39.78 & \fontsize{21pt}{21pt}\selectfont\ 24.86 & \fontsize{21pt}{21pt}\selectfont\ 38.02 \\

& \fontsize{20pt}{20pt}\selectfont\ SPECTER2-Base-\textbf{A2A} & \fontsize{21pt}{21pt}\selectfont\ 25.82 & \fontsize{21pt}{21pt}\selectfont\ 40.48 & \fontsize{21pt}{21pt}\selectfont\ 27.61 & \fontsize{21pt}{21pt}\selectfont\ 41.19 & \fontsize{21pt}{21pt}\selectfont\ 24.32 & \fontsize{21pt}{21pt}\selectfont\ 37.55 & \fontsize{21pt}{21pt}\selectfont\ 26.09 & \fontsize{20pt}{20pt}\selectfont\ 39.22 \\

& \fontsize{20pt}{20pt}\selectfont\ SPECTER2-Adapter-MTL CTRL-\textbf{A2A} & \fontsize{21pt}{21pt}\selectfont\ 25.76 & \fontsize{20pt}{20pt}\selectfont\ 40.13 & \fontsize{21pt}{21pt}\selectfont\ 26.31 & \fontsize{20pt}{20pt}\selectfont\ 39.61 & \fontsize{21pt}{21pt}\selectfont\ 24.87 & \fontsize{20pt}{20pt}\selectfont\ 37.59 & \fontsize{21pt}{21pt}\selectfont\ 24.65 & \fontsize{20pt}{20pt}\selectfont\ 37.47 \\

& \fontsize{20pt}{20pt}\selectfont\ SciMult-MHAExpert-\textbf{A2A} & \fontsize{21pt}{21pt}\selectfont\ 24.19 & \fontsize{21pt}{21pt}\selectfont\ 37.49 & \fontsize{21pt}{21pt}\selectfont\ 23.76 & \fontsize{21pt}{21pt}\selectfont\ 35.31 & \fontsize{21pt}{21pt}\selectfont\ 22.02 & \fontsize{21pt}{21pt}\selectfont\ 34.75 & \fontsize{21pt}{21pt}\selectfont\ 21.38 & \fontsize{21pt}{21pt}\selectfont\ 32.26 \\

\midrule\midrule
\noalign{\vskip 0.5ex} 
& \multicolumn{9}{l}{\textbf{\fontsize{20pt}{20pt}\selectfont Jina-Embeddings-v2-BASE-EN}} \\
\multirow{12}{*}{\raisebox{-9ex}[0pt][0pt]{\rotatebox[origin=c]{90}
{\textbf{\fontsize{26pt}{26pt}\selectfont Domain-Agnostic Retriever}}}} & 
\fontsize{20pt}{20pt}\selectfont\ A2A (Abstract-to-Abstract)& \fontsize{20pt}{21pt}\selectfont\ 25.50 & \fontsize{21pt}{21pt}\selectfont\ 38.80 & \fontsize{21pt}{21pt}\selectfont\ 27.66 & \fontsize{21pt}{21pt}\selectfont\ 40.52 & \fontsize{21pt}{21pt}\selectfont\ 24.41 & \fontsize{21pt}{21pt}\selectfont\ 37.00 & \fontsize{21pt}{21pt}\selectfont\ 25.23 & \fontsize{21pt}{21pt}\selectfont\ 37.32 \\

& \fontsize{20pt}{20pt}\selectfont\ F2F (Full-to-Full) & \fontsize{21pt}{21pt}\selectfont\ 27.27 & \fontsize{21pt}{21pt}\selectfont\ 41.66 & \fontsize{21pt}{21pt}\selectfont\ 28.80 & \fontsize{21pt}{21pt}\selectfont\ 40.95 & \fontsize{21pt}{21pt}\selectfont\ 27.83 & \fontsize{21pt}{21pt}\selectfont\ 41.73 & \fontsize{21pt}{21pt}\selectfont\ 27.87 & \fontsize{21pt}{21pt}\selectfont\ 40.34 \\

\noalign{\vskip 0.25ex}\cdashline{2-10}\noalign{\vskip 1.5ex}
& \cellcolor{blue!5} \fontsize{20pt}{20pt}\selectfont\ \textbf{CoR w/ Llama-3.2-3B-Instruct (w/ DPO)} & \cellcolor{blue!5} \fontsize{21pt}{21pt}\selectfont\ \textbf{28.52} & \cellcolor{blue!5} \fontsize{21pt}{21pt}\selectfont\ \textbf{44.92} & \cellcolor{blue!5} \fontsize{21pt}{21pt}\selectfont\ \textbf{32.76} & \cellcolor{blue!5} \fontsize{21pt}{21pt}\selectfont\ \textbf{48.05} & \cellcolor{blue!5} \fontsize{21pt}{21pt}\selectfont\ \textbf{28.15} & \cellcolor{blue!5} \fontsize{21pt}{21pt}\selectfont\ \textbf{43.96} & \cellcolor{blue!5} \fontsize{21pt}{21pt}\selectfont\ \textbf{30.84} & \cellcolor{blue!5} \fontsize{21pt}{21pt}\selectfont\ \textbf{46.00} \\

& \cellcolor{blue!5} \fontsize{20pt}{20pt}\selectfont\ \textbf{CoR w/ QWEN-2.5-3B-Instruct (w/ DPO)} & \cellcolor{blue!5} \fontsize{21pt}{21pt}\selectfont\ \textbf{28.62} & \cellcolor{blue!5} \fontsize{21pt}{21pt}\selectfont\ \textbf{45.67} & \cellcolor{blue!5} \fontsize{21pt}{21pt}\selectfont\ \textbf{33.40} & \cellcolor{blue!5} \fontsize{21pt}{21pt}\selectfont\ \textbf{49.00} & \cellcolor{blue!5} \fontsize{21pt}{21pt}\selectfont\ \textbf{27.80} & \cellcolor{blue!5} \fontsize{21pt}{21pt}\selectfont\ \textbf{43.99} & \cellcolor{blue!5} \fontsize{21pt}{21pt}\selectfont\ \textbf{31.68} & \cellcolor{blue!5} \fontsize{21pt}{21pt}\selectfont\ \textbf{47.60} \\

\noalign{\vskip 0.75ex}\cline{2-10}\noalign{\vskip 1.5ex} 
&\multicolumn{9}{l}{\textbf{\fontsize{20pt}{20pt}\selectfont BGE-M3 }} \\
\rule{0pt}{2ex}%
& \fontsize{20pt}{20pt}\selectfont\ A2A (Abstract-to-Abstract) & \fontsize{21pt}{21pt}\selectfont\ 23.65 & \fontsize{21pt}{21pt}\selectfont\ 35.81 & \fontsize{21pt}{21pt}\selectfont\ 25.95 & \fontsize{21pt}{21pt}\selectfont\ 38.40 & \fontsize{21pt}{21pt}\selectfont\ 22.25 & \fontsize{21pt}{21pt}\selectfont\ 33.29 & \fontsize{21pt}{21pt}\selectfont\ 22.87 & \fontsize{21pt}{21pt}\selectfont\ 34.15 \\

& \fontsize{20pt}{20pt}\selectfont\ F2F (Full-to-Full) & \fontsize{21pt}{21pt}\selectfont\ 24.39 & \fontsize{21pt}{21pt}\selectfont\ 36.67 & \fontsize{21pt}{21pt}\selectfont\ 25.78 & \fontsize{21pt}{21pt}\selectfont\ 37.01 & \fontsize{21pt}{21pt}\selectfont\ 23.88 & \fontsize{21pt}{21pt}\selectfont\ 35.08 & \fontsize{21pt}{21pt}\selectfont\ 23.92 & \fontsize{21pt}{21pt}\selectfont\ 34.73 \\

\noalign{\vskip 0.25ex}\cdashline{2-10}\noalign{\vskip 1.5ex}
& \cellcolor{blue!5} \fontsize{20pt}{20pt}\selectfont\ \textbf{CoR w/ Llama-3.2-3B-Instruct (w/ DPO)}  & \cellcolor{blue!5} \fontsize{21pt}{21pt}\selectfont\ \textbf{25.12} & \cellcolor{blue!5} \fontsize{21pt}{21pt}\selectfont\ \textbf{40.19} & \cellcolor{blue!5} \fontsize{21pt}{21pt}\selectfont\ \textbf{30.18} & \cellcolor{blue!5} \fontsize{21pt}{21pt}\selectfont\ \textbf{44.46} & \cellcolor{blue!5} \fontsize{21pt}{21pt}\selectfont\ \textbf{24.12} & \cellcolor{blue!5} \fontsize{21pt}{21pt}\selectfont\ \textbf{38.24} & \cellcolor{blue!5} \fontsize{21pt}{21pt}\selectfont\ \textbf{27.55} & \cellcolor{blue!5} \fontsize{21pt}{21pt}\selectfont\ \textbf{40.87} \\

& \cellcolor{blue!5} \fontsize{20pt}{20pt}\selectfont\ \textbf{CoR w/ QWEN-2.5-3B-Instruct (w/ DPO)} & \cellcolor{blue!5} \fontsize{21pt}{21pt}\selectfont\ \textbf{25.77} & \cellcolor{blue!5} \fontsize{21pt}{21pt}\selectfont\ \textbf{40.71} & \cellcolor{blue!5} \fontsize{21pt}{21pt}\selectfont\ \textbf{30.44} & \cellcolor{blue!5} \fontsize{21pt}{21pt}\selectfont\ \textbf{45.29} & \cellcolor{blue!5} \fontsize{21pt}{21pt}\selectfont\ \textbf{24.66} & \cellcolor{blue!5} \fontsize{21pt}{21pt}\selectfont\ \textbf{38.97} & \cellcolor{blue!5} \fontsize{21pt}{21pt}\selectfont\ \textbf{28.22} & \cellcolor{blue!5} \fontsize{21pt}{21pt}\selectfont\ \textbf{42.14} \\

\noalign{\vskip 0.75ex}\cline{2-10}\noalign{\vskip 1.5ex} 
&\multicolumn{9}{l}{\textbf{\fontsize{20pt}{20pt}\selectfont Inf-Retriever-v1-1.5B }} \\
\rule{0pt}{2ex}%
& \fontsize{20pt}{20pt}\selectfont\ A2A (Abstract-to-Abstract) & \fontsize{21pt}{21pt}\selectfont\ 34.17 & \fontsize{21pt}{21pt}\selectfont\ 51.51 & \fontsize{21pt}{21pt}\selectfont\ 32.20 & \fontsize{21pt}{21pt}\selectfont\ 47.33 & \fontsize{21pt}{21pt}\selectfont\ 32.68 & \fontsize{21pt}{21pt}\selectfont\ 48.68 & \fontsize{21pt}{21pt}\selectfont\ 29.68 & \fontsize{21pt}{21pt}\selectfont\ 43.72 \\

& \fontsize{20pt}{20pt}\selectfont\ F2F (Full-to-Full) & \fontsize{21pt}{21pt}\selectfont\ 33.35 & \fontsize{21pt}{21pt}\selectfont\ 49.59 & \fontsize{21pt}{21pt}\selectfont\ 30.61 & \fontsize{21pt}{21pt}\selectfont\ 43.72 & \fontsize{21pt}{21pt}\selectfont\ 32.37 & \fontsize{21pt}{21pt}\selectfont\ 46.40 & \fontsize{21pt}{21pt}\selectfont\ 30.02 & \fontsize{21pt}{21pt}\selectfont\ 43.57 \\

\noalign{\vskip 0.25ex}\cdashline{2-10}\noalign{\vskip 1.5ex}
& \cellcolor{blue!5} \fontsize{20pt}{20pt}\selectfont\ \textbf{CoR w/ Llama-3.2-3B-Instruct (w/ DPO)} & \cellcolor{blue!5} \fontsize{21pt}{21pt}\selectfont\ \textbf{35.09} & \cellcolor{blue!5} \fontsize{21pt}{21pt}\selectfont\ \textbf{55.14} & \cellcolor{blue!5} \fontsize{21pt}{21pt}\selectfont\ \textbf{36.98} & \cellcolor{blue!5} \fontsize{21pt}{21pt}\selectfont\ \textbf{54.13} & \cellcolor{blue!5} \fontsize{21pt}{21pt}\selectfont\ \textbf{33.86} & \cellcolor{blue!5} \fontsize{21pt}{21pt}\selectfont\ \textbf{52.11} & \cellcolor{blue!5} \fontsize{21pt}{21pt}\selectfont\ \textbf{35.37} & \cellcolor{blue!5} \fontsize{21pt}{21pt}\selectfont\ \textbf{51.77} \\

& \cellcolor{blue!5} \fontsize{20pt}{20pt}\selectfont\ \textbf{CoR w/ QWEN-2.5-3B-Instruct (w/ DPO)} & \cellcolor{blue!5} \fontsize{21pt}{21pt}\selectfont\ \textbf{35.16} & \cellcolor{blue!5} \fontsize{21pt}{21pt}\selectfont\ \textbf{54.68} & \cellcolor{blue!5} \fontsize{21pt}{21pt}\selectfont\ \textbf{37.74} & \cellcolor{blue!5} \fontsize{21pt}{21pt}\selectfont\ \textbf{54.78} & \cellcolor{blue!5} \fontsize{21pt}{21pt}\selectfont\ \textbf{33.65} & \cellcolor{blue!5} \fontsize{21pt}{21pt}\selectfont\ \textbf{51.91} & \cellcolor{blue!5} \fontsize{21pt}{21pt}\selectfont\ \textbf{35.99} & \cellcolor{blue!5} \fontsize{21pt}{21pt}\selectfont\ \textbf{52.40}
\\

\midrule
\bottomrule
\end{tabular}
}
\caption{Main results with additional metrics.}
\vspace{-0.1in}
\label{tab:main_final_expanded}
\vspace{-0.075in}
\end{table*}
\begin{table*}[t!]
%\caption{Retrieval Performance Analysis on additional metrics with lower cutoff k values on \textsc{SciFullBench}.}
%\vspace{-0.075in}
\setlength{\tabcolsep}{18pt}
\resizebox{\linewidth}{!}{
\renewcommand{\arraystretch}{0.45}
\begin{tabular}{llcccc}
\toprule

\midrule
& \textbf{Retrieval Method} & \textbf{nDCG@20} & \textbf{Recall@20} & \textbf{nDCG@30} & \textbf{Recall@30} \\
\midrule
\multicolumn{5}{l}{\textbf{JEmb-v2-Base-EN}} \\
& A2A & 16.82 & 13.05 & 17.38 & 15.85 \\
& F2F & 18.52 & 14.20 & 19.27 & 17.41 \\
\cellcolor{blue!5} & 
\cellcolor{blue!5}\textbf{CoR w/ Llama-3.2-3B-Instruct} & 
\cellcolor{blue!5}\textbf{18.86} & \cellcolor{blue!5}\textbf{14.84} & \cellcolor{blue!5}\textbf{19.91} & \cellcolor{blue!5}\textbf{18.64} \\
\cellcolor{blue!5} & \cellcolor{blue!5}\textbf{CoR w/ QWEN-2.5-3B-Instruct} & \cellcolor{blue!5}\textbf{18.77} & \cellcolor{blue!5}\textbf{14.82} & \cellcolor{blue!5}\textbf{19.87} & \cellcolor{blue!5}\textbf{18.66} \\
\noalign{\vskip 0.2ex} 
\midrule
\multicolumn{5}{l}{\textbf{BGE-M3}} \\
& A2A & 15.36 & 11.80 & 16.00 & 14.59 \\
& F2F & 16.57 & 12.61 & 17.13 & 15.36 \\
\cellcolor{blue!5} & \cellcolor{blue!5}\textbf{CoR w/ Llama-3.2-3B-Instruct} & \cellcolor{blue!5}\textbf{16.87} & \cellcolor{blue!5}\textbf{13.05} & \cellcolor{blue!5}\textbf{17.73} & \cellcolor{blue!5}\textbf{16.39} \\
\cellcolor{blue!5}& \cellcolor{blue!5}\textbf{CoR w/ QWEN-2.5-3B-Instruct} & \cellcolor{blue!5}\textbf{17.13} & \cellcolor{blue!5}\cellcolor{blue!5}\textbf{13.21} & \cellcolor{blue!5}\textbf{17.98} & \cellcolor{blue!5}\textbf{16.60} \\
\noalign{\vskip 0.2ex} 
\midrule
\multicolumn{5}{l}{\textbf{Inf-Retriever-v1-1.5B}} \\
& A2A & 20.85 & 16.43 & 21.86 & 20.31 \\
& F2F & 21.16 & 16.75 & 22.29 & 20.80 \\
\cellcolor{blue!5} & \cellcolor{blue!5}\textbf{CoR w/ Llama-3.2-3B-Instruct} & \cellcolor{blue!5}\textbf{22.47} & \cellcolor{blue!5}\textbf{17.88} & \cellcolor{blue!5}\textbf{23.75} & \cellcolor{blue!5}\textbf{22.41} \\
\cellcolor{blue!5} & \cellcolor{blue!5}\textbf{CoR w/ QWEN-2.5-3B-Instruct} & \cellcolor{blue!5}\textbf{22.83} & \cellcolor{blue!5}\textbf{18.14} & \cellcolor{blue!5}\textbf{24.15} & \cellcolor{blue!5}\textbf{22.82}  \\
\noalign{\vskip 0.2ex} 
\midrule
\bottomrule
\end{tabular}
}

\caption{Retrieval Performance Analysis on additional metrics with lower cutoff k values on \textsc{SciFullBench}.}
\label{tab:main_result_w_lower_k_metrics}
\end{table*}
\begin{table*}[t!]
%\caption{Main results with additional metrics.}
%\vspace{-0.1in}
\renewcommand{\arraystretch}{1.6}
\resizebox{\linewidth}{!}{
\begin{tabular}{llcccccccc}

\midrule \toprule \rule{0pt}{3.5ex}%
& & \multicolumn{4}{c}{\fontsize{22pt}{22pt}\selectfont\ \bf ICLR} & \multicolumn{4}{c}{\fontsize{22pt}{22pt}\selectfont\ \bf NeurIPS} \\
\cmidrule(l{2pt}r{2pt}){3-6} \cmidrule(l{2pt}r{2pt}){7-10}
\rule{0pt}{3.5ex}%
& & \multicolumn{2}{c}{\fontsize{22pt}{22pt}\selectfont \bf References}& \multicolumn{2}{c}{\fontsize{22pt}{22pt}\selectfont\ \bf Citations} 
  & \multicolumn{2}{c}{\fontsize{22pt}{22pt}\selectfont \bf References} & \multicolumn{2}{c}{\fontsize{22pt}{22pt}\selectfont \bf Citations} \\
\cmidrule(l{2pt}r{2pt}){3-4} \cmidrule(l{2pt}r{2pt}){5-6}
\cmidrule(l{2pt}r{2pt}){7-8} \cmidrule(l{2pt}r{2pt}){9-10}
& \fontsize{22pt}{20pt}\selectfont\ \textbf{IR Method} 
& \fontsize{20pt}{20pt}\selectfont\ Recall@100 & 
\fontsize{20pt}{20pt}\selectfont Recall@200 & 
\fontsize{20pt}{20pt}\selectfont\ nDCG@200 &
\fontsize{20pt}{20pt}\selectfont\ Recall@100 & 
\fontsize{20pt}{20pt}\selectfont\ Recall@100 & 
\fontsize{20pt}{20pt}\selectfont\ Recall@200 & 
\fontsize{20pt}{20pt}\selectfont\ nDCG@200 & 
\fontsize{20pt}{20pt}\selectfont\ Recall@100 \\
\midrule
\noalign{\vskip 0.5ex}  
& \multicolumn{9}{l}{\textbf{\fontsize{20pt}{20pt}\selectfont Lexical-Based Retrievers}} \\
\rule{0pt}{2ex}%
& \fontsize{20pt}{20pt}\selectfont\ BM-25 (A2A) & \fontsize{21pt}{21pt}\selectfont\ 24.51 & \fontsize{21pt}{21pt}\selectfont\ 30.77 & \fontsize{21pt}{21pt}\selectfont\ 21.83 & \fontsize{21pt}{21pt}\selectfont\ 22.57 & \fontsize{21pt}{21pt}\selectfont\ 30.86 & \fontsize{21pt}{21pt}\selectfont\ 37.60 & \fontsize{21pt}{21pt}\selectfont\ 30.08 & \fontsize{21pt}{21pt}\selectfont\ 31.17 \\
& \fontsize{20pt}{20pt}\selectfont\ BM-25 (F2F) & \fontsize{21pt}{21pt}\selectfont\ 31.66 & \fontsize{21pt}{21pt}\selectfont\ 39.59 & \fontsize{21pt}{21pt}\selectfont\ 32.72 & \fontsize{21pt}{21pt}\selectfont\ 33.01 & \fontsize{21pt}{21pt}\selectfont\ 20.19 & \fontsize{21pt}{21pt}\selectfont\ 25.07 & \fontsize{21pt}{21pt}\selectfont\ 36.84 & \fontsize{21pt}{21pt}\selectfont\ 37.57 \\
\midrule\midrule
& \multicolumn{9}{l}{\textbf{\fontsize{20pt}{20pt}\selectfont Domain-Specific Retriever}} \\
\rule{0pt}{2ex}%
& \fontsize{20pt}{20pt}\selectfont\ SciNCL-\textbf{A2A} & \fontsize{21pt}{21pt}\selectfont\ 34.92 & \fontsize{21pt}{21pt}\selectfont\ 44.55 & \fontsize{21pt}{21pt}\selectfont\ 27.33 & \fontsize{21pt}{21pt}\selectfont\ 29.49 & \fontsize{21pt}{21pt}\selectfont\ 38.10 & \fontsize{21pt}{21pt}\selectfont\ 47.37 & \fontsize{21pt}{21pt}\selectfont\ 34.40 & \fontsize{21pt}{21pt}\selectfont\ 36.90 \\

& \fontsize{20pt}{20pt}\selectfont\ SPECTER2-Base-\textbf{A2A} & \fontsize{21pt}{21pt}\selectfont\ 33.47 & \fontsize{21pt}{21pt}\selectfont\ 43.43  & \fontsize{21pt}{21pt}\selectfont\ 28.66 & \fontsize{21pt}{21pt}\selectfont\ 30.32 & \fontsize{21pt}{21pt}\selectfont\ 36.65 & \fontsize{21pt}{21pt}\selectfont\ 46.15 & \fontsize{21pt}{21pt}\selectfont\ 36.05 & \fontsize{20pt}{20pt}\selectfont\ 37.74 \\

& \fontsize{20pt}{20pt}\selectfont\ SPECTER2-Adapter-MTL CTRL-\textbf{A2A} & \fontsize{21pt}{21pt}\selectfont\ 34.51 & \fontsize{20pt}{20pt}\selectfont\ 43.72 & \fontsize{21pt}{21pt}\selectfont\ 27.43 & \fontsize{20pt}{20pt}\selectfont\ 29.13 & \fontsize{21pt}{21pt}\selectfont\ 37.30 & \fontsize{20pt}{20pt}\selectfont\ 46.60 & \fontsize{21pt}{21pt}\selectfont\ 34.70 & \fontsize{20pt}{20pt}\selectfont\ 37.24 \\

& \fontsize{20pt}{20pt}\selectfont\ SciMult-MHAExpert-\textbf{A2A} & \fontsize{21pt}{21pt}\selectfont\ 28.44 & \fontsize{21pt}{21pt}\selectfont\ 37.13 & \fontsize{21pt}{21pt}\selectfont\ 22.57 & \fontsize{21pt}{21pt}\selectfont\ 23.51 & \fontsize{21pt}{21pt}\selectfont\ 33.01 & \fontsize{21pt}{21pt}\selectfont\ 41.98 & \fontsize{21pt}{21pt}\selectfont\ 30.14 & \fontsize{21pt}{21pt}\selectfont\ 31.66 \\

\midrule\midrule
\noalign{\vskip 0.5ex} 
& \multicolumn{9}{l}{\textbf{\fontsize{20pt}{20pt}\selectfont Jina-Embeddings-v2-BASE-EN}} \\
\multirow{12}{*}{\raisebox{-9ex}[0pt][0pt]{\rotatebox[origin=c]{90}{\textbf{\fontsize{26pt}{26pt}\selectfont Domain-Agnostic Retriever}}}} & 
\fontsize{20pt}{20pt}\selectfont\ A2A (Abstract-to-Abstract)& \fontsize{20pt}{21pt}\selectfont\ 34.71 & \fontsize{21pt}{21pt}\selectfont\ 43.35 & \fontsize{21pt}{21pt}\selectfont\ 28.38 & \fontsize{21pt}{21pt}\selectfont\ 30.09 & \fontsize{21pt}{21pt}\selectfont\ 37.91 & \fontsize{21pt}{21pt}\selectfont\ 46.30 & \fontsize{21pt}{21pt}\selectfont\ 35.16 & \fontsize{21pt}{21pt}\selectfont\ 36.85 \\

& \fontsize{20pt}{20pt}\selectfont\ F2F (Full-to-Full) & \fontsize{21pt}{21pt}\selectfont\ 35.17 & \fontsize{21pt}{21pt}\selectfont\ 44.29 & \fontsize{21pt}{21pt}\selectfont\ 31.60 & \fontsize{21pt}{21pt}\selectfont\ 33.25  & \fontsize{21pt}{21pt}\selectfont\ 37.25 & \fontsize{21pt}{21pt}\selectfont\ 46.15 & \fontsize{21pt}{21pt}\selectfont\ 35.88 & \fontsize{21pt}{21pt}\selectfont\ 37.43\\

\noalign{\vskip 0.25ex}\cdashline{2-10}\noalign{\vskip 1.5ex}
& \cellcolor{blue!5} \fontsize{20pt}{20pt}\selectfont\ \textbf{CoR w/ Llama-3.2-3B-Instruct (w/ DPO)}  & \cellcolor{blue!5} \fontsize{21pt}{21pt}\selectfont\ \textbf{38.36} & \cellcolor{blue!5} \fontsize{21pt}{21pt}\selectfont\ \textbf{48.71} & \cellcolor{blue!5} \fontsize{21pt}{21pt}\selectfont\ \textbf{33.94} & \cellcolor{blue!5} \fontsize{21pt}{21pt}\selectfont\ \textbf{36.50} & \cellcolor{blue!5} \fontsize{21pt}{21pt}\selectfont\ \textbf{41.16} & \cellcolor{blue!5} \fontsize{21pt}{21pt}\selectfont\ \textbf{50.57} & \cellcolor{blue!5} \fontsize{21pt}{21pt}\selectfont\ \textbf{38.57} & \cellcolor{blue!5} \fontsize{21pt}{21pt}\selectfont\ \textbf{42.55} \\

& \cellcolor{blue!5} \fontsize{20pt}{20pt}\selectfont\ \textbf{CoR w/ QWEN-2.5-3B-Instruct (w/ DPO)} & \cellcolor{blue!5} \fontsize{21pt}{21pt}\selectfont\ \textbf{38.52} & \cellcolor{blue!5} \fontsize{21pt}{21pt}\selectfont\ \textbf{48.33} & \cellcolor{blue!5} \fontsize{21pt}{21pt}\selectfont\ \textbf{34.16} & \cellcolor{blue!5} \fontsize{21pt}{21pt}\selectfont\ \textbf{36.52} & \cellcolor{blue!5} \fontsize{21pt}{21pt}\selectfont\ \textbf{41.36} & \cellcolor{blue!5} \fontsize{21pt}{21pt}\selectfont\ \textbf{51.07} & \cellcolor{blue!5} \fontsize{21pt}{21pt}\selectfont\ \textbf{39.91} & \cellcolor{blue!5} \fontsize{21pt}{21pt}\selectfont\ \textbf{43.81} \\

\noalign{\vskip 0.75ex}\cline{2-10}\noalign{\vskip 1.5ex} 
&\multicolumn{9}{l}{\textbf{\fontsize{20pt}{20pt}\selectfont BGE-M3 }} \\
\rule{0pt}{2ex}%
& \fontsize{20pt}{20pt}\selectfont\ A2A (Abstract-to-Abstract) & \fontsize{21pt}{21pt}\selectfont\ 29.69 & \fontsize{21pt}{21pt}\selectfont\ 37.23 & \fontsize{21pt}{21pt}\selectfont\ 25.44 & \fontsize{21pt}{21pt}\selectfont\ 26.66 & \fontsize{21pt}{21pt}\selectfont\ 33.28 & \fontsize{21pt}{21pt}\selectfont\ 41.65 & \fontsize{21pt}{21pt}\selectfont\ 31.69 & \fontsize{21pt}{21pt}\selectfont\ 33.40 \\

& \fontsize{20pt}{20pt}\selectfont\ F2F (Full-to-Full) & \fontsize{21pt}{21pt}\selectfont\ 32.46 & \fontsize{21pt}{21pt}\selectfont\ 39.41 & \fontsize{21pt}{21pt}\selectfont\ 27.12 & \fontsize{21pt}{21pt}\selectfont\ 28.50 & \fontsize{21pt}{21pt}\selectfont\ 34.81 & \fontsize{21pt}{21pt}\selectfont\ 42.98 & \fontsize{21pt}{21pt}\selectfont\ 33.57 & \fontsize{21pt}{21pt}\selectfont\ 34.31 \\

\noalign{\vskip 0.25ex}\cdashline{2-10}\noalign{\vskip 1.5ex}
& \cellcolor{blue!5} \fontsize{20pt}{20pt}\selectfont\ \textbf{CoR w/ Llama-3.2-3B-Instruct (w/ DPO)} & \cellcolor{blue!5} \fontsize{21pt}{21pt}\selectfont\ \textbf{32.88} & \cellcolor{blue!5} \fontsize{21pt}{21pt}\selectfont\ \textbf{42.47} & \cellcolor{blue!5} \fontsize{21pt}{21pt}\selectfont\ \textbf{29.99} & \cellcolor{blue!5} \fontsize{21pt}{21pt}\selectfont\ \textbf{32.13} & \cellcolor{blue!5} \fontsize{21pt}{21pt}\selectfont\ \textbf{36.47} & \cellcolor{blue!5} \fontsize{21pt}{21pt}\selectfont\ \textbf{46.17} & \cellcolor{blue!5} \fontsize{21pt}{21pt}\selectfont\ \textbf{35.29} & \cellcolor{blue!5} \fontsize{21pt}{21pt}\selectfont\ \textbf{37.58} \\

& \cellcolor{blue!5} \fontsize{20pt}{20pt}\selectfont\ \textbf{CoR w/ QWEN-2.5-3B-Instruct (w/ DPO)} & \cellcolor{blue!5} \fontsize{21pt}{21pt}\selectfont\ \textbf{33.76} & \cellcolor{blue!5} \fontsize{21pt}{21pt}\selectfont\ \textbf{43.28} & \cellcolor{blue!5} \fontsize{21pt}{21pt}\selectfont\ \textbf{30.95} & \cellcolor{blue!5} \fontsize{21pt}{21pt}\selectfont\ \textbf{33.07} & \cellcolor{blue!5} \fontsize{21pt}{21pt}\selectfont\ \textbf{37.34} & \cellcolor{blue!5} \fontsize{21pt}{21pt}\selectfont\ \textbf{47.40} & \cellcolor{blue!5} \fontsize{21pt}{21pt}\selectfont\ \textbf{36.50} & \cellcolor{blue!5} \fontsize{21pt}{21pt}\selectfont\ \textbf{39.19} \\

\noalign{\vskip 0.75ex}\cline{2-10}\noalign{\vskip 1.5ex} 
&\multicolumn{9}{l}{\textbf{\fontsize{20pt}{20pt}\selectfont Inf-Retriever-v1-1.5B }} \\
\rule{0pt}{2ex}%
& \fontsize{20pt}{20pt}\selectfont\ A2A (Abstract-to-Abstract) & \fontsize{21pt}{21pt}\selectfont\ 44.49 & \fontsize{21pt}{21pt}\selectfont\ 53.91 & \fontsize{21pt}{21pt}\selectfont\ 33.03 & \fontsize{21pt}{21pt}\selectfont\ 35.03 & \fontsize{21pt}{21pt}\selectfont\ 48.20  & \fontsize{21pt}{21pt}\selectfont\ 57.89 & \fontsize{21pt}{21pt}\selectfont\ 39.88 & \fontsize{21pt}{21pt}\selectfont\ 42.97 \\

& \fontsize{20pt}{20pt}\selectfont\ F2F (Full-to-Full) & \fontsize{21pt}{21pt}\selectfont\ 39.21 & \fontsize{21pt}{21pt}\selectfont\ 48.37 & \fontsize{21pt}{21pt}\selectfont\ 32.15 & \fontsize{21pt}{21pt}\selectfont\ 34.65 & \fontsize{21pt}{21pt}\selectfont\ 22.95 & \fontsize{21pt}{21pt}\selectfont\ 28.06 & \fontsize{21pt}{21pt}\selectfont\ 35.35 & \fontsize{21pt}{21pt}\selectfont\ 37.46\\

\noalign{\vskip 0.25ex}\cdashline{2-10}\noalign{\vskip 1.5ex}
& \cellcolor{blue!5} \fontsize{20pt}{20pt}\selectfont\ \textbf{CoR w/ Llama-3.2-3B-Instruct (w/ DPO)} & \cellcolor{blue!5} \fontsize{21pt}{21pt}\selectfont\ \textbf{47.15} & \cellcolor{blue!5} \fontsize{21pt}{21pt}\selectfont\ \textbf{57.14} & \cellcolor{blue!5} \fontsize{21pt}{21pt}\selectfont\ \textbf{38.26} & \cellcolor{blue!5} \fontsize{21pt}{21pt}\selectfont\ \textbf{41.44} & \cellcolor{blue!5} \fontsize{21pt}{21pt}\selectfont\ \textbf{50.87} & \cellcolor{blue!5} \fontsize{21pt}{21pt}\selectfont\ \textbf{61.38} & \cellcolor{blue!5} \fontsize{21pt}{21pt}\selectfont\ \textbf{44.45} & \cellcolor{blue!5} \fontsize{21pt}{21pt}\selectfont\ \textbf{48.48} \\

& \cellcolor{blue!5} \fontsize{20pt}{20pt}\selectfont\ \textbf{CoR w/ QWEN-2.5-3B-Instruct (w/ DPO)} & \cellcolor{blue!5} \fontsize{21pt}{21pt}\selectfont\ \textbf{46.81} & \cellcolor{blue!5} \fontsize{21pt}{21pt}\selectfont\ \textbf{57.22} & \cellcolor{blue!5} \fontsize{21pt}{21pt}\selectfont\ \textbf{38.83} & \cellcolor{blue!5} \fontsize{21pt}{21pt}\selectfont\ \textbf{41.99} & \cellcolor{blue!5} \fontsize{21pt}{21pt}\selectfont\ \textbf{50.78} & \cellcolor{blue!5} \fontsize{21pt}{21pt}\selectfont\ \textbf{60.62} & \cellcolor{blue!5} \fontsize{21pt}{21pt}\selectfont\ \textbf{44.91} & \cellcolor{blue!5} \fontsize{21pt}{21pt}\selectfont\ \textbf{49.11} \\
\midrule\midrule
& & \multicolumn{4}{c}{\fontsize{22pt}{22pt}\selectfont\ \bf ACL} & \multicolumn{4}{c}{\fontsize{22pt}{22pt}\selectfont\ \bf EMNLP} \\
\cmidrule(l{2pt}r{2pt}){3-6} \cmidrule(l{2pt}r{2pt}){7-10}
\rule{0pt}{3.5ex}%
& & \multicolumn{2}{c}{\fontsize{22pt}{22pt}\selectfont \bf References}& \multicolumn{2}{c}{\fontsize{22pt}{22pt}\selectfont\ \bf Citations} 
  & \multicolumn{2}{c}{\fontsize{22pt}{22pt}\selectfont \bf References} & \multicolumn{2}{c}{\fontsize{22pt}{22pt}\selectfont \bf Citations} \\
\cmidrule(l{2pt}r{2pt}){3-4} \cmidrule(l{2pt}r{2pt}){5-6}
\cmidrule(l{2pt}r{2pt}){7-8} \cmidrule(l{2pt}r{2pt}){9-10}
& \fontsize{22pt}{20pt}\selectfont\ \textbf{IR Method} 
& \fontsize{20pt}{20pt}\selectfont\ Recall@100 & 
\fontsize{20pt}{20pt}\selectfont Recall@200 & 
\fontsize{20pt}{20pt}\selectfont\ mAP@30 &
\fontsize{20pt}{20pt}\selectfont\ Recall@200 & 
\fontsize{20pt}{20pt}\selectfont\ Recall@100 & 
\fontsize{20pt}{20pt}\selectfont\ Recall@200 & 
\fontsize{20pt}{20pt}\selectfont\ mAP@30 & 
\fontsize{20pt}{20pt}\selectfont\ Recall@200 \\
\midrule
\noalign{\vskip 0.5ex} 
& \multicolumn{9}{l}{\textbf{\fontsize{20pt}{20pt}\selectfont Lexical-Based Retrievers}} \\
\rule{0pt}{2ex}%
& \fontsize{20pt}{20pt}\selectfont\ BM-25 (A2A) & \fontsize{21pt}{21pt}\selectfont\ 23.24 & \fontsize{21pt}{21pt}\selectfont\ 28.45 & \fontsize{21pt}{21pt}\selectfont\ 4.82 & \fontsize{21pt}{21pt}\selectfont\ 27.61 & \fontsize{21pt}{21pt}\selectfont\ 21.50 & \fontsize{21pt}{21pt}\selectfont\ 26.44 & \fontsize{21pt}{21pt}\selectfont\ 4.36 & \fontsize{21pt}{21pt}\selectfont\ 24.64 \\
& \fontsize{20pt}{20pt}\selectfont\ BM-25 (F2F) & \fontsize{21pt}{21pt}\selectfont\ 23.68 & \fontsize{21pt}{21pt}\selectfont\ 30.25 & \fontsize{21pt}{21pt}\selectfont\ 7.24 & \fontsize{21pt}{21pt}\selectfont\ 33.81 & \fontsize{21pt}{21pt}\selectfont\ 24.92 & \fontsize{21pt}{21pt}\selectfont\ 31.12 & \fontsize{21pt}{21pt}\selectfont\ 7.44 & \fontsize{21pt}{21pt}\selectfont\ 34.90 \\
\midrule\midrule
& \multicolumn{9}{l}{\textbf{\fontsize{20pt}{20pt}\selectfont Domain-Specific Retriever}} \\
\rule{0pt}{2ex}%
& \fontsize{20pt}{20pt}\selectfont\ SciNCL-\textbf{A2A} & \fontsize{21pt}{21pt}\selectfont\ 29.18 & \fontsize{21pt}{21pt}\selectfont\ 36.62 & \fontsize{21pt}{21pt}\selectfont\ 5.67 & \fontsize{21pt}{21pt}\selectfont\ 34.69 & \fontsize{21pt}{21pt}\selectfont\ 27.23 & \fontsize{21pt}{21pt}\selectfont\ 34.55 & \fontsize{21pt}{21pt}\selectfont\ 4.87 & \fontsize{21pt}{21pt}\selectfont\ 33.05  \\

& \fontsize{20pt}{20pt}\selectfont\ SPECTER2-Base-\textbf{A2A} & \fontsize{21pt}{21pt}\selectfont\ 28.28 & \fontsize{21pt}{21pt}\selectfont\ 35.81 & \fontsize{21pt}{21pt}\selectfont\ 6.21 & \fontsize{21pt}{21pt}\selectfont\ 36.38 & \fontsize{21pt}{21pt}\selectfont\ 25.47 & \fontsize{21pt}{21pt}\selectfont\ 33.12 & \fontsize{21pt}{21pt}\selectfont\ 5.33 & \fontsize{20pt}{20pt}\selectfont\ 34.14  \\

& \fontsize{20pt}{20pt}\selectfont\ SPECTER2-Adapter-MTL CTRL-\textbf{A2A} & \fontsize{21pt}{21pt}\selectfont\ 28.42 & \fontsize{20pt}{20pt}\selectfont\ 35.73 & \fontsize{21pt}{21pt}\selectfont\ 5.81 & \fontsize{20pt}{20pt}\selectfont\ 34.77 & \fontsize{21pt}{21pt}\selectfont\ 26.28 & \fontsize{20pt}{20pt}\selectfont\ 33.14 & \fontsize{21pt}{21pt}\selectfont\ 4.88 & \fontsize{20pt}{20pt}\selectfont\ 32.36 \\

& \fontsize{20pt}{20pt}\selectfont\ SciMult-MHAExpert-\textbf{A2A} & \fontsize{21pt}{21pt}\selectfont\ 25.96 & \fontsize{21pt}{21pt}\selectfont\ 33.07 & \fontsize{21pt}{21pt}\selectfont\ 5.11 & \fontsize{21pt}{21pt}\selectfont\ 30.78 & \fontsize{21pt}{21pt}\selectfont\ 23.37 & \fontsize{21pt}{21pt}\selectfont\ 30.32 & \fontsize{21pt}{21pt}\selectfont\ 4.13 & \fontsize{21pt}{21pt}\selectfont\ 27.35 \\

\midrule\midrule
\noalign{\vskip 0.5ex} 
& \multicolumn{9}{l}{\textbf{\fontsize{20pt}{20pt}\selectfont Jina-Embeddings-v2-BASE-EN}} \\
\multirow{12}{*}{\raisebox{-9ex}[0pt][0pt]{\rotatebox[origin=c]{90}
{\textbf{\fontsize{26pt}{26pt}\selectfont Domain-Agnostic Retriever}}}} & 
\fontsize{20pt}{20pt}\selectfont\ A2A (Abstract-to-Abstract)& \fontsize{20pt}{21pt}\selectfont\ 27.75 & \fontsize{21pt}{21pt}\selectfont\ 34.49 & \fontsize{21pt}{21pt}\selectfont\ 6.20 & \fontsize{21pt}{21pt}\selectfont\ 35.33 & \fontsize{21pt}{21pt}\selectfont\ 26.08 & \fontsize{21pt}{21pt}\selectfont\ 32.64 & \fontsize{21pt}{21pt}\selectfont\ 5.24 & \fontsize{21pt}{21pt}\selectfont\ 32.43 \\

& \fontsize{20pt}{20pt}\selectfont\ F2F (Full-to-Full) & \fontsize{21pt}{21pt}\selectfont\ 28.78 & \fontsize{21pt}{21pt}\selectfont\ 36.46 & \fontsize{21pt}{21pt}\selectfont\ 7.03 & \fontsize{21pt}{21pt}\selectfont\ 35.88 & \fontsize{21pt}{21pt}\selectfont\ 29.51 & \fontsize{21pt}{21pt}\selectfont\ 36.98 & \fontsize{21pt}{21pt}\selectfont\ 6.32 & \fontsize{21pt}{21pt}\selectfont\ 35.51 \\

\noalign{\vskip 0.25ex}\cdashline{2-10}\noalign{\vskip 1.5ex}
& \cellcolor{blue!5} \fontsize{20pt}{20pt}\selectfont\ \textbf{CoR w/ Llama-3.2-3B-Instruct (w/ DPO)} & \cellcolor{blue!5} \fontsize{21pt}{21pt}\selectfont\ \textbf{31.96} & \cellcolor{blue!5} \fontsize{21pt}{21pt}\selectfont\ \textbf{39.81} & \cellcolor{blue!5} \fontsize{21pt}{21pt}\selectfont\ \textbf{7.67} & \cellcolor{blue!5} \fontsize{21pt}{21pt}\selectfont\ \textbf{42.93} & \cellcolor{blue!5} \fontsize{21pt}{21pt}\selectfont\ \textbf{30.76} & \cellcolor{blue!5} \fontsize{21pt}{21pt}\selectfont\ \textbf{39.21} & \cellcolor{blue!5} \fontsize{21pt}{21pt}\selectfont\ \textbf{6.57} & \cellcolor{blue!5} \fontsize{21pt}{21pt}\selectfont\ \textbf{40.31} \\

& \cellcolor{blue!5} \fontsize{20pt}{20pt}\selectfont\ \textbf{CoR w/ QWEN-2.5-3B-Instruct (w/ DPO)} & \cellcolor{blue!5} \fontsize{21pt}{21pt}\selectfont\ \textbf{32.02} & \cellcolor{blue!5} \fontsize{21pt}{21pt}\selectfont\ \textbf{40.41}  & \cellcolor{blue!5} \fontsize{21pt}{21pt}\selectfont\ \textbf{7.92}  & \cellcolor{blue!5} \fontsize{21pt}{21pt}\selectfont\ \textbf{43.75} & \cellcolor{blue!5} \fontsize{21pt}{21pt}\selectfont\ \textbf{30.83} & \cellcolor{blue!5} \fontsize{21pt}{21pt}\selectfont\ \textbf{38.96} & \cellcolor{blue!5} \fontsize{21pt}{21pt}\selectfont\ \textbf{6.66} & \cellcolor{blue!5} \fontsize{21pt}{21pt}\selectfont\ \textbf{41.86} \\

\noalign{\vskip 0.75ex}\cline{2-10}\noalign{\vskip 1.5ex} 
&\multicolumn{9}{l}{\textbf{\fontsize{20pt}{20pt}\selectfont BGE-M3 }} \\
\rule{0pt}{2ex}%
& \fontsize{20pt}{20pt}\selectfont\ A2A (Abstract-to-Abstract) & \fontsize{21pt}{21pt}\selectfont\ 25.88 & \fontsize{21pt}{21pt}\selectfont\ 32.01 & \fontsize{21pt}{21pt}\selectfont\ 5.56 & \fontsize{21pt}{21pt}\selectfont\ 33.76 & \fontsize{21pt}{21pt}\selectfont\ 23.50 & \fontsize{21pt}{21pt}\selectfont\ 29.36 & \fontsize{21pt}{21pt}\selectfont\ 4.57 & \fontsize{21pt}{21pt}\selectfont\ 29.63 \\

& \fontsize{20pt}{20pt}\selectfont\ F2F (Full-to-Full) & \fontsize{21pt}{21pt}\selectfont\ 25.15 & \fontsize{21pt}{21pt}\selectfont\ 32.21 & \fontsize{21pt}{21pt}\selectfont\ 5.98 & \fontsize{21pt}{21pt}\selectfont\ 32.64 & \fontsize{21pt}{21pt}\selectfont\ 24.81 & \fontsize{21pt}{21pt}\selectfont\ 30.84 & \fontsize{21pt}{21pt}\selectfont\ 5.24 & \fontsize{21pt}{21pt}\selectfont\ 30.08 \\

\noalign{\vskip 0.25ex}\cdashline{2-10}\noalign{\vskip 1.5ex}
& \cellcolor{blue!5} \fontsize{20pt}{20pt}\selectfont\ \textbf{CoR w/ Llama-3.2-3B-Instruct (w/ DPO)}  & \cellcolor{blue!5} \fontsize{21pt}{21pt}\selectfont\ \textbf{28.43} & \cellcolor{blue!5} \fontsize{21pt}{21pt}\selectfont\ \textbf{35.75} & \cellcolor{blue!5} \fontsize{21pt}{21pt}\selectfont\ \textbf{6.84} & \cellcolor{blue!5} \fontsize{21pt}{21pt}\selectfont\ \textbf{39.48} & \cellcolor{blue!5} \fontsize{21pt}{21pt}\selectfont\ \textbf{26.19} & \cellcolor{blue!5} \fontsize{21pt}{21pt}\selectfont\ \textbf{33.31} & \cellcolor{blue!5} \fontsize{21pt}{21pt}\selectfont\ \textbf{5.78} & \cellcolor{blue!5} \fontsize{21pt}{21pt}\selectfont\ \textbf{35.63} \\

& \cellcolor{blue!5} \fontsize{20pt}{20pt}\selectfont\ \textbf{CoR w/ QWEN-2.5-3B-Instruct (w/ DPO)} & \cellcolor{blue!5} \fontsize{21pt}{21pt}\selectfont\ \textbf{28.65} & \cellcolor{blue!5} \fontsize{21pt}{21pt}\selectfont\ \textbf{36.33} & \cellcolor{blue!5} \fontsize{21pt}{21pt}\selectfont\ \textbf{6.75} & \cellcolor{blue!5} \fontsize{21pt}{21pt}\selectfont\ \textbf{40.22} & \cellcolor{blue!5} \fontsize{21pt}{21pt}\selectfont\ \textbf{26.77} & \cellcolor{blue!5} \fontsize{21pt}{21pt}\selectfont\ \textbf{34.28} & \cellcolor{blue!5} \fontsize{21pt}{21pt}\selectfont\ \textbf{6.01}& \cellcolor{blue!5} \fontsize{21pt}{21pt}\selectfont\ \textbf{37.01} \\

\noalign{\vskip 0.75ex}\cline{2-10}\noalign{\vskip 1.5ex} 
&\multicolumn{9}{l}{\textbf{\fontsize{20pt}{20pt}\selectfont Inf-Retriever-v1-1.5B }} \\
\rule{0pt}{2ex}%
& \fontsize{20pt}{20pt}\selectfont\ A2A (Abstract-to-Abstract) & \fontsize{21pt}{21pt}\selectfont\ 37.55 & \fontsize{21pt}{21pt}\selectfont\ 46.24 & \fontsize{21pt}{21pt}\selectfont\ 7.65 & \fontsize{21pt}{21pt}\selectfont\ 41.77 & \fontsize{21pt}{21pt}\selectfont\ 34.73 & \fontsize{21pt}{21pt}\selectfont\ 43.40 & \fontsize{21pt}{21pt}\selectfont\ 6.35 & \fontsize{21pt}{21pt}\selectfont\ 39.00 \\

& \fontsize{20pt}{20pt}\selectfont\ F2F (Full-to-Full) & \fontsize{21pt}{21pt}\selectfont\ 36.35 & \fontsize{21pt}{21pt}\selectfont\ 44.37 & \fontsize{21pt}{21pt}\selectfont\ 7.81 & \fontsize{21pt}{21pt}\selectfont\ 38.86 & \fontsize{21pt}{21pt}\selectfont\ 34.65 & \fontsize{21pt}{21pt}\selectfont\ 41.83 & \fontsize{21pt}{21pt}\selectfont\ 6.98 & \fontsize{21pt}{21pt}\selectfont\ 38.92 \\

\noalign{\vskip 0.25ex}\cdashline{2-10}\noalign{\vskip 1.5ex}
& \cellcolor{blue!5} \fontsize{20pt}{20pt}\selectfont\ \textbf{CoR w/ Llama-3.2-3B-Instruct (w/ DPO)} & \cellcolor{blue!5} \fontsize{21pt}{21pt}\selectfont\ \textbf{39.44} & \cellcolor{blue!5} \fontsize{21pt}{21pt}\selectfont\ \textbf{49.3} & \cellcolor{blue!5} \fontsize{21pt}{21pt}\selectfont\ \textbf{9.13} & \cellcolor{blue!5} \fontsize{21pt}{21pt}\selectfont\ \textbf{48.23} & \cellcolor{blue!5} \fontsize{21pt}{21pt}\selectfont\ \textbf{37.26} & \cellcolor{blue!5} \fontsize{21pt}{21pt}\selectfont\ \textbf{46.81} & \cellcolor{blue!5} \fontsize{21pt}{21pt}\selectfont\ \textbf{8.05} & \cellcolor{blue!5} \fontsize{21pt}{21pt}\selectfont\ \textbf{46.29} \\

& \cellcolor{blue!5} \fontsize{20pt}{20pt}\selectfont\ \textbf{CoR w/ QWEN-2.5-3B-Instruct (w/ DPO)} & \cellcolor{blue!5} \fontsize{21pt}{21pt}\selectfont\ \textbf{39.73} & \cellcolor{blue!5} \fontsize{21pt}{21pt}\selectfont\ \textbf{49.31} & \cellcolor{blue!5} \fontsize{21pt}{21pt}\selectfont\ \textbf{9.60} & \cellcolor{blue!5} \fontsize{21pt}{21pt}\selectfont\ \textbf{48.82} & \cellcolor{blue!5} \fontsize{21pt}{21pt}\selectfont\ \textbf{36.80} & \cellcolor{blue!5} \fontsize{21pt}{21pt}\selectfont\ \textbf{46.37} & \cellcolor{blue!5} \fontsize{21pt}{21pt}\selectfont\ \textbf{8.38} & \cellcolor{blue!5} \fontsize{21pt}{21pt}\selectfont\ \textbf{46.51} \\

\midrule
\bottomrule
\end{tabular}
}
\caption{Main results with additional metrics.}
\label{tab:main_final_w_additional_metrics}
\vspace{-0.075in}
\end{table*}
\clearpage
\begin{figure*}[ht]
    \renewcommand{\arraystretch}{0.8}
    \centering

    \begin{minipage}{0.49\linewidth}
        \centering
        \includegraphics[width=0.98\columnwidth]{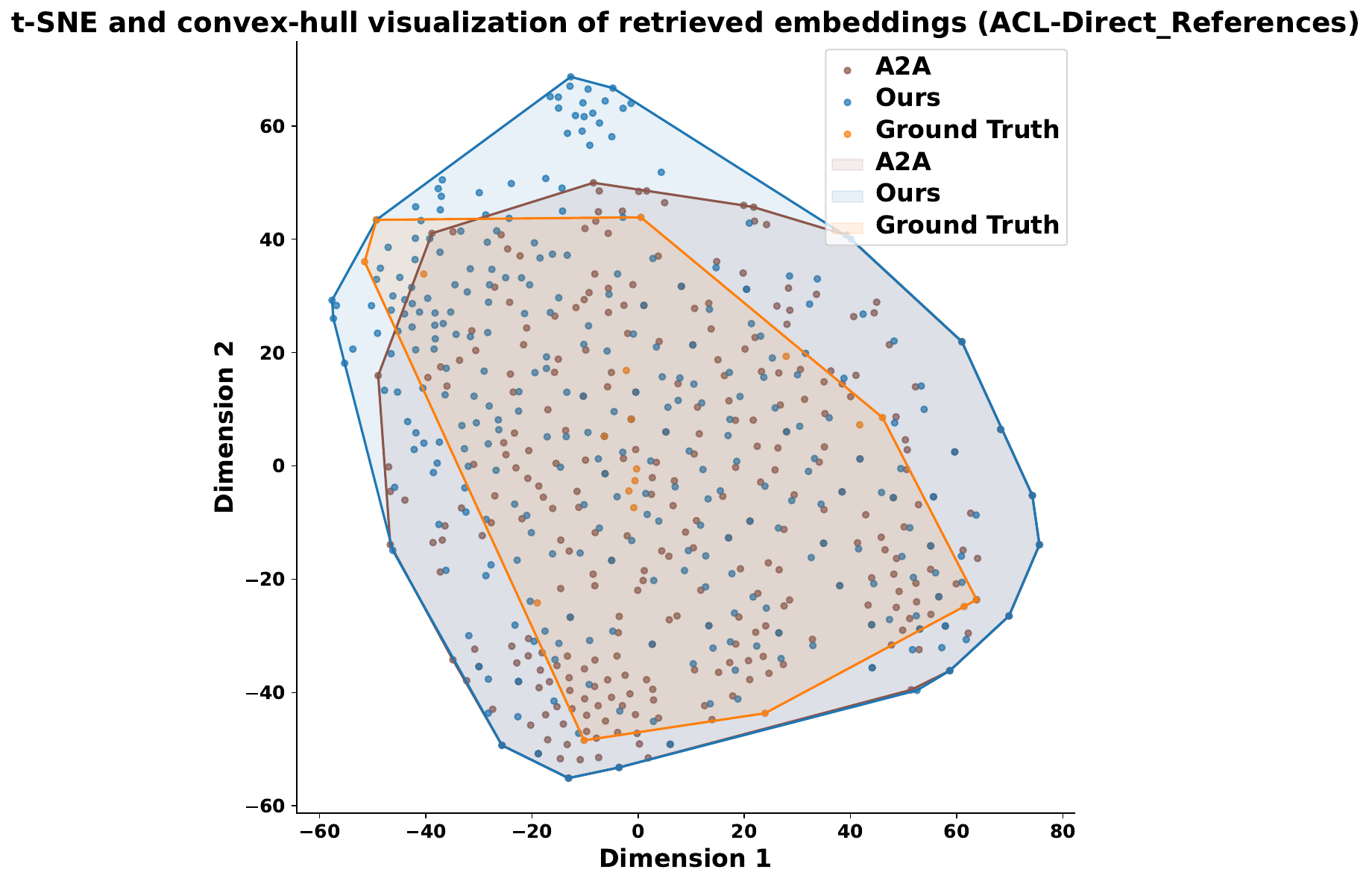}
        \label{fig:tsne1}
    \end{minipage}\hfill
    \begin{minipage}{0.49\linewidth}
        \centering
        \includegraphics[width=0.98\columnwidth]{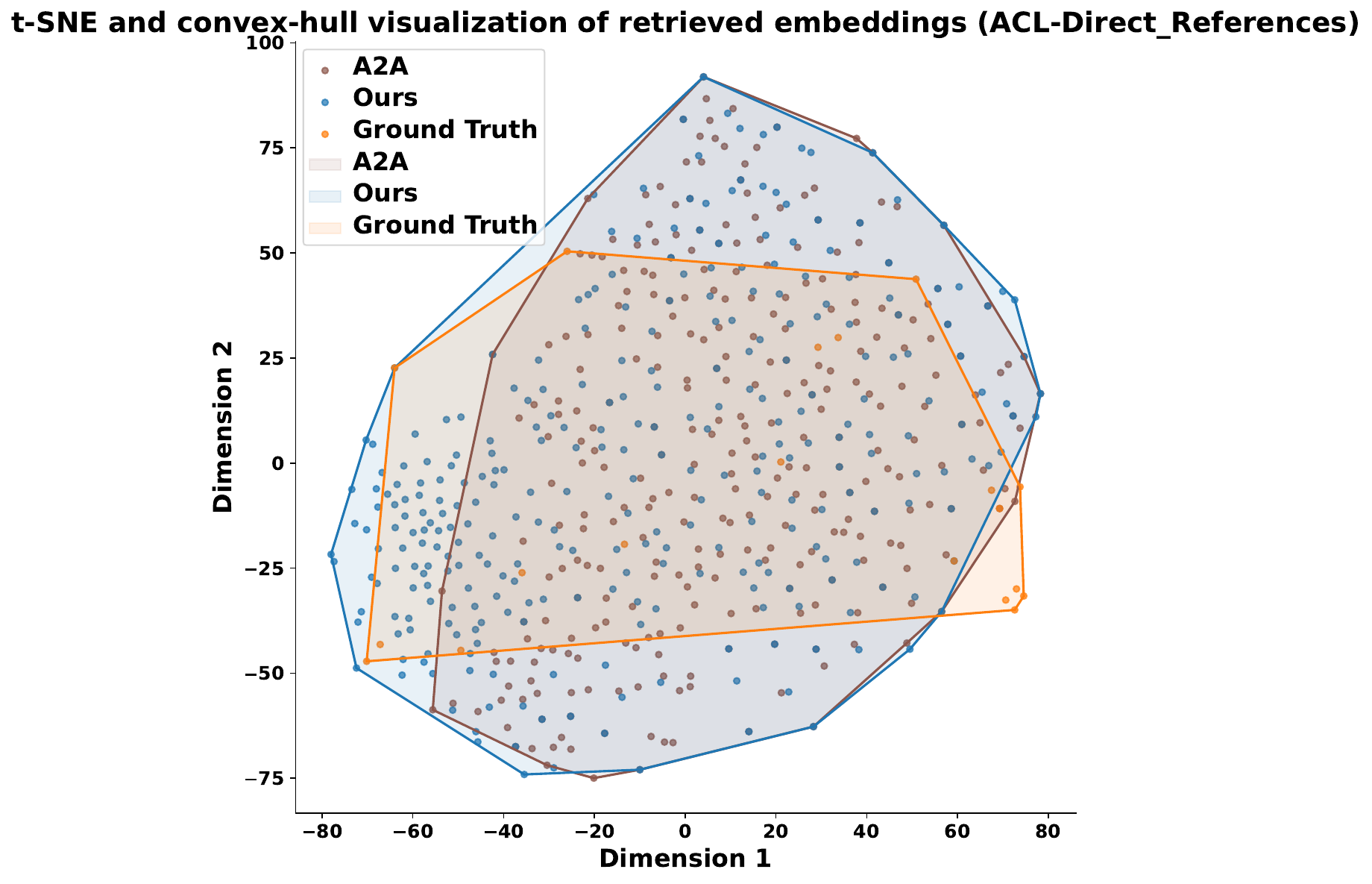}
        \label{fig:tsne2}
    \end{minipage}

    \vspace{0.12in}

    \begin{minipage}{0.49\linewidth}
        \centering
        \includegraphics[width=0.98\columnwidth]{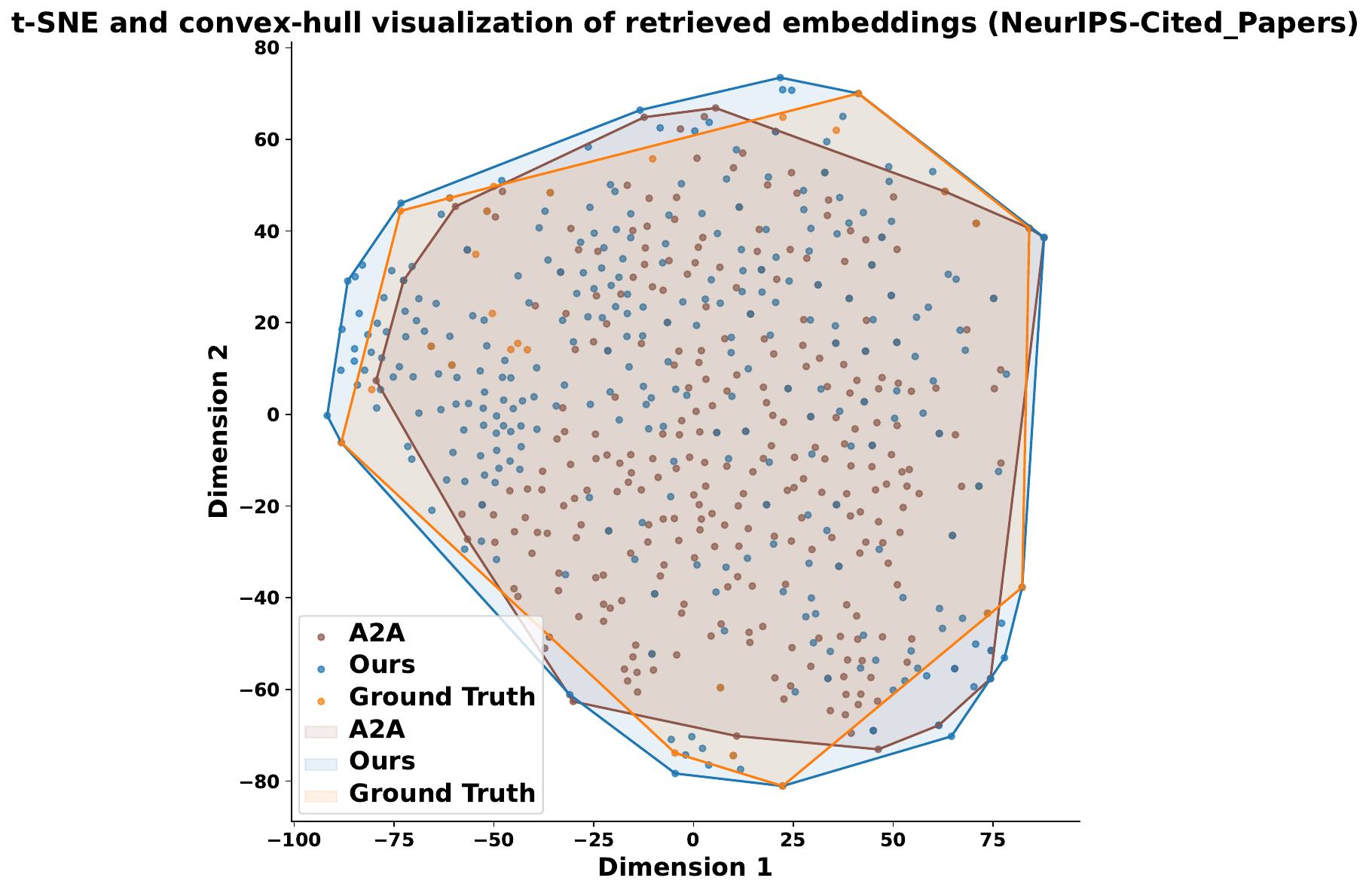}
        \label{fig:tsne3}
    \end{minipage}\hfill
    \begin{minipage}{0.49\linewidth}
        \centering
        \includegraphics[width=0.98\columnwidth]{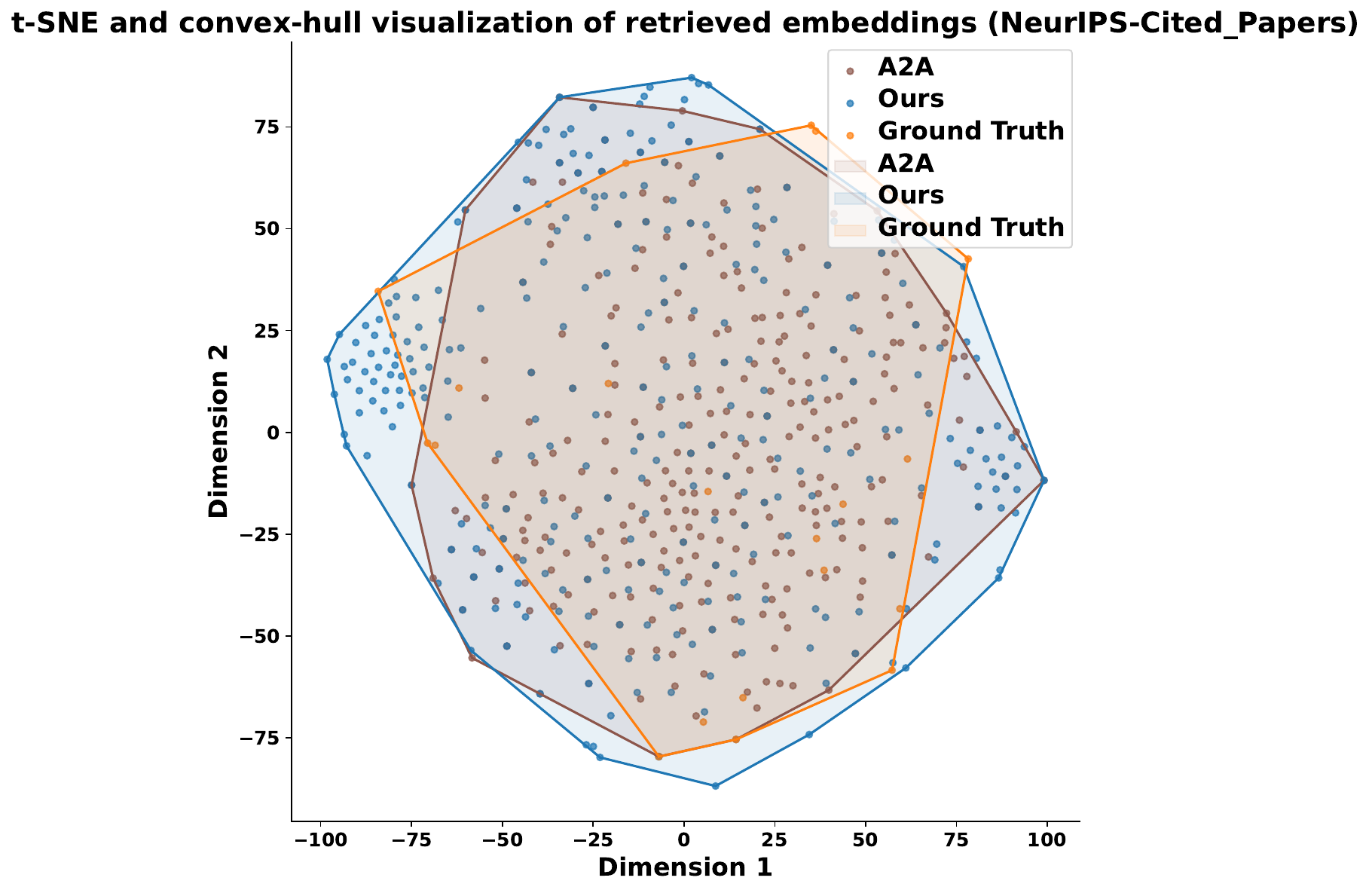}
        \label{fig:tsne4}
    \end{minipage}

    \vspace{0.12in}

    \begin{minipage}{0.49\linewidth}
        \centering
        \includegraphics[width=0.98\columnwidth]{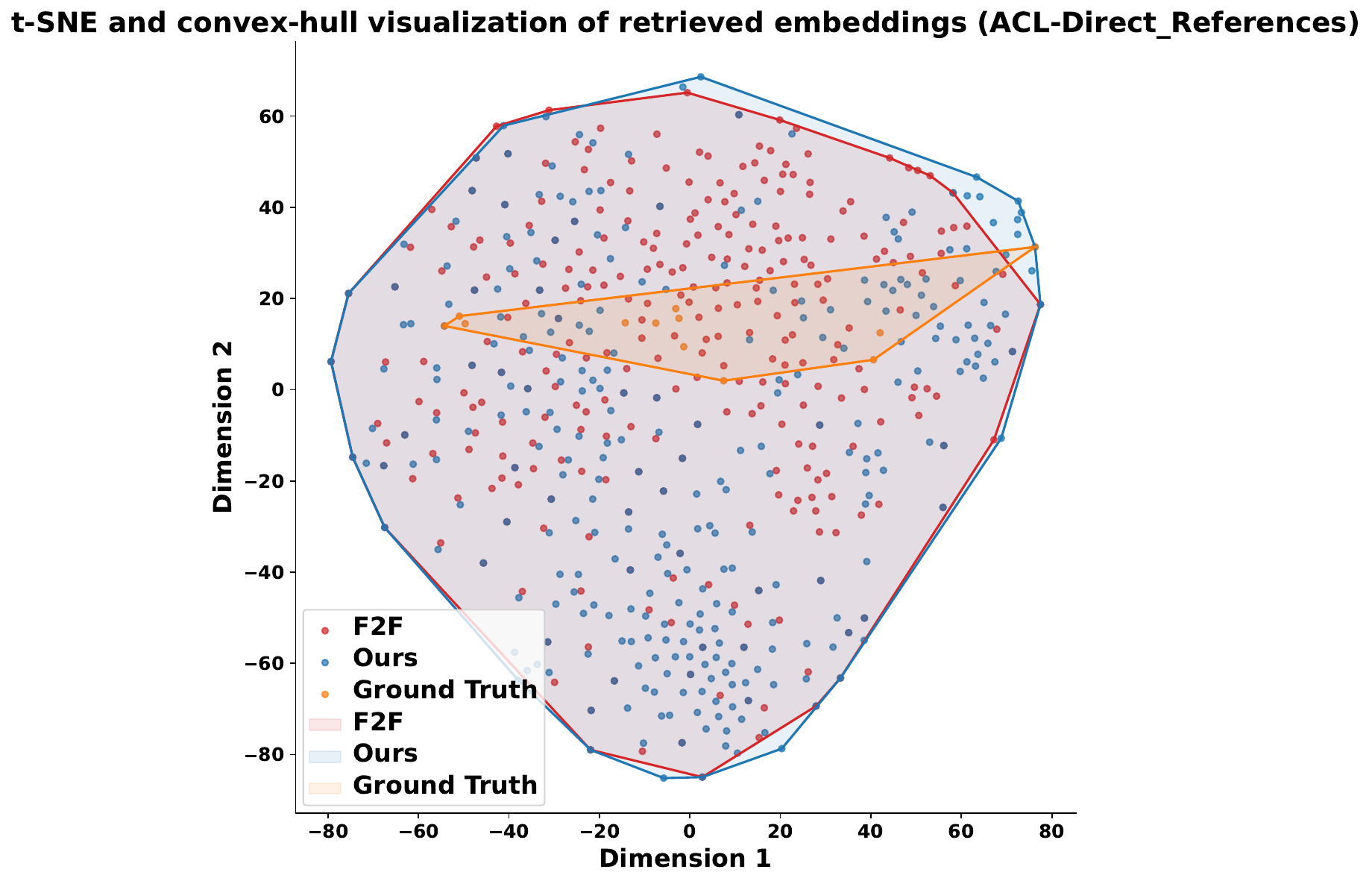}
        \label{fig:tsne5}
    \end{minipage}\hfill
    \begin{minipage}{0.49\linewidth}
        \centering
        \includegraphics[width=0.98\columnwidth]{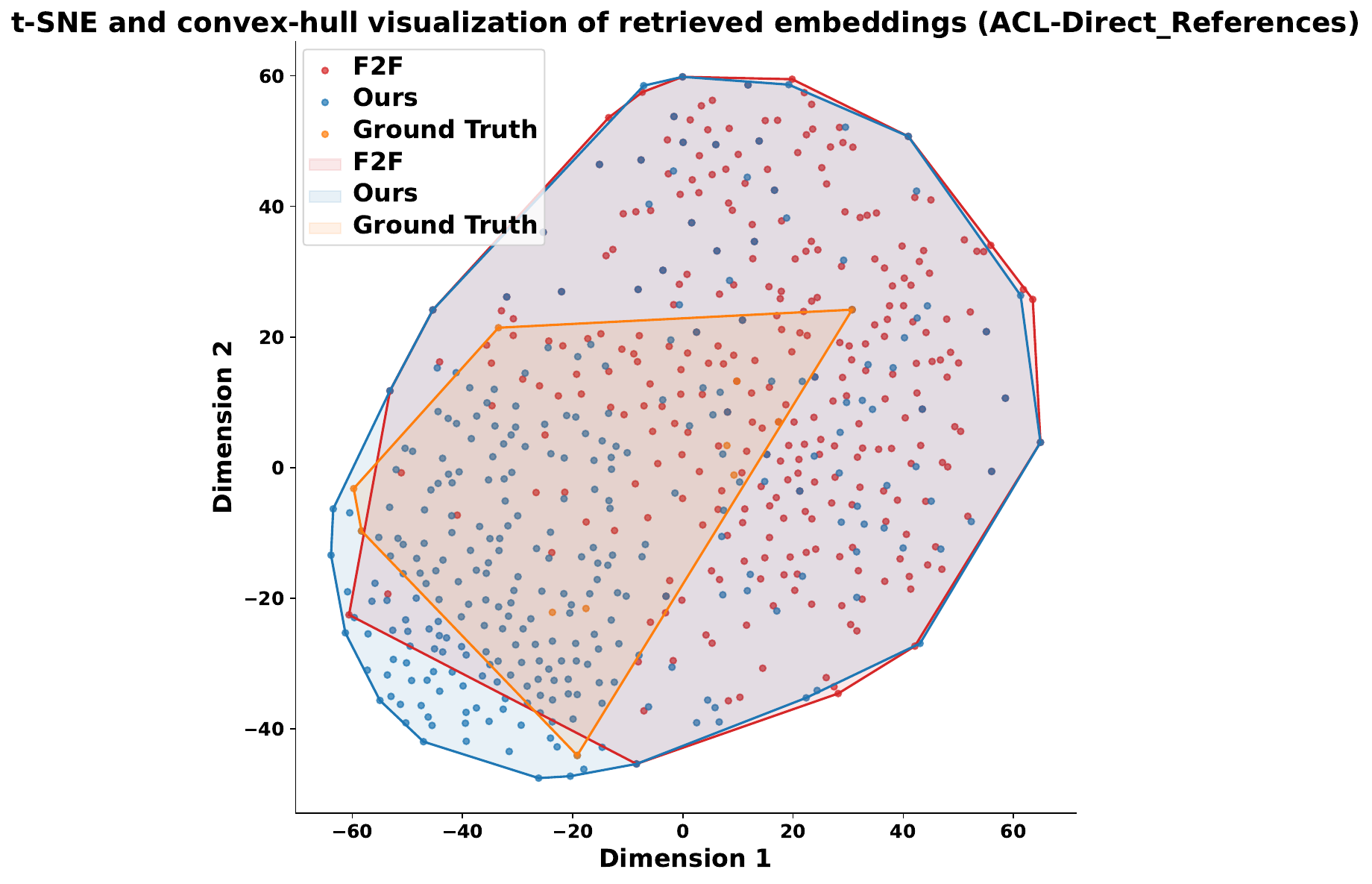}
        \label{fig:tsne6}
    \end{minipage}

    \vspace{0.12in}

    \begin{minipage}{0.49\linewidth}
        \centering
        \includegraphics[width=0.98\columnwidth]{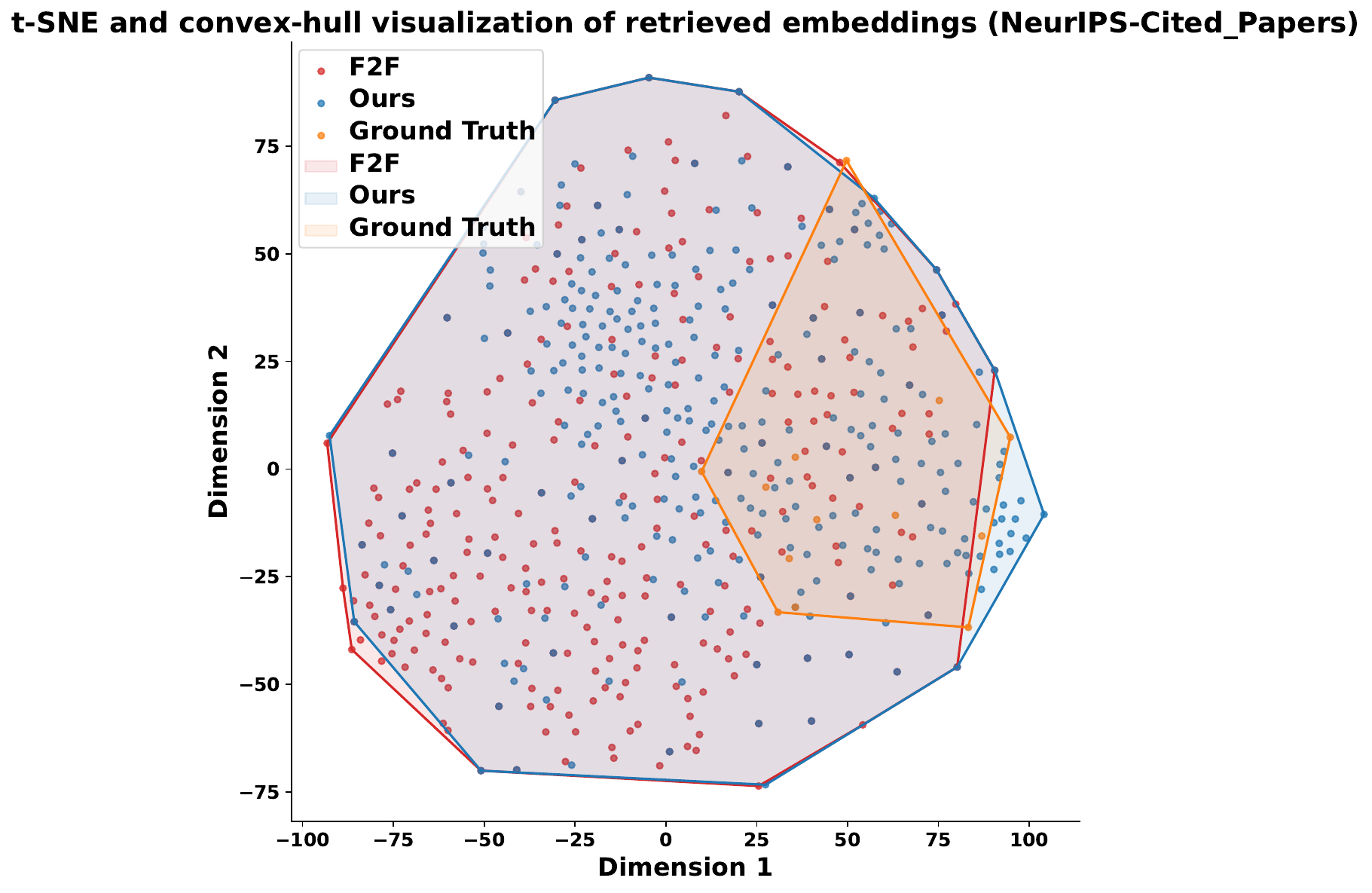}
        \label{fig:tsne7}
    \end{minipage}\hfill
    \begin{minipage}{0.49\linewidth}
        \centering
        \includegraphics[width=0.98\columnwidth]{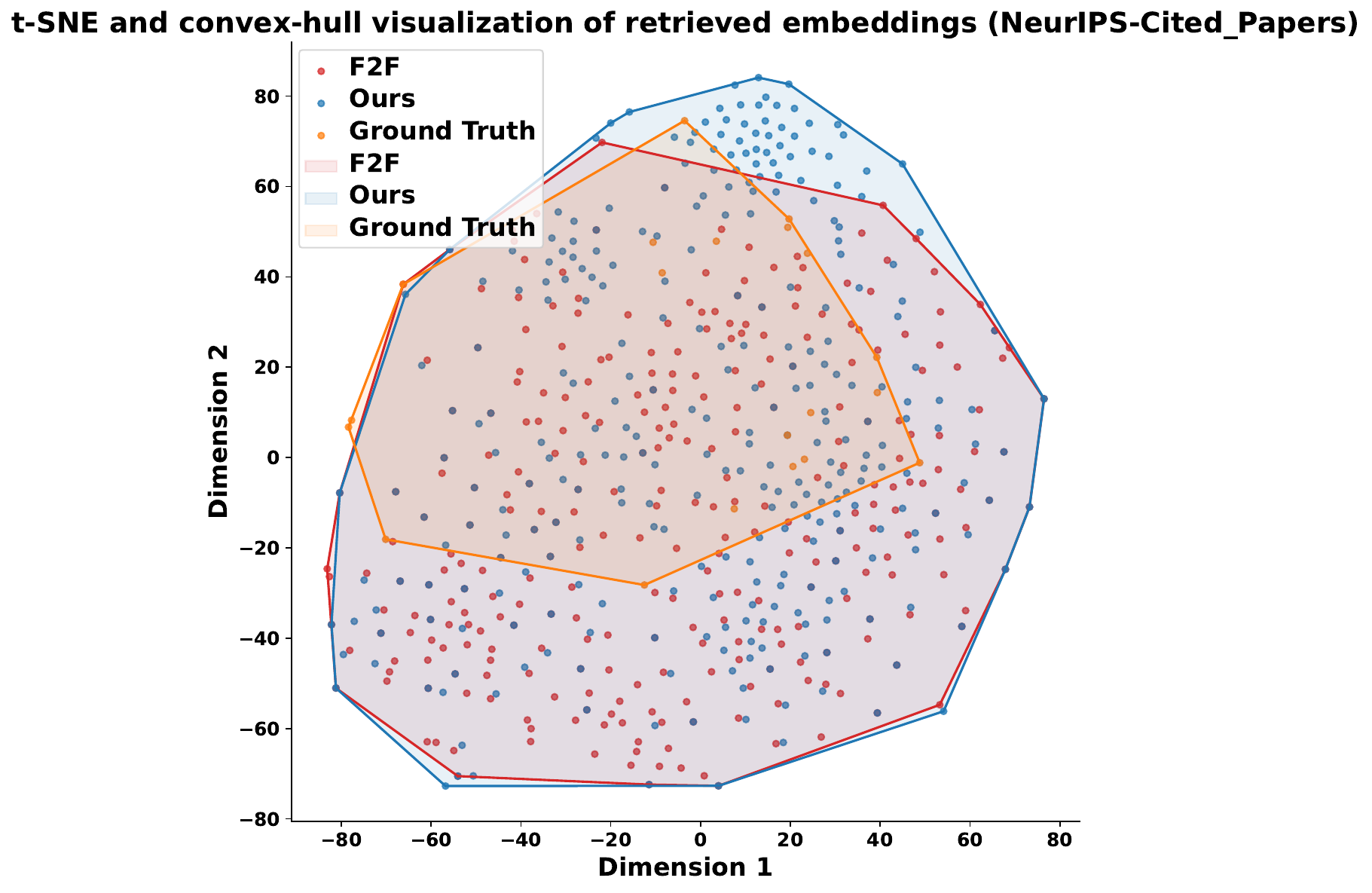}
        \label{fig:tsne8}
    \end{minipage}

    \vspace{-0.1in}
    \captionsetup{justification=justified, singlelinecheck=false}
    \caption{t-SNE and convex hull boundary visualization of Top-300 retrieved document embeddings across eight queries from respective benchmark splits. We report the results with Jina-Embeddings-v2-Base-EN as the embedding model.}
    \label{fig:embedding_diversity_tsne_visualization}
\end{figure*}

\clearpage
\section{Algorithm Details}
In this section, we provide further details on the functional roles of our \textsc{CoR} framework with formal algorithms. Specifically, Algorithm~\ref{alg:Aspect_Aware_Retrieval} defines an aspect-aware retrieval process from Section~\ref{method:aspect}, where the query optimizers decompose the input document into aspect-aware queries for exploration and then the papers are retrieved from multi-vector corpora. Algorithm~\ref{alg:NextQuerySelection} describes the aspect-aware next query selection process in Section~\ref{method:iterative_chain}. Algorithm~\ref{alg:chain_ret} shows the overall exploration process of \textsc{CoR}, using Algorithm~\ref{alg:Aspect_Aware_Retrieval} and Algorithm~\ref{alg:NextQuerySelection}. Finally, Algorithm~\ref{alg:Post_Merging} explains the Post-Order Aggregation algorithm of \textsc{CoR} explained in Section~\ref{method:aggregation}.

\label{appendix:algorithm_details}
\begin{algorithm*}[!t]
\caption{Aspect-Aware Multi-Vector Retrieval}
\label{alg:Aspect_Aware_Retrieval}
\begin{algorithmic}[1]

\State \textbf{Require:} Input paper $D$; Input paper abstract $D_\text{abstract}$; Top-$k$ per query $K$; \\Name of the Parent Retrieved Results \textsc{parent}; 
Corpora $\mathcal{C} \gets \{\mathcal{C}_\text{abstract}, \mathcal{C}_\text{chunked}\}$;\\

\State \textbf{Initialize:} Functional aspect-aware LLM query optimizer agents $\mathcal{F} = \{f_\text{M}, f_\text{E}, f_\text{R}\}$\\

\Function{OneHopRetrieval}{$D,\textsc{parent}$, $\mathcal{C}$, $K$}
  \State $R\gets\text{[ ]}$ \Comment{\textcolor{submission}{Memory to save aspect-aware retrieved results per document}}
  \ForAll{$f_{i}\in\mathcal{F}$} 
    \State $\textsc{Name}\gets\textsc{parent}\Vert f_{i}$ \Comment{\textcolor{submission}{Update name for aspect-aware retrieval for document D}}
    \State $q\gets f_{i}(D)$
    \State $T\gets\texttt{Retrieve}(q,C_\text{chunked},K)$\Comment{ \textcolor{submission}{Top-K Retrieved Results from chunked corpus}}
    \State $\mathcal{R}_{q}\gets[\,h(x)\mid x\in T\,]$\Comment{\textcolor{submission}{Retrieved Results mapped back to paper}}
    \State $R.\textsc{Append}((\mathcal{R}_{q},\ {\textsc{Parent},\textsc{Name}, f_i))}$
  \EndFor

  \State $\textsc{Name}\gets\textsc{parent}\Vert \text{abstract}$
  \State $q\gets D_\text{abstract}$
  \State $T\gets\texttt{Retrieve}(q, \mathcal{C}_\text{abstract},K)$\Comment{\textcolor{submission}{Top-K Retrieved Results from abstract corpus}}
  \State $\mathcal{R}_{q}\gets[\,h(x)\mid x\in T\,]$\Comment{\textcolor{submission}{Retrieved Results mapped to paper}}
  \State $R.\textsc{Append}((\mathcal{R}_{q},\ {\textsc{Parent},\textsc{Name}, \text{abstract}))}$
  
  \State \Return $R$
\EndFunction
\label{alg:aspect_aware_retrieval}
\end{algorithmic}
\end{algorithm*}

\begin{algorithm*}[t]
\caption{Aspect-Aware Next Query Selection}
\label{alg:NextQuerySelection}
\begin{algorithmic}[1]
\State \textbf{Require:} Root paper $D$; Max depth $R$; selection starting index $\gamma$; Top-$k$ per query $K$\\
Aspect Aware Cache \textsc{Cache}; memory to save query paper for next round retrieval $Q_{\text{next}}$
\vspace{0.5em}\\

\Function{InitialAspect}{$\textsc{Name}$} \Comment{\textcolor{submission}{Obtain Initial Branching Aspect of the root document}}
  \State Parse $\textsc{Name}$ as $(\textsc{Root} \,\Vert\, a_1 \,\Vert\, a_2 \,\Vert\, \cdots)$
  \State \Return $a_1$
\EndFunction
\\
\Function{SelectNextQuery}{$P$, $\textsc{Cache}$, $\gamma$}
   \ForAll{$r \in P$}
      \State $\textsc{TopK},\textsc{Parent},\textsc{Name}, \textsc{Previous Aspect}  \gets\ r$
      \If {$\textsc{Previous Aspect} = \text{abstract}$} 
          \State  $\textsc{continue}$ \\\Comment{\textcolor{submission}{Skip selection if previous retrieved results were from abstract query}}\\
      \EndIf
      \State $\textsc{Initial Aspect} \gets\ \textsc{InitialAspect}(\textsc{Name}) $
      \For{$j=\gamma$ \textbf{to} $K$}
        \State $\textsc{Doc}\gets \textsc{TopK}[j]$
        \If{$\textsc{Doc}\notin \textsc{Cache}[\textsc{Initial Aspect}]$}
            \State $\textsc{Cache}[\textsc{Initial Aspect}].\text{append}(\textsc{Doc})$
            \State $Q_{\text{next}}.\textsc{Append}((\textsc{Doc},\textsc{Name}))$
            \State \textbf{break}
        \EndIf
      \EndFor\\\\
    \Comment{\textcolor{submission}{Select most similar document not in aspect-aware cache starting from the $\gamma$ index.}}
   \EndFor
\EndFunction
\\

\end{algorithmic}
\end{algorithm*}
\begin{algorithm*}[t]
\caption{Chain-of-Retrieval}
\label{alg:chain_ret}
\begin{algorithmic}[1]
\State \textbf{Require:} Root paper $D$; Max depth $R$; selection starting index $\gamma$;  Top-k per query \textsc{K}; Corpora $\mathcal{C} \gets \{\mathcal{C}_\text{abstract}, \mathcal{C}_\text{chunked}\}$\\

\vspace{0.5em}
\State \textbf{Initialize: } \\
query queue $Q \gets [(D,\textsc{Root})]$\\ 
memory to save retrieved results $\mathcal{M} \gets [\,]$
\State $\textsc{Cache}[a] \gets \emptyset \quad \forall a \in \{\mathrm{M},\mathrm{E}, \mathrm{R}\}$ \Comment{\textcolor{submission}{Initialize Aspect-Aware Cache for Method, Experiment, and Research Question Aspect}}\\

\For{$r = 0$ \textbf{to} $R-1$}
    \State $\mathcal{M}[r] \gets [\,]$, $Q_{\text{next}} \gets [\,]$
    \ForAll{$(d,\textsc{parent})$ in $Q$}
        \State $P \gets$ \Call{OneHopRetrieval}{$d,\textsc{parent}$, $\mathcal{C}$, \textsc{K}} 
        \State $\mathcal{M}[r].\textsc{Extend}(P)$
        \State $Q_{\text{next}} \gets$ \Call{SelectNextQuery}{$P$, \textsc{Cache}, $\gamma$} 
    \EndFor
    \State $Q \gets Q_{\text{next}}$
\EndFor
\end{algorithmic}
\end{algorithm*}

\begin{algorithm*}
\caption{Recursive Post-Order Merging}
\begin{algorithmic}[1]
\State \textbf{Require:} Root paper $D$; Max depth $R$; All Retrieved Results after depth $R$ of retrieval $\mathcal{M}$; \\

\State \textbf{Ensure:} Final top-$k$ candidates for $D$\\

\Function{Group}{$L$}\Comment{\textcolor{submission}{Group list of per round retrieved results with same parent nodes}}
   \State \Return $\{\, p \mapsto [\,x \in L : \mathrm{Parent}(x)=p\,] \,\}$
\EndFunction
\\
\Function{MergeSiblings}{$L$} \Comment{\textcolor{submission}{Merge retrieved results with same parent nodes}}
   \State $S \gets \{\}$
   \For{$(\textsc{Parent},G) \in \textsc{Group}(L)$}
      \State $S[\textsc{Parent}] \gets \textsc{RRF}(G)$
   \EndFor
   \State \Return $S$
\EndFunction
\\
\Function{MergeEdges}{$P, S$} \Comment{\textcolor{submission}{Merge retrieved results from parent nodes and merged child nodes}}
   \State $O \gets [\ ]$
   \For{$(\text{TopK},\textsc{Parent}, \textsc{Name}) \in P$}
      \If{$\textsc{Parent} \in Keys(S)$}
         \State $O.\text{append}(\textsc{RRF}(\text{TopK},\, S[\textsc{Parent}]))$
      \EndIf
   \EndFor
   \State \Return $O$
\EndFunction
\\
\State $r \gets R-1$ 
\\
\While{$r > 0$}\Comment{\textcolor{submission}{Post-Order Merge, starting from the bottom level nodes}}
   \State $S \gets \textsc{MergeSiblings}(\mathcal{M}[r])$
   \State $\mathcal{M}[r-1] \gets \textsc{MergeEdges}(\mathcal{M}[r-1], S)$
   \State $r \gets r-1$
\EndWhile

\State \Return $\textsc{RRF}(\mathcal{M}[0])$
\end{algorithmic}
\label{alg:Post_Merging}
\end{algorithm*}

\clearpage

\section{Case Study}

\label{appendix:Case_Study}
\vspace*{\fill}
\parbox{\textwidth}{
    \small
    \centering
    %\captionsetup{justification=justified, singlelinecheck=false}
    %\captionof{table}{Example of the retrieved documents for the input document with \textsc{CoR} from a EMNLP-Citations split in \textsc{SciFullBench}. Due to the length of input documents, we provide only the title and abstract for the input, as well as the titles for the retrieved documents. Retrieved documents that belong to ground truth are highlighted in blue.}
    \resizebox{\textwidth}{!}{
    %\label{tab:retrieved_case_analysis}
    \renewcommand{\arraystretch}{1.1}
    \begin{tabular}{l c}
    \toprule
    \multicolumn{2}{c}{} \\ \midrule
    \textbf{\textsf{Input Document Meta Data}} & 
    \multicolumn{1}{p{.9\textwidth}}{\textsf{\textbf{[Title]} Stance Detection on Social Media with Background Knowledge}
    \par \textsf{\textbf{[Abstract]} Identifying users\u2019 stances regarding specific targets/topics is a significant route to learning public opinion from social media platforms. Most existing studies of stance detection strive to learn stance information about specific targets from the context, in order to determine the user\u2019s stance on the target. However, in real-world scenarios, we usually have a certain understanding of a target when we express our stance on it. In this paper, we investigate stance detection from a novel perspective, where the background knowledge of the targets is taken into account for better stance detection. To be specific, we categorize background knowledge into two categories: episodic knowledge and discourse knowledge, and propose a novel Knowledge-Augmented Stance Detection (KASD) framework. For episodic knowledge, we devise a heuristic retrieval algorithm based on the topic to retrieve the Wikipedia documents relevant to the sample. Further, we construct a prompt for ChatGPT to filter the Wikipedia documents to derive episodic knowledge. For discourse knowledge, we construct a prompt for ChatGPT to paraphrase the hashtags, references, etc., in the sample, thereby injecting discourse knowledge into the sample. Experimental results on four benchmark datasets demonstrate that our KASD achieves state-of-the-art performance in in-target and zero-shot stance detection.}} \\ \noalign{\vskip 0.25ex}\cdashline{1-2}\noalign{\vskip 0.75ex}\textbf{\textsf{Ground-Truth Document Meta Data}} & 
    \multicolumn{1}{p{.9\textwidth}}{
    \par \textsf{\textbf{[Title]} A More Advanced Group Polarization Measurement Approach Based on LLM-Based Agents and Graphs }
    \par \textsf{\textbf{[Title]} A Survey of Stance Detection on Social Media: New Directions and Perspectives}
    \par \textsf{\textbf{[Title]} Chain of Stance: Stance Detection with Large Language Models}
    \par \textsf{\textbf{[Title]} A Challenge Dataset and Effective Models for Conversational Stance Detection}
    \par \textsf{\textbf{[Title]} Mitigating Biases of Large Language Models in Stance Detection with Counterfactual Augmented Calibration}
    \par \textsf{\textbf{[Title]} Multi-modal Stance Detection: New Datasets and Model}
    \par \textsf{\textbf{[Title]} A Logically Consistent Chain-of-Thought Approach for Stance Detection}
    \par \textsf{\textbf{[Title]} Stance Detection with Collaborative Role-Infused LLM-Based Agents}
    \par \textsf{\textbf{[Title]} Prompting and Fine-Tuning Open-Sourced Large Language Models for Stance Classification} 
    \par \textsf{\textbf{[Title]} Ladder-of-Thought: Using Knowledge as Steps to Elevate Stance Detection}
    \par \textsf{\textbf{[Title]} Investigating Chain-of-thought with ChatGPT for Stance Detection on Social Media}
    }
    \\\noalign{\vskip 0.25ex}\cdashline{1-2}\noalign{\vskip 0.75ex}\textbf{\textsf{Top@30 Retrieved Document Meta Data}} & 
    \multicolumn{1}{p{.9\textwidth}}{
    \textsf{\textbf{[Title]} \textcolor{submission} {A Survey of Stance Detection on Social Media: New Directions and Perspectives}}
    \par \textsf{\textbf{[Title]} \textcolor{submission} {Stance Detection with Collaborative Role-Infused LLM-Based Agents}}
    \par \textsf{\textbf{[Title]} A Survey on Stance Detection for Mis- and Disinformation Identification}
    \par \textsf{\textbf{[Title]} \textcolor{submission} {Prompting and Fine-Tuning Open-Sourced Large Language Models for Stance Classification}}
    \par \textsf{\textbf{[Title]} \textcolor{submission} {Chain of Stance: Stance Detection with Large Language Models}}
    \par \textsf{\textbf{[Title]} \textcolor{submission} {A Challenge Dataset and Effective Models for Conversational Stance Detection}}
    \par \textsf{\textbf{[Title]} Enabling Contextual Soft Moderation on Social Media through Contrastive Textual Deviation}
    \par \textsf{\textbf{[Title]} Stance Detection on Social Media with Fine-Tuned Large Language Models}
    \par \textsf{\textbf{[Title]} Stance Detection in Web and Social Media: A Comparative Study}
    \par \textsf{\textbf{[Title]} \textcolor{submission} {Multi-modal Stance Detection: New Datasets and Model}}
    \par \textsf{\textbf{[Title]} TATA: Stance Detection via Topic-Agnostic and Topic-Aware Embeddings}
    \par \textsf{\textbf{[Title]} A Benchmark for Cross-Domain Argumentative Stance Classification on Social Media}
    \par \textsf{\textbf{[Title]} \textcolor{submission} {Mitigating Biases of Large Language Models in Stance Detection with Counterfactual Augmented Calibration}}
    \par \textsf{\textbf{[Title]} Advancing Annotation of Stance in Social Media Posts: A Comparative Analysis of Large Language Models and Crowd Sourcing}
    \par \textsf{\textbf{[Title]} DEEM: Dynamic Experienced Expert Modeling for Stance Detection}
    \par \textsf{\textbf{[Title]} FarExStance: Explainable Stance Detection for Farsi}
    \par \textsf{\textbf{[Title]} \textcolor{submission} {Investigating Chain-of-thought with ChatGPT for Stance Detection on Social Media}}
    \par \textsf{\textbf{[Title]} Relative Counterfactual Contrastive Learning for Mitigating Pretrained Stance Bias in Stance Detection}
    \par \textsf{\textbf{[Title]} \textcolor{submission} {A Logically Consistent Chain-of-Thought Approach for Stance Detection}}
    \par \textsf{\textbf{[Title]} Embeddings-Based Clustering for Target Specific Stances: The Case of a Polarized Turkey}
    \par \textsf{\textbf{[Title]} Reinforcement Tuning for Detecting Stances and Debunking Rumors Jointly with Large Language Models}
    \par \textsf{\textbf{[Title]} Examining the Influence of Political Bias on Large Language Model Performance in Stance Classification}
    \par \textsf{\textbf{[Title]} KCD: Knowledge Walks and Textual Cues Enhanced Political Perspective Detection in News Media}
    \par \textsf{\textbf{[Title]} Deciphering Political Entity Sentiment in News with Large Language Models: Zero-Shot and Few-Shot Strategies}
    \par \textsf{\textbf{[Title]} Argumentative Stance Prediction: An Exploratory Study on Multimodality and Few-Shot Learning}
    \par \textsf{\textbf{[Title]} \textcolor{submission} {A More Advanced Group Polarization Measurement Approach Based on LLM-Based Agents and Graphs}}
    \par \textsf{\textbf{[Title]} Beneath the Tip of the Iceberg: Current Challenges and New Directions in Sentiment Analysis Research}
    \par \textsf{\textbf{[Title]} WIBA: What Is Being Argued? A Comprehensive Approach to Argument Mining}
    \par \textsf{\textbf{[Title]} KGAP: Knowledge Graph Augmented Political Perspective Detection in News Media}
    \par \textsf{\textbf{[Title]} "We Demand Justice!": Towards Social Context Grounding of Political Texts}
    }\\
    \bottomrule
    \end{tabular}
    }
    \captionsetup{justification=justified, singlelinecheck=false}
    \captionof{table}{Example of the retrieved documents for the input document with \textsc{CoR} from a EMNLP-Citations split in \textsc{SciFullBench}. Due to the length of input documents, we provide only the title and abstract for the input, as well as the titles for the retrieved documents. Retrieved documents that belong to ground truth are highlighted in blue.}
    \label{tab:retrieved_case_analysis}
}
\vspace*{\fill}
\clearpage
\vspace*{\fill}
\parbox{\textwidth}{
    \small
    \centering
    %\captionsetup{justification=justified, singlelinecheck=false}
    %\captionof{table}{Example of the retrieved documents for the input document with $\textsc{CoR}$ from a Citations split in \textsc{PatentFullBench}. Due to the length of input documents, we provide only the title and abstract for the input, as well as the titles for the retrieved documents. Retrieved documents that belong to ground truth are highlighted in blue.}
    \resizebox{\textwidth}{!}{
    %\label{tab:retrieved_case_analysis_patents}
    \renewcommand{\arraystretch}{1.1}
    \begin{tabular}{l c}
    \toprule
    \multicolumn{2}{c}{} \\ \midrule
    \textbf{\textsf{Input Document Meta Data}} & 
    \multicolumn{1}{p{.9\textwidth}}{\textsf{\textbf{[Abstract]} The present disclosure describes methods and systems directed towards providing scaled engagement and views of an e-sports event. Instead of providing the same distribution of live e-sport event data to all remote viewers of a live e-sports event, features associated with e-sports gaming network could be used to customize the distribution of live e-sport event data to promote immersive viewer experience. The enhanced immersion can also be carried out in a virtual reality or augmented reality setting. The features would be capable of providing additional information, different views, and a variety of different commentators for the e-sports event so that the viewer can be more engaged when viewing the particular e-sports event. With the increased engagement from remote viewers, the distribution of live e-sports event data can also be further modified for monetization by incorporating advertisements as well.}} \\ \noalign{\vskip 0.25ex}\cdashline{1-2}\noalign{\vskip 0.75ex}\textbf{\textsf{Ground-Truth Document Meta Data}} & 
    \multicolumn{1}{p{.9\textwidth}}{
    \par \textsf{\textbf{[Title]} Statistical driven tournaments}
    \par \textsf{\textbf{[Title]} Scaled VR engagement and views in an e-sports event}
    \par \textsf{\textbf{[Title]} Player to spectator handoff and other spectator controls}
    \par \textsf{\textbf{[Title]} Creation of winner tournaments with fandom influence}
    \par \textsf{\textbf{[Title]} User-driven spectator channel for live game play in multi-player games}
    \par \textsf{\textbf{[Title]} Methods and systems to increase interest in and viewership of content before, during and after a live event}
    \par \textsf{\textbf{[Title]} Discovery and detection of events in interactive content}
    \par \textsf{\textbf{[Title]} Integrating commentary content and gameplay content over a multi-user platform}
    \par \textsf{\textbf{[Title]} Statistically defined game channels} 
    \par \textsf{\textbf{[Title]} Online tournament integration}
    \par \textsf{\textbf{[Title]} De-interleaving gameplay data}
    }
    \\\noalign{\vskip 0.25ex}\cdashline{1-2}\noalign{\vskip 0.75ex}\textbf{\textsf{Top@30 Retrieved Document Meta Data}} & 
    \multicolumn{1}{p{.9\textwidth}}{
    \textsf{\textbf{[Title]} \textcolor{submission}{Methods and systems to increase interest in and viewership of content before, during and after a live event}}
    \par \textsf{\textbf{[Title]} \textcolor{submission}{Scaled VR engagement and views in an e-sports event}}
    \par \textsf{\textbf{[Title]} \textcolor{submission} {Creation of winner tournaments with fandom influence}}
    \par \textsf{\textbf{[Title]} \textcolor{submission}{Player to spectator handoff and other spectator controls}}
    \par \textsf{\textbf{[Title]} \textcolor{submission} {Discovery and detection of events in interactive content}}
    \par \textsf{\textbf{[Title]} Real-time modifications in augmented reality experiences}
    \par \textsf{\textbf{[Title]} \textcolor{submission}{User-driven spectator channel for live game play in multi-player games}}
    \par \textsf{\textbf{[Title]} Integrating commentary content and gameplay content over a multi-user platform}
    \par \textsf{\textbf{[Title]} \textcolor{submission}{Online tournament integration}}
    \par \textsf{\textbf{[Title]} \textcolor{submission}{Integrating augmented reality experiences with other components}}
    \par \textsf{\textbf{[Title]} AR-based connected portal shopping}
    \par \textsf{\textbf{[Title]} External screen streaming for an eyewear device}
    \par \textsf{\textbf{[Title]} Systems and methods for generating and facilitating access to a personalized augmented rendering of a user}
    \par \textsf{\textbf{[Title]} Systems, methods and apparatuses of digital assistants in an augmented reality environment and local determination of virtual object placement and apparatuses of single or multi-directional lens as portals between a physical world and a digital world component of the augmented reality environment}
    \par \textsf{\textbf{[Title]} \textcolor{submission}{Statistically defined game channels}}
    \par \textsf{\textbf{[Title]} AR/VR enabled contact lens}
    \par \textsf{\textbf{[Title]} Shared augmented reality unboxing experience}
    \par \textsf{\textbf{[Title]} \textcolor{submission}{De-interleaving gameplay data}}
    \par \textsf{\textbf{[Title]} \textcolor{submission}{Statistical driven tournaments}}
    \par \textsf{\textbf{[Title]} Automated augmented reality experience creation based on sample source and target images}
    \par \textsf{\textbf{[Title]} Augmented reality unboxing experience}
    \par \textsf{\textbf{[Title]} Real-time video dimensional transformations of video for presentation in mixed reality-based virtual spaces}
    \par \textsf{\textbf{[Title]} Systems and methods for pinning content items to locations in an augmented reality display based on user preferences}
    \par \textsf{\textbf{[Title]} Avatar customization system}
    \par \textsf{\textbf{[Title]} Controlling interactive fashion based on body gestures}
    \par \textsf{\textbf{[Title]} Systems, methods, and apparatus for enhanced headsets}
    \par \textsf{\textbf{[Title]} User interfaces for wide angle video conference}
    \par \textsf{\textbf{[Title]} User interface for multi-user communication session}
    \par \textsf{\textbf{[Title]} Real-time upper-body garment exchange}
    \par \textsf{\textbf{[Title]} Applying animated 3D avatar in AR experiences}
    }\\
    \bottomrule
    \end{tabular}
    }
    \captionsetup{justification=justified, singlelinecheck=false}
    \captionof{table}{Example of the retrieved documents for the input document with $\textsc{CoR}$ from a Citations split in \textsc{PatentFullBench}. Due to the length of input documents, we provide only the title and abstract for the input, as well as the titles for the retrieved documents. Retrieved documents that belong to ground truth are highlighted in blue.}
    \label{tab:retrieved_case_analysis_patents}
}
\vspace*{\fill}
\clearpage
\newtcolorbox{userboxstar8}[1][]{
  width=\linewidth,
  breakable,
  colback=teal!4,
  colframe=teal!50, 
  title={Query Case Analysis for Paper-to-Paper Retrieval},
  fonttitle=\bfseries,
  left=8pt, right=8pt, top=6pt, bottom=6pt,
}
  
\begin{userboxstar8}
\textbf{Document Meta-Data}\\
\textbf{Abstract:} We introduce GAIA, a benchmark for General AI Assistants that, if solved, would represent a milestone in AI research. GAIA proposes real-world questions that require a set of fundamental abilities such as reasoning, multi-modality handling, web browsing, and generally tool-use proficiency. GAIA questions are conceptually simple for humans yet challenging for most advanced AIs: we show that human respondents obtain 92\% vs. 15\% for GPT-4 equipped with plugins. This notable performance disparity contrasts with the recent trend of LLMs outperforming humans on tasks requiring professional skills in e.g. law or chemistry. GAIA’s philosophy departs from the current trend in AI benchmarks suggesting to target tasks that are ever more difficult for humans. We posit that the advent of Artificial General Intelligence (AGI) hinges on a system’s capability to exhibit similar robustness as the average human does on such questions. Using GAIA’s methodology, we devise 466 questions and their answer. We release our questions while retaining answers to 300 of them to power a leader-board hereby accessible.\\
 
\textbf{Method-Focused Query:}\par
{\leftskip=1em GAIA, a benchmark for General AI Assistants, presents a novel approach to evaluating the capabilities of large language models (LLMs) in performing real-world tasks that require a range of fundamental abilities, including reasoning, multi-modality handling, web browsing, and tool-use proficiency. The methodology employed in developing GAIA involves creating a set of 466 carefully crafted questions and their answers, covering various assistant use cases such as daily personal tasks, science, and general knowledge. Each question is designed to admit a short, single correct answer, enabling simple and robust automatic evaluation.\par
\medskip
To address the limitations of traditional LLM benchmarks, GAIA focuses on targeting tasks that are conceptually simple for humans yet challenging for most advanced AIs. The questions are intended to be easy to create, yet vary, root in the real world, and pose a unique combination of factual and inferential demands. Furthermore, GAIA's design principles prioritize interpretability, robustness against memorization, and easiness of use, aiming to provide a more nuanced evaluation framework for assessing the capabilities of LLMs.\par
\medskip
The development of GAIA involved several stages, starting with the creation of a seed set of questions by the authors themselves, followed by extensive annotation and validation phases involving human curators and evaluators. Two independent annotators assessed each question for ambivalence, and discrepancies were resolved through iterative refinements. This rigorous process ensured that the questions remained unambiguous and suitable for evaluation.\par
\medskip
Upon releasing the benchmark, GAIA evaluates LLMs' performance across various capabilities, including web browsing, multi-modality processing, coding, and diverse filetype reading. The evaluation procedure leverages a zero-shot prompt strategy, utilizing a prefix prompt to guide the model toward producing a factual answer. Results indicate significant disparities between human and LLM performances, highlighting the need for more sophisticated evaluation frameworks that account for the complexities inherent in real-world scenarios.\par
\medskip
Notably, GAIA's methodology diverges from previous benchmarks by focusing on the overall performance of AI assistants rather than individual tasks or narrow skill sets. This shift underscores the importance of considering the broader cognitive landscape encompassed by general intelligence, particularly in light of the growing emphasis on achieving true Artificial General Intelligence (AGI). Future directions for GAIA entail exploring extensions to accommodate emerging topics, incorporating human-in-the-loop feedback, and investigating novel methodologies for tackling issues such as linguistic and cultural diversity.\par
\medskip
Overall, GAIA represents a pioneering endeavor in establishing a standardized benchmark for evaluating the capabilities of General AI Assistants, offering valuable insights into the strengths and weaknesses of contemporary LLM architectures. By fostering a deeper understanding of the intricacies underlying effective AI assistance, GAIA paves the way for continued innovation in the pursuit of AGI.\par}
\medskip
\medskip
\textbf{Experiment-Focused Query:}\par
{\leftskip=1em Here's a detailed summary of the experiment setup, datasets, evaluation metrics, baselines, and key findings of the paper "GAIA: A Benchmark for General AI Assistants":\par\medskip}

{\leftskip=1em \textbf{Experiment Setup:}\par}
{\leftskip=2em The authors propose a novel benchmark called GAIA (General AI Assistant) consisting of 466 carefully crafted questions and their answers. Each question requires a series of steps to arrive at a factual answer, mimicking real-world scenarios where AI assistants interact with humans. The questions cover various domains, including science, literature, and everyday life, and are designed to be conceptually simple yet challenging for current AI systems.\par\medskip}

{\leftskip=1em \textbf{Datasets:}\par}
{\leftskip=2em The authors develop a custom dataset of 466 questions, divided into three levels of difficulty: Level 1 (simplest), Level 2 (moderate), and Level 3 (most challenging). The questions are paired with their answers, reasoning traces, and metadata describing the required capabilities and tools.\par\medskip}

{\leftskip=1em \textbf{Evaluation Metrics:}\par}
{\leftskip=2em The authors employ a combination of human evaluation and automated scoring to assess the performance of AI systems on the GAIA benchmark. Specifically:\par\medskip}

{\leftskip=2em \textbf{* Human Score:} \par}
{\leftskip=3em The authors annotate a subset of questions with their expected answers and measure the similarity between the predicted answer and the true answer.\par\medskip}

{\leftskip=2em \textbf{* Automated Scoring:} \par} 
{\leftskip=3em The authors implement a scoring mechanism that evaluates the correctness of the predicted answer based on the presence of certain keywords and phrases in the answer.\par\medskip}

{\leftskip=1em \textbf{Baselines:} \par}
{\leftskip=2em The authors compare the performance of several popular large language models (LLMs) on the GAIA benchmark, including GPT-4, AutoGPT, and human evaluators. The baseline performances serve as a reference point for assessing the effectiveness of the proposed GAIA benchmark.\par\medskip}

{\leftskip=1em \textbf{Key Findings:}\par\medskip}

{\leftskip=2em * Despite achieving impressive results on various NLP tasks, current LLMs struggle significantly on the GAIA benchmark, particularly on Level 3 questions.\par}
{\leftskip=2em * The authors observe that equipping LLMs with external tools and APIs can significantly enhance their performance on the GAIA benchmark, highlighting the importance of integrating external resources into AI systems.\par}
{\leftskip=2em * The GAIA benchmark demonstrates the limitations of traditional evaluation metrics, such as perplexity and BLEU score, in capturing the nuances of human-AI interaction.\par}
{\leftskip=2em The authors conclude that solving the GAIA benchmark represents a significant milestone toward developing truly general-purpose AI assistants that can effectively assist humans in a wide range of contexts.\par\medskip}

{\leftskip=1em Overall, the GAIA benchmark offers a comprehensive framework for evaluating the capabilities of AI systems in simulating real-world human-AI interactions, emphasizing the importance of integrating external resources, improving reasoning and decision-making, and enhancing overall human-AI alignment.\par}
\medskip
\medskip
\textbf{Research-Question Focused Query:}\par
{\leftskip=1em Here's a detailed explanation of the scientific paper, focusing on its motivation, research questions, and contributions: \par}
\medskip
{\leftskip=1em \textbf{Motivation}: \par}
{\leftskip=2em The paper introduces a novel benchmark called GAIA (General AI Assistant) aimed at assessing the capabilities of General AI Assistants. The author argues that traditional benchmarks, such as GLUE and MMLU, are insufficient for evaluating AI systems' ability to generalize across various tasks and domains. Current benchmarks tend to focus on narrow tasks, whereas real-world applications require AI systems to tackle complex, multifaceted problems. GAIA addresses this limitation by presenting a comprehensive framework for evaluating AI assistents' reasoning, multi-modality handling, web browsing, and tool-use proficiency.\par}
\medskip
{\leftskip=1em \textbf{Research Questions:}\par}
{\leftskip=2em The primary research question driving this paper is: "Can we develop a benchmark that comprehensively evaluates the capabilities of General AI Assistants?" Specifically, the author seeks to investigate:\par}

{\leftskip=3em * Can AI systems generalize across various tasks and domains?\par}
{\leftskip=3em * How effective are current benchmarks in capturing the complexities of real-world applications?\par}

{\leftskip=2em To address these questions, the author designs and implements GAIA, a benchmark comprising 466 carefully crafted questions covering various aspects of AI assistance, including reasoning, multi-modality handling, web browsing, and tool-use proficiency.\par}
\medskip
{\leftskip=1em \textbf{Contributions}: The paper contributes several significant ideas to the field of Natural Language Processing (NLP):\par}

{\leftskip=2em 1. \textbf{Comprehensive Benchmark for General AI Assistants}: GAIA offers a standardized framework for evaluating AI assistents' capabilities, filling the gap left by existing benchmarks.\par}
{\leftskip=2em 2. \textbf{Realistic Task Set}: The benchmark features questions that mimic real-world scenarios, encouraging AI systems to apply their knowledge and skills in context-dependent settings. \par}
{\leftskip=2em 3. \textbf{Focus on Reasoning and Multi-Modality Handling}: GAIA emphasizes the importance of reasoning and multi-modality handling in AI assistents, highlighting areas where current models fall short. \par}
{\leftskip=2em 4. \textbf{Evaluation Methodology}: The paper presents a systematic approach to evaluating AI assistents, incorporating human evaluation, model-based evaluation, and automatic scoring techniques. \par}
\medskip
{\leftskip=1em Overall, the paper demonstrates the feasibility of developing a comprehensive benchmark for General AI Assistants, paving the way for improved evaluation methodologies and more sophisticated AI systems.}
…
\end{userboxstar8}
\vspace{-0.15in}
{\captionsetup{justification=justified, singlelinecheck=false}
\captionof{figure}{Query Case Analysis for Scientific Paper Retrieval}
\label{app:fig:Query_Case_Analysis}
}
\newtcolorbox{userboxstar9}[1][]{
  width=\linewidth,
  breakable,
  colback=purple!5,
  colframe=purple!40,
  title={Query Case Analysis for Patent-to-Patent Retrieval},
  fonttitle=\bfseries,
  left=8pt, right=8pt, top=6pt, bottom=6pt,
}

\begin{userboxstar9}

\textbf{Document Meta-Data}\\
\textbf{Abstract:} The present disclosure describes methods and systems directed towards providing scaled engagement and views of an e-sports event. Instead of providing the same distribution of live e-sport event data to all remote viewers of a live e-sports event, features associated with e-sports gaming network could be used to customize the distribution of live e-sport event data to promote immersive viewer experience. The enhanced immersion can also be carried out in a virtual reality or augmented reality setting. The features would be capable of providing additional information, different views, and a variety of different commentators for the e-sports event so that the viewer can be more engaged when viewing the particular e-sports event. With the increased engagement from remote viewers, the distribution of live e-sports event data can also be further modified for monetization by incorporating advertisements as well.\par
\medskip
\textbf{Method-Focused Query:} The disclosed method for providing scaled engagement and views of a live electronic sports (e-sport) event involves a systematic approach to enhance viewer immersion through customized data distribution. Initially, the method collects comprehensive information from the live e-sport event, including in-game elements, player statistics, and audience reactions. This data is processed to generate detailed statistics about the players, which are then used to create a dynamic display of the event from an in-game camera perspective. The generated display is further modified based on individual viewer preferences, which may include adjustments for augmented reality (AR) environments, different camera angles, and the integration of advertisements. The modifications are tailored to enhance the viewer's experience by providing relevant statistics, commentary, and interactive features, such as the ability to follow specific players or engage with other viewers. The final modified display is distributed to the viewer's device, ensuring a personalized and immersive viewing experience that can also facilitate monetization through targeted advertisements. The system architecture includes an event server for data collection, a gaming server for processing and display generation, and a public network for distribution, all of which work together to deliver a rich, engaging e-sport viewing experience. \par
\medskip
\textbf{Claim-Focused Query:} The claims of this patent outline a method and system for enhancing viewer engagement and experience during live electronic sports (e-sports) events through customized data distribution. The independent claim details a method that includes collecting event data, generating player statistics, and creating a display that presents the event from an in-game camera perspective. This display is then modified based on viewer preferences, which may include augmented reality integration and personalized camera angles, before being distributed to remote viewers. The dependent claims specify refinements such as the inclusion of game-related information from player devices, audience reactions, advertisement incorporation, and user profile-based modifications that tailor the viewing experience to individual preferences. Additional claims address the provision of multiple viewing perspectives, interactive functions for viewers, and audio from other viewers, enhancing the immersive experience. The system claims describe an architecture comprising an event server and a gaming server that collectively manage data collection, processing, and distribution, while also allowing for the integration of advertisements and user interactions. Overall, the claims provide broad protection for methods and systems that facilitate a customizable and engaging viewing experience for remote participants in e-sports events.\par
\medskip
\textbf{Background Focused Query:} The present invention pertains to the field of electronic sports (e-sports), specifically focusing on enhancing viewer engagement and experience during live e-sports events through customized data distribution. Existing methods of broadcasting e-sports events typically provide a uniform viewing experience to all remote viewers, lacking personalization and immersive features that could significantly enhance engagement. Current streaming platforms do not adequately leverage the wealth of data generated during live competitions, such as player statistics, audience reactions, and in-game dynamics, which limits the depth of viewer interaction and understanding of the event. Furthermore, the integration of advanced technologies like virtual reality (VR) and augmented reality (AR) remains underutilized, preventing viewers from experiencing events in a more immersive manner akin to being physically present. There is a pressing need for systems that can dynamically modify the presentation of e-sports data based on individual viewer preferences, including customizable perspectives, additional commentary, and interactive features, while also addressing challenges related to monetization through targeted advertisements. Operating constraints include the need for real-time data processing, compatibility with various user devices, and the ability to handle high viewer throughput without compromising latency or quality. The application context spans competitive gaming leagues, online streaming services, and event organizers, all of whom seek to enhance viewer satisfaction and engagement while navigating the complexities of data management and user interaction in a rapidly evolving digital landscape.\par
\label{case_study:query_case_analysis_patents}
\end{userboxstar9}
\vspace{-0.15in}
{\captionsetup{justification=justified, singlelinecheck=false}
\captionof{figure}{Query Case Analysis for Patent-to-Patent Retrieval}
\label{app:fig:Query_Case_Analysis_Patent}
}
%\twocolumn
\section{Human Evaluation Guideline}
In this section, we provide the guidelines for human evaluation provided to annotators in Table~\ref{tab:Human_Evaluation}.
\label{appendix:HumanEvaluation}
\newtcolorbox{userboxstar13}[1][]{
  width=\linewidth,
  breakable,
  colback=black!1!white,
  colframe=black!55!black,
  colbacktitle=black!20!black,
  coltitle=white,
  title={Human Evaluation Guideline},
  fonttitle=\bfseries,
  boxrule=0.5pt, arc=2.5pt,
  left=8pt, right=8pt, top=6pt, bottom=6pt,
}

\begin{userboxstar13}
\textbf{1. Evaluation Guidelines}\par
Each task contains Title, Abstract information of Input Query Document, and has 3 different randomly shuffled retrieval variants with 5 retrieved documents with their titles and abstracts respectively.
You will annotate each Query–Document pair with \textbf{EXACTLY ONE} of the following labels:\par
\medskip
\textbf{Method}\par
\textbf{Experiment}\par
\textbf{Research Question/Motivation}\par
\textbf{Irrelevant}\par
\medskip
A total of 15 retrieved documents must be labeled per Input paper.\par
\medskip
The evaluation unit is one retrieved document for one query.\par
For each retrieved document, Exactly One label can be assigned.\par
\medskip
\textbf{2. Evaluation Instruction} \par
{\leftskip=1em \textbf{1. Evaluation Criteria}\par}
{\leftskip=2em \textbf{1.1 Research Question Definition:}\par}

{\leftskip=3em Choose this label when the retrieved document addresses the same or a highly similar research problem, task, or scientific question as the query.\par}
\medskip
{\leftskip=3em \textbf{Criteria:} \par}
{\leftskip=3em Same or very similar task/problem/research objectives
Overlapping motivation or problem definition
High-level goals align strongly
Do not consider overly vague and broad similarities as Motivation-Wise Related.\par}
\medskip
{\leftskip=3em \textbf{Examples:}\par}
\medskip
{\leftskip=3em \textbf{Correct:} Both works aim to address the fundamental challenge of retrieving relevant information from long-context documents, focusing on the issues of efficiently utilizing the important content within the query and candidate documents, and mitigate the effect of irrelevant noise information.\par}
\medskip
{\leftskip=3em \textbf{Incorrect:} Both papers focus on Large Language Models.\par}
\medskip
{\leftskip=2em \textbf{1.2 Method
Definition:}\par}

{\leftskip=3em Choose this label when the retrieved document is relevant mainly because it provides similar methodology, to solve its respective problem. Focus on the similarity between Algorithms, Model Architectures, Training Methods, Optimization Strategies, or Theoretical Frameworks.\par}
\medskip
{\leftskip=3em \textbf{Criteria:} \par}
\medskip
{\leftskip=3em 1) Similar algorithm or model architecture \par}
{\leftskip=3em 2) Methodological or optimization insights overlap \par}
{\leftskip=3em 3) Technical design relevant to the query \par}
{\leftskip=3em 4) Try to Avoid Overly Vague Methodological Similarities \par}
\medskip
{\leftskip=3em \textbf{Examples:} \par}
\medskip
{\leftskip=3em \textbf{Correct} Although the query document A focuses on research agents for solution generation and the retrieved document B addresses automated web agents, their core methodological frameworks are highly similar. Both frameworks implement a training-free, self-evolving memory architecture in which past experiences with the highest rewards are stored during an offline stage and later retrieved during inference. This mechanism enables the policy model to generalize to unseen tasks and effectively perform a form of curriculum learning through experience reuse.\par}
\medskip
{\leftskip=3em \textbf{Incorrect} Both Paper A and Paper B leverage large language models as autonomous agents and employ multi-agent frameworks to address their respective problem settings.\par}
\medskip
{\leftskip=2em \textbf{1.3 Experiment
Definition:}\par}
\medskip
{\leftskip=3em Choose this label when the retrieved document is primarily relevant due to its experimental framework: datasets, benchmarks, evaluation methodologies, or baseline setups.\par}
\medskip
{\leftskip=3em \textbf{Criteria:}\par}
\medskip
{\leftskip=3em 1) Introduces a similar dataset or benchmark used in the query \par}
{\leftskip=3em 2) Uses similar evaluation protocol \par}
{\leftskip=3em 3) Provides relevant baselines or experiment structures\par}
{\leftskip=3em 4) Try to Avoid Overly Vague Experimental Similarities\par}
\medskip
{\leftskip=3em \textbf{Examples:}\par}
\medskip
{\leftskip=3em \textbf{Correct} Both paper A and paper B attempts to solve the long-horizon forgetting problem of automated Agents with different methodologies and research objectives. Both paper A and paper B performs evaluation on OsWorld benchmark and VisualWebArena benchmark, while CoACT and UITars as their strong baselines.\par}
\medskip
{\leftskip=3em \textbf{Incorrect} Paper A and Paper B both utilize benchmarks to evaluate agents.\par}
\medskip
{\leftskip=2em \textbf{1.4 Irrelevant Definition:}\par}
\medskip
{\leftskip=3em Choose this label when the retrieved document does not meaningfully relate to the query in terms of research question, method, or experimental setup.\par}
\medskip
{\leftskip=3em \textbf{Criteria:} \par}
\medskip
{\leftskip=3em 1) Only superficial keyword overlap (e.g., both mention “LLM”)\par}
{\leftskip=3em 2) Different domain or unrelated task\par}
{\leftskip=3em 3) No substantial conceptual or technical relevance\par}
\medskip
{\leftskip=3em \textbf{Examples:}\par}
\medskip
{\leftskip=3em Vision Language Model layer compression for GUI Agents for efficiency improvement <-> Preference Optimization for Vision Language Models\par}
\medskip
{\leftskip=1em \textbf{2. How to Resolve Ambiguous Cases
When a retrieved document could belong to multiple categories, use the following priority:}\par}
\medskip
{\leftskip=2em \textbf{Phase 1: Check the aspect that aligns strongly among three given aspects.}\par}
\medskip
{\leftskip=3em \textbf{Research Question} → if the research problem aligns strongly\par}
{\leftskip=3em \textbf{Method} → if the alignment is mainly methodological\par}
{\leftskip=3em \textbf{Experiment} → if the relevance is primarily datasets/benchmarks/evaluation\par}
\medskip
{\leftskip=2em \textbf{Phase 2
Irrelevant → if none of the above apply}\par}
\medskip
{\leftskip=1em \textbf{3. Quality Checklist
Before submitting, verify the following:}\par}
\medskip
{\leftskip=2em Exactly one label selected per document\par}
{\leftskip=2em Strong task alignment → \textbf{Research Question}\par}
{\leftskip=2em Methodological overlap → \textbf{Method}\par}
{\leftskip=2em Dataset/evaluation relevance → \textbf{Experiment}\par}
{\leftskip=2em Unrelated documents → \textbf{irrelevant}\par}
{\leftskip=2em There are duplicate candidate documents across variants. Please make sure that the labels are applied consistently across all query, document pairs.\par}

\end{userboxstar13}
\vspace{-0.15in}
\captionof{figure}{Human Evaluation Guideline}
\label{app:fig:Human_Evalaution_Guideline}
\section{Prompts}
Here, we provide prompts used for experiments.
\label{appendix:Prompts}
\newtcolorbox{userboxstar1}[1][]{     
  width=\linewidth,
  breakable,
  colback=green!4!white,
  colframe=green!55!black,
  colbacktitle=green!20!white,
  coltitle=black,
  title={Method-Focused Agent Prompt for Paper-to-Paper Retrieval},
  fonttitle=\bfseries,
  boxrule=0.5pt, arc=2.5pt,
  left=8pt, right=8pt, top=6pt, bottom=6pt,
}
 
%\onecolumn

\begin{userboxstar1} 
\label{box:sci_method}
\textbf{Instruction:}\par
You are a specialized research assistant tasked with generating a structured, detailed explanation of a scientific paper based on its \textbf{Methodology}. Your goal is to provide a clear yet comprehensive summary that makes it easy to identify relevant papers by emphasizing the \textbf{methodology} of given paper.\par
\medskip
\textbf{IMPORTANT}\par
{\leftskip=1em - Your explanation is going to be used as a query to retrieve similar papers \textbf{METHOD} wise.\par} 
{\leftskip=1em - Make sure that your explanation can retrieve highly relevant papers easily.\par}
{\leftskip=1em - There are also other agents who are tasked with generating explanation on given paper. Unlike you, they are focused on experiments, and research questions of given paper. You must try to avoid overlap with possible explanations that the other two agents might generate.\par}
\medskip
\textbf{Input:} \par
You will be given the full text of a scientific paper. Carefully analyze its content, with a particular focus on the \textbf{METHODOLOGY} section, to extract its main approaches.\par
\medskip

\textbf{Key Considerations:} \par
Highlight specific method/approach details, avoiding vague or overly general descriptions. Use precise language to ensure clarity while maintaining depth.\par
\medskip

\textbf{Output Format:}\par
Generate a well-structured, detailed and yet clear paragraph that effectively captures the paper’s approach, and key concepts in a concise yet informative manner. Focus on high-level insights rather than excessive detail.\par

You must not include title and abstract of given paper in your answer, and try to put it into your own words with high level reasoning after reading the paper.

\end{userboxstar1}
\vspace{-0.2in}
{
\captionsetup{justification=justified, singlelinecheck=false}
\captionof{figure}{Method-Focused Agent Prompt for Paper-to-Paper Retrieval}
\label{app:fig:method_focused_query_prompt}
}

\newtcolorbox{userboxstar2}[1][]{
  width=\linewidth,
  breakable,
  colback=blue!4!white,
  colframe=blue!55!black,
  colbacktitle=blue!20!white,
  coltitle=black,
  title={Experiment-Focused Agent Prompt for Paper-to-Paper Retrieval},
  fonttitle=\bfseries,
  boxrule=0.5pt, arc=2.5pt,
  left=8pt, right=8pt, top=6pt, bottom=6pt,
}
%\vspace*{\fill}  \
\begin{userboxstar2}
You are a specialized research assistant tasked with generating a structured, detailed explanation of a scientific paper’s experimental setup. \par
Your goal is to clearly outline the \textbf{datasets, evaluation metrics, baselines, and key experimental findings}, making it easy to understand how the paper validates its approach.\par
\medskip

\textbf{IMPORTANT}\par
{\leftskip=1em - Your explanation is going to be used as a query to retrieve similar papers \textbf{EXPERIMENT} wise.\par}
{\leftskip=1em - Make sure that your explanation can retrieve highly relevant papers easily.\par}
{\leftskip=1em - There are also other agents who are tasked with generating explanation on given paper. Unlike you, they are focused on methods, and research questions of given paper. You must try to avoid overlap with possible explanations that the other two agents might generate.\par}
\medskip
\textbf{Input:}\\
You will be provided with the full text of a scientific paper. Carefully analyze its content, paying particular attention to the \textbf{Experiments, Results, and Evaluation} sections to extract the key experimental details.\\

\textbf{Key Considerations:}

{\leftskip=1em \textbf{Datasets \& Benchmarks:} Clearly specify the datasets and benchmarks used for evaluation.\par}
{\leftskip=1em \textbf{Baselines \& Comparisons:} Identify what methods or models the paper compares against.\par}
{\leftskip=1em \textbf{Key Results \& Insights:} Summarize the main experimental findings without excessive detail.\par}
\medskip

\textbf{Output Format:}\\
Generate a clear, well structured and detailed paragraph that highlights the experimental methodology, datasets, evaluation metrics, baselines, and key results. Focus on high-level insights rather than excessive detail.\\

You must not include title and abstract of given paper in your answer, and try to put it into your own words with high level reasoning after reading the paper.

\end{userboxstar2}
{
\vspace{-0.2in}
\captionsetup{justification=justified, singlelinecheck=false}
\captionof{figure}{Experiment-Focused Agent Prompt for Paper-to-Paper Retrieval}
\label{app:fig:experiment_focused_query_prompt}
}
%\vspace*{\fill}     
%\clearpage 
\newtcolorbox{userboxstar3}[1][]{
  width=\linewidth,
  breakable,
  colback=red!4!white,
  colframe=red!55!black,
  colbacktitle=red!20!white,
  coltitle=black,
  title={Research Question-Focused Agent Prompt for Paper-to-Paper Retrieval},
  fonttitle=\bfseries,
  boxrule=0.5pt, arc=2.5pt,
  left=8pt, right=8pt, top=6pt, bottom=6pt,
}

\begin{userboxstar3}
\textbf{Instruction:}\\
You are a specialized research assistant tasked with generating a structured, detailed explanation of a scientific paper based on its \textbf{Motivation, Research Questions, and Contributions}. Your goal is to provide a clear yet comprehensive summary that makes it easy to identify relevant papers by emphasizing the \textbf{core problem, key contributions, and research objectives.}\par
\medskip

\textbf{IMPORTANT}\par
{\leftskip=1em - Your explanation is going to be used as a query to retrieve similar papers \textbf{RESEARCH QUESTION} wise.\par}
{\leftskip=1em - Make sure that your explanation can retrieve highly relevant papers easily.\par}
{\leftskip=1em - There are also other agents who are tasked with generating explanation on given paper. Unlike you, they are focused on experiments, and methods of given paper. You must try to avoid overlap with possible explanations that the other two agents might generate.\par}
\medskip

\textbf{Input:}\\
You will be given the full text of a scientific paper. Carefully analyze its content, with a particular focus on the Introduction and Conclusion, to extract its main contributions, research questions, and motivations.\\

\textbf{Key Considerations:}\\
Highlight specific motivations, research questions, and contributions, avoiding vague or overly general descriptions.\\
Use precise language to ensure clarity while maintaining depth.\\
  
\textbf{Output Format:}
Generate a well-structured, detailed and yet clear paragraph that effectively captures the paper’s motivation, problem statement, research questions, and key contributions in a concise yet informative manner. Focus on high-level insights rather than excessive detail\\

You must not include title and abstract of given paper in your answer, and try to put it into your own words with high level reasoning after reading the paper.\\
  
\end{userboxstar3}
{
\vspace{-0.2in}
\captionsetup{justification=justified, singlelinecheck=false}
\captionof{figure}{Research Question-Focused Agent Prompt for Paper-to-Paper Retrieval}
\label{app:fig:research_question_focused_query_prompt}
}
\newtcolorbox{userboxstar5}[1][]{     
  width=\linewidth,
  breakable,
  colback=orange!4!white,
  colframe=orange!55!black,
  colbacktitle=orange!20!white,
  coltitle=black,
  title={Method-Focused Agent Prompt for Patent-to-Patent Retrieval},
  fonttitle=\bfseries,
  boxrule=0.5pt, arc=2.5pt,
  left=8pt, right=8pt, top=6pt, bottom=6pt,
}

%\vspace*{\fill}
\begin{userboxstar5}  
You are a specialized research assistant tasked with generating a structured, detailed explanation of a patent’s \textbf{METHOD}.\\ 

Your goal is to summarize how the invention works in practice: its core technical principles, implementation procedures, system architecture, and functional mechanisms. This explanation will be used as a query to retrieve patents with similar methods or implementation techniques.\\

\textbf{IMPORTANT}\par
{\leftskip=1em - Focus \textbf{ONLY} on the Detailed Description and Embodiments sections.\par}
{\leftskip=1em - Capture the technical processes, structures, system flows, or algorithms.\par}
{\leftskip=1em - Paraphrase at a high-level while keeping enough technical detail for retrieval.\par}
{\leftskip=1em - Other agents will generate explanations about the invention claims and the background/problems of the given patent. You must avoid overlap with those aspects.\par}
\medskip

\textbf{Input}\\
You will be provided with the full text of a patent. Carefully analyze its Detailed Description, Examples, and Figures to extract the methodological details.\par
\medskip

\textbf{Key Considerations}\par
Here are some key aspects that you may focus on.\\
{\leftskip=1em \textbf{- Core technical principle:} what mechanism enables the invention to function?\par}
{\leftskip=1em \textbf{- Implementation structure:} key components, modules, or subsystems.\par}
{\leftskip=1em \textbf{- Operational flow:} how the method proceeds step by step.\par}  
{\leftskip=1em \textbf{- Variants or embodiments:} different configurations or modes of execution.\par}  
{\leftskip=1em \textbf{- Integration context:} how it interacts with external systems or environments.\par}
\medskip

\textbf{Output Format}\\
Produce a single, clear, and well-structured paragraph that captures the \textbf{METHOD} of the invention. Write in concise, retrieval-friendly technical language that highlights implementation strategies and system functionality.\\

\end{userboxstar5}
{
\vspace{-0.15in}
\captionsetup{justification=justified, singlelinecheck=false}
\captionof{figure}{Method-Focused Agent Prompt for Patent-to-Patent Retrieval}
\label{app:fig:patent_method_focused_query_prompt}
}
%\vspace*{\fill}
%\clearpage  
\newtcolorbox{userboxstar6}[1][]{     
  width=\linewidth,
  breakable,
  colback=pink!4!white,
  colframe=pink!55!black,
  colbacktitle=pink!20!white,
  coltitle=black,
  title={Claim-Focused Agent Prompt for Patent-to-Patent Retrieval},
  fonttitle=\bfseries,
  boxrule=0.5pt, arc=2.5pt,
  left=8pt, right=8pt, top=6pt, bottom=6pt,
}

%\vspace*{\fill}
\begin{userboxstar6}  
You are a specialized research assistant tasked with generating a structured, detailed explanation of a patent’s \textbf{CLAIMS}.\\  
  
Your goal is to clearly outline the legal protection scope: what is being claimed, how broad the claims are, and what technical features or components are covered.\\  
This explanation will be used as a query to retrieve patents with similar \textbf{CLAIM} structures and protection scopes.\\

\textbf{IMPORTANT}\par
{\leftskip=1em - Focus \textbf{ONLY} on the \textbf{CLAIMS} section.\\  
- Emphasize the scope of protection, claimed components, processes, and relationships among them.\par}
{\leftskip=1em - Avoid background information, prior art, or detailed embodiments (those are handled by other agents).\par}  
{\leftskip=1em - Do not directly copy claim sentences; paraphrase into concise, high-level, retrieval-friendly language.\par}  
{\leftskip=1em - Include both the broad independent claims (core invention scope) and notable dependent claims (specific refinements).\par}  
{\leftskip=1em - Other agents will generate explanations about the invention \textbf{details/method} and the \textbf{background/problems} of the given patent. You must avoid overlap with those aspects.\par}
\medskip

\textbf{Input}\\
You will be provided with the full text of a patent. Carefully analyze the \textbf{CLAIMS} section.\\

\textbf{Output Format}\\
Produce a clear, well-structured paragraph that captures the scope and focus of the \textbf{CLAIMS}. Use neutral, technical language suitable for retrieval so that patents with overlapping protection scopes can be surfaced.\\

\end{userboxstar6}

{
\vspace{-0.2in}
\captionsetup{justification=justified, singlelinecheck=false}
\captionof{figure}{Claim-Focused Agent Prompt for Patent-to-Patent Retrieval}
\label{app:fig:patent_claim_focused_query_prompt}
}
%\vspace*{\fill}
%\clearpage  
\newtcolorbox{userboxstar7}[1][]{     
  width=\linewidth,
  breakable,
  colback=purple!4!white,
  colframe=purple!55!black,
  colbacktitle=purple!20!white,
  coltitle=black,
  title={Background-Focused Agent Prompt for Patent-to-Patent Retrieval},
  fonttitle=\bfseries,
  boxrule=0.5pt, arc=2.5pt,
  left=8pt, right=8pt, top=6pt, bottom=6pt,
}

%\vspace*{\fill}
\begin{userboxstar7}  
You are a specialized research assistant tasked with generating a structured, detailed explanation of a patent’s \textbf{Problem / Background} (i.e., the technical field and the shortcomings of prior art).\\ 

Your goal is to clearly state the problem space—technical domain, prior-art limitations, unresolved challenges, operating constraints, and desired (non-solution) performance objectives—so that similar patents can be retrieved by shared problem patterns rather than specific solutions.\\

\textbf{IMPORTANT}\par
{\leftskip=1em - Your explanation is going to be used as a query to retrieve patents that are similar in the \textbf{PROBLEM SPACE.}\par}
{\leftskip=1em - Include concrete domain terms (components, materials, data types), relevant standards/regulations, operating conditions (e.g., throughput, latency, power, temperature), failure modes, and application contexts that characterize the problem.\par}
{\leftskip=1em - Other agents will generate explanations about the invention details/method and the claims made from the input patent. You must avoid overlap with those aspects.\par}
\medskip

\textbf{Input}\\
You will be provided with the full text of a patent.\\

\textbf{Key Considerations}\par
Here are some key aspects that you may focus on.\\
{\leftskip=1em - \textbf{Technical Field:} the domain and subdomain (e.g., “wireless edge inference for medical imaging”).\par}
{\leftskip=1em - \textbf{Prior Art \& Limitations:} concrete bottlenecks, inefficiencies, failure cases, safety/privacy concerns, interoperability issues.\par}
{\leftskip=1em - \textbf{Unmet Technical Objectives:} what must be achieved (targets or constraints).\par}
{\leftskip=1em - \textbf{Operating Constraints \& Edge Cases:} data distributions, environmental/thermal limits, network conditions, power/memory budgets, size/weight/cost constraints, lifecycle/maintenance issues.\par}
{\leftskip=1em - \textbf{Application Contexts \& Stakeholders:} deployment scenarios, industries, users, and integration boundaries.\par}
\medskip
\textbf{Output Format}\\
Generate a single, clear, well-structured paragraph in neutral technical language that captures \textbf{ONLY} the \textbf{Problem / Background}. Paraphrase in your own words at a high level; do not copy text or include any solution, embodiment, or claim content. The paragraph should be information-dense and retrieval-friendly so that it can surface patents that confront the same technical challenges.\\

\end{userboxstar7}

{
\vspace{-0.2in}
\captionsetup{justification=justified, singlelinecheck=false}
\captionof{figure}{Background-Focused Agent Prompt for Patent-to-Patent Retrieval}
\label{app:fig:patent_background_focused_query_prompt}
}
%\vspace*{\fill}
%\clearpage  
\newtcolorbox{userboxstar4}[1][]{
  width=\linewidth,
  breakable,
  colback=yellow!4!white,
  colframe=yellow!55!black,
  colbacktitle=yellow!20!white,
  coltitle=black,
  title={Base Agent Prompt for Paper-to-Paper Retrieval},
  fonttitle=\bfseries,
  boxrule=0.5pt, arc=2.5pt,
  left=8pt, right=8pt, top=6pt, bottom=6pt,
}

\begin{userboxstar4}
\textbf{Instruction:} You are a specialized research assistant tasked with generating a structured,detailed explanation of a scientific paper based three different aspects.\\
  
\textbf{1. Method-Specific Queries:}\par
{\leftskip=1em Generate a structured, detailed explanation of a scientific paper based on its Methodology Your goal is to provide a clear yet comprehensive summary that makes it easy to identify relevant papers by emphasizing the methodology of given paper.\par}
\medskip
{\leftskip=1em \textbf{IMPORTANT} \par}
{\leftskip=2em - Your explanation is going to be used as a query to retrieve similar papers METHOD wise.\par}
{\leftskip=2em - Make sure that your explanation can retrieve highly relevant papers easily.\par}
\medskip
{\leftskip=1em \textbf{Input:}\par}
{\leftskip=2em You will be given the full text of a scientific paper. Carefully analyze its content, with a particular focus on the \textbf{METHODOLOGY} section, to extract its main approaches.\par}
\medskip    
{\leftskip=1em \textbf{Key Considerations:} \par}
{\leftskip=2em Highlight specific method/approach details, avoiding vague or overly general descriptions. Use precise language to ensure clarity while maintaining depth.\par}
\medskip
{\leftskip=1em \textbf{Output Format:} \par}
{\leftskip=2em Generate a well-structured, detailed and yet clear paragraph that effectively captures the paper’s approach, and key concepts in a concise yet informative manner. Focus on high-level insights rather than excessive detail You must not include title and abstract of given paper in your answer, and try to put it into your own words with high level reasoning after reading the paper. \par}
\medskip
\textbf{2. Experiment-Specific Queries:}\par
{\leftskip=1em Generate a structured, detailed explanation of a scientific paper’s experimental setup Your goal is to clearly outline the datasets, evaluation metrics, baselines, and key experimental findings, making it easy to understand how the paper validates its approach.\par}
\medskip
{\leftskip=1em \textbf{IMPORTANT} \par}
{\leftskip=2em - Your explanation is going to be used as a query to retrieve similar papers \textbf{EXPERIMENT} wise.\par}
{\leftskip=2em - Make sure that your explanation can retrieve highly relevant papers easily.\par}
\medskip
{\leftskip=1em You will be provided with the full text of a scientific paper. Carefully analyze its content, paying particular attention to the Experiments, Results, and Evaluation sections to extract the key experimental details.\par}
\medskip
{\leftskip=1em \textbf{Key Considerations:}\par}
{\leftskip=2em \textbf{Datasets \& Benchmarks:} Clearly specify the datasets and benchmarks used for evaluation.\par}
{\leftskip=2em \textbf{Baselines \& Comparisons:} Identify what methods or models the paper compares against.\par}
{\leftskip=2em \textbf{Key Results \& Insights:} Summarize the main experimental findings without excessive detail.\par}
\medskip
{\leftskip=1em \textbf{Output Format:}\par}
{\leftskip=2em Generate a clear, well structured and detailed paragraph that highlights the experimental methodology, datasets, evaluation metrics, baselines, and key results. Focus on high-level insights rather than excessive detail. You must not include title and abstract of given paper in your answer, and try to put it into your own words with high level reasoning after reading the paper.\par}
\medskip
\textbf{3. Research Question-Specific Queries:}

{\leftskip=1em Generate a structured, detailed explanation of a scientific paper based on its Motivation, Research Questions, and Contributions. Your goal is to provide a clear yet comprehensive summary that makes it easy to identify relevant papers by emphasizing the core problem, key contributions, and research objectives.\par}
\medskip
{\leftskip=1em \textbf{IMPORTANT} \par}
{\leftskip=2em - Your explanation is going to be used as a query to retrieve similar papers \textbf{RESEARCH QUESTION} wise. \par}
{\leftskip=2em - Make sure that your explanation can retrieve highly relevant papers easily. \par}
\medskip
{\leftskip=1em \textbf{Input:} \par}

{\leftskip=2em You will be given the full text of a scientific paper. Carefully analyze its content, with a particular focus on the Introduction and Conclusion, to extract its main contributions, research questions, and motivations.\par}
\medskip
{\leftskip=1em \textbf{Key Considerations:} \par}

{\leftskip=2em Highlight specific motivations, research questions, and contributions, avoiding vague or overly general descriptions. Use precise language to ensure clarity while maintaining depth.\par}
\medskip    
{\leftskip=1em \textbf{Output Format:} \par}
{\leftskip=2em Generate a well-structured, detailed and yet clear paragraph that effectively captures the paper’s \quad motivation, problem statement, research questions, and key contributions in a concise yet informative manner. Focus on high-level insights rather than excessive detail You must not include title and abstract of given paper in your answer, and try to put it into your own words with high level reasoning after reading the paper.\par}
\medskip
\textbf{Output:}\par

{\leftskip=1em Return a structured json file for respective Method-Specific Queries, Experiment-Specific Queries, and Research Question-Specific queries, with the respective keys as "method\_query", "experiment\_query", and "research\_question\_query". Each key should contain the generated explanation as a string.\par}

\end{userboxstar4}
{
\vspace{-0.15in}
\captionsetup{justification=justified, singlelinecheck=false}
\captionof{figure}{Base Agent Prompt for Paper-to-Paper Retrieval}
\label{app:fig:base_query_prompt}
}

\newtcolorbox{userboxstar10}[1][]{
  width=\linewidth,
  breakable,
  colback=gray!4!white,
  colframe=gray!55!black,
  colbacktitle=gray!20!white,
  coltitle=black,
  title={Prompt for Single Comprehensive Query for Paper-to-Paper Retrieval},
  fonttitle=\bfseries,
  boxrule=0.5pt, arc=2.5pt,
  left=8pt, right=8pt, top=6pt, bottom=6pt,
}

\begin{userboxstar10}
\textbf{Instruction:} \\
You are a specialized research assistant tasked with generating a structured, detailed explanation of a scientific paper based three different aspects. You should generate a single query covering method, experiment, and research question aspects. Carefully read the below instructions and generate a single comprehensive query that covers below three aspects.\\
  
\textbf{1. Method-Specific Queries:}\par 
{\leftskip=1em Generate a structured, detailed explanation of a scientific paper based on its Methodology. Your goal is to provide a clear yet comprehensive summary that makes it easy to identify relevant papers by emphasizing the methodology of given paper.\par}
\medskip
{\leftskip=1em \textbf{IMPORTANT}\par}
{\leftskip=1em- Your explanation is going to be used as a query to retrieve similar papers \textbf{METHOD} wise.\par}
{\leftskip=1em- Make sure that your explanation can retrieve highly relevant papers easily.\par}
\medskip
{\leftskip=1em \textbf{Input:}\par}
{\leftskip=1em You will be given the full text of a scientific paper. Carefully analyze its content, with a particular focus on the \textbf{METHODOLOGY} section, to extract its main approaches.\par}
\medskip    
{\leftskip=1em \textbf{Key Considerations:}\par}
{\leftskip=1em Highlight specific method/approach details, avoiding vague or overly general descriptions. Use precise language to ensure clarity while maintaining depth.\par}
\medskip
{\leftskip=1em \textbf{Output Format:} \par}
{\leftskip=1em Generate a well-structured, detailed and yet clear paragraph that effectively captures the paper’s approach, and key concepts in a concise yet informative manner. Focus on high-level insights rather than excessive detail You must not include title and abstract of given paper in your answer, and try to put it into your own words with high level reasoning after reading the paper.\par}
\medskip
\textbf{2. Experiment-Specific Queries:}\par
{\leftskip=1em Generate a structured, detailed explanation of a scientific paper’s experimental setupYour goal is to clearly outline the datasets, evaluation metrics, baselines, and key experimental findings, making it easy to understand how the paper validates its approach.\par}
\medskip
{\leftskip=1em \textbf{IMPORTANT}\par}
{\leftskip=1em - Your explanation is going to be used as a query to retrieve similar papers \textbf{EXPERIMENT} wise.\par}
{\leftskip=1em - Make sure that your explanation can retrieve highly relevant papers easily.\par}
\medskip
{\leftskip=1em You will be provided with the full text of a scientific paper. Carefully analyze its content, paying particular attention to the Experiments, Results, and Evaluation sections to extract the key experimental details.\par}
\medskip
{\leftskip=1em \textbf{Key Considerations:}\par}
{\leftskip=1em \textbf{Datasets \& Benchmarks:} Clearly specify the datasets and benchmarks used for evaluation.\par}
{\leftskip=1em \textbf{Baselines \& Comparisons:} Identify what methods or models the paper compares against.\par}
{\leftskip=1em \textbf{Key Results \& Insights:} Summarize the main experimental findings without excessive detail.\par}
\medskip
{\leftskip=1em \textbf{Output Format:}\par}
{\leftskip=1em Generate a clear, well structured and detailed paragraph that highlights the experimental methodology, datasets, evaluation metrics, baselines, and key results. Focus on high-level insights rather than excessive detail. You must not include title and abstract of given paper in your answer, and try to put it into your own words with high level reasoning after reading the paper.\par}
\medskip
\textbf{3. Research Question-Specific Queries:}\par
{\leftskip=1em Generate a structured, detailed explanation of a scientific paper based on its Motivation, Research Questions, and Contributions. Your goal is to provide a clear yet comprehensive summary that makes it easy to identify relevant papers by emphasizing the core problem, key contributions, and research objectives.\par}
\medskip
{\leftskip=1em \textbf{IMPORTANT}\par}
{\leftskip=1em - Your explanation is going to be used as a query to retrieve similar papers \textbf{RESEARCH QUESTION wise.}\par}
{\leftskip=1em- Make sure that your explanation can retrieve highly relevant papers easily.\par}
\medskip
{\leftskip=1em \textbf{Input:}\par}
{\leftskip=1em You will be given the full text of a scientific paper. Carefully analyze its content, with a particular focus on the Introduction and Conclusion, to extract its main contributions, research questions, and motivations.\par}
\medskip
{\leftskip=1em \textbf{Key Considerations:}\par}
{\leftskip=1em Highlight specific motivations, research questions, and contributions, avoiding vague or overly general descriptions. Use precise language to ensure clarity while maintaining depth.\par}
\medskip    
{\leftskip=1em \textbf{Output Format:} Generate a well-structured, detailed and yet clear paragraph that effectively captures the paper’s motivation, problem statement, research questions, and key contributions in a concise yet informative manner. Focus on high-level insights rather than excessive detail You must not include title and abstract of given paper in your answer, and try to put it into your own words with high level reasoning after reading the paper.\par}
\medskip
\textbf{Output:}\par
{\leftskip=1em Return a comprehensive query that covers the Methodology, Research Questions, and Experimental Details within a single paper.\par}

\end{userboxstar10}
{
\vspace{-0.15in}
\captionsetup{justification=justified, singlelinecheck=false}
\captionof{figure}{Prompt for Single Comprehensive Query for Paper-to-Paper Retrieval}
\label{app:fig:single_comprehensive_query_prompt}
}
\twocolumn
%\clearpage

\end{document}